\documentclass[fleqn,usenatbib]{mnras}
\usepackage{newtxtext,newtxmath}
\usepackage{xcolor}

\usepackage[T1]{fontenc}

\DeclareRobustCommand{\VAN}[3]{#2}
\let\VANthebibliography\thebibliography
\def\thebibliography{\DeclareRobustCommand{\VAN}[3]{##3}\VANthebibliography}

\usepackage{graphicx}	
\usepackage{amsmath}	
\usepackage{threeparttable,booktabs}
\usepackage{float}
\usepackage{orcidlink}

\newcommand{\nwprp}{\bar{n}_2w_{{\rm{p}}}(r_{\rm{p}})}
\newcommand{\nwp}{\bar{n}_2w_{{\rm{p}}}}
\newcommand{\rp}{r_{\rm p}}
\newcommand{\n}{\bar{n}_2}
\newcommand{\Wp}{w_{{\rm{p}}}}
\newcommand{\ms}{{\rm M}_{\sun}}
\newcommand{\msh}{h^{-1}{\rm M}_{\sun}}
\newcommand{\mpch}{h^{-1}{\rm Mpc}}

\usepackage{xcolor}
\usepackage[normalem]{ulem} 

\definecolor{revadd}{RGB}{0,140,90}     
\definecolor{revdel}{RGB}{200,40,40}    
\definecolor{revcom}{RGB}{140,0,180}    

\newif\ifrev
\revtrue          

\ifrev
  
  \newcommand{\del}[1]{\textcolor{revdel}{\sout{#1}}}
  
  \newcommand{\com}[1]{\textcolor{revcom}{\textbf{[}#1\textbf{]}}}
\else
  
  \newcommand{\del}[1]{}
  
  \newcommand{\com}[1]{}
\fi

\usepackage[normalem]{ulem}

\defcitealias{2025MNRAS.540.1635X}{DESI PAC I}
\defcitealias{2022ApJ...925...31X}{SDSS PAC I}
\defcitealias{2023ApJ...944..200X}{SDSS PAC IV}
\defcitealias{2020MNRAS.498.4887B}{BF20}
\defcitealias{2023MNRAS.524.2290N}{NGM23}

\title[DESI Y1 SHMR]{PAC in DESI. II. Galaxy-halo connection into the $10^{6}{\rm M}_{\odot}$ frontier}

\author[Kun Xu et al.]{
\parbox{\textwidth}{
\Large
Kun Xu\orcidlink{0000-0002-7697-3306},$^{1,2,3}$\thanks{E-mail: kunxu@sas.upenn.edu}
Carlos~S.~Frenk\orcidlink{0000-0002-2338-716X},$^{2}$
Y.~P.~Jing\orcidlink{0000-0002-4534-3125},$^{3}$
Shaun~Cole\orcidlink{0000-0002-5954-7903},$^{2}$
Sownak~Bose\orcidlink{0000-0002-0974-5266},$^{2}$
J.~Aguilar,$^{4}$
S.~Ahlen\orcidlink{0000-0001-6098-7247},$^{5}$
D.~Bianchi\orcidlink{0000-0001-9712-0006},$^{6,7}$
D.~Brooks,$^{8}$
F.~J.~Castander\orcidlink{0000-0001-7316-4573},$^{9,10}$
T.~Claybaugh,$^{4}$
A.~de~la~Macorra\orcidlink{0000-0002-1769-1640},$^{11}$
P.~Doel,$^{8}$
J.~E.~Forero-Romero\orcidlink{0000-0002-2890-3725},$^{12,13}$
E.~Gazta\~naga\orcidlink{0000-0001-9632-0815},$^{9,14,10}$
S.~Gontcho~A~Gontcho\orcidlink{0000-0003-3142-233X},$^{15}$
G.~Gutierrez,$^{16}$
C.~Hahn\orcidlink{0000-0003-1197-0902},$^{17}$
R.~Joyce\orcidlink{0000-0003-0201-5241},$^{18}$
S.~Juneau\orcidlink{0000-0002-0000-2394},$^{18}$
R.~Kehoe,$^{19}$
T.~Kisner\orcidlink{0000-0003-3510-7134},$^{4}$
M.~Landriau\orcidlink{0000-0003-1838-8528},$^{4}$
L.~Le~Guillou\orcidlink{0000-0001-7178-8868},$^{20}$
M.~Manera\orcidlink{0000-0003-4962-8934},$^{21,22}$
R.~Miquel,$^{23,22}$
J.~Moustakas\orcidlink{0000-0002-2733-4559},$^{24}$
S.~Nadathur\orcidlink{0000-0001-9070-3102},$^{14}$
W.~J.~Percival\orcidlink{0000-0002-0644-5727},$^{25,26,27}$
F.~Prada\orcidlink{0000-0001-7145-8674},$^{28}$
I.~P\'erez-R\`afols\orcidlink{0000-0001-6979-0125},$^{29}$
G.~Rossi,$^{30}$
L.~Samushia\orcidlink{0000-0002-1609-5687},$^{31,32,33}$
E.~Sanchez\orcidlink{0000-0002-9646-8198},$^{34}$
D.~Schlegel,$^{4}$
J.~H.~Silber\orcidlink{0000-0002-3461-0320},$^{4}$
D.~Sprayberry,$^{18}$
G.~Tarl\'e\orcidlink{0000-0003-1704-0781},$^{35}$
B.~A.~Weaver,$^{18}$
and H.~Zou\orcidlink{0000-0002-6684-3997}$^{36}$
}
\vspace{0.4cm}
\\
\parbox{\textwidth}{Affiliations are listed in Appendix~\ref{sec:author}}
}

\date{Accepted XXX. Received YYY; in original form ZZZ}

\pubyear{\the\year{}}

\begin{document}
\label{firstpage}
\pagerange{\pageref{firstpage}-\pageref{lastpage}}
\maketitle

\begin{abstract}
Understanding dwarf galaxy formation is crucial for testing dark matter models and reionization physics. However, constructing stellar-mass complete spectroscopic samples at low masses is increasingly difficult, and the potential existence of a local void complicates studies of dwarf galaxies in an average environment. The Photometric object Around Cosmic webs (PAC) method, which combines deep photometric and spectroscopic data to measure the excess surface density $\bar{n}_2w_{{\rm{p}}}(r_{\rm{p}})$ of photometric objects around spectroscopic tracers, offers a promising path forward. We model 349 $\bar{n}_2w_{{\rm{p}}}(r_{\rm{p}})$ measurements from DESI Y1 BGS and DECaLS, reaching $M_*=10^{6.4}\,{\rm M}_{\odot}$, using a stellar mass–halo mass relation (SHMR)-based subhalo abundance matching framework applied to two high-resolution $N$-body simulations from the Jiutian suite. The resulting SHMR is constrained down to $M_{\rm h}\simeq10^{8.0}\,h^{-1}{\rm M}_{\odot}$, revealing a clear upturn at $\sim10^{10.0}\,h^{-1}{\rm M}_{\odot}$ toward lower masses, indicating rising star-formation efficiency (SFE) in small haloes. This feature persists under extensions of the model that allow mass-dependent scatter, reionization-induced suppression of the halo occupation fraction, galaxy assembly bias, and alternative cosmologies. Combining $f_{\rm sat}(M_*)$ measured in this work with $f_{\rm red}(M_*)$ from Paper I, we find that {\it central red galaxies} dominate the low-mass regime. Our results motivate a hypothesis in which SFE is significantly higher than previously thought prior to reionization, enabling relatively massive galaxies to form in small haloes. These systems are subsequently quenched by the UV background, producing the central red dwarf galaxies observed, whose measured masses are much larger than those formed in current galaxy formation models. Finally, we obtain $3\sigma$ and $5\sigma$ upper mass bounds of $10^{8.80}\,h^{-1}{\rm M}_{\odot}$ and $10^{10.24}\,h^{-1}{\rm M}_{\odot}$ on the smallest haloes required to exist by the data.
\end{abstract}




\section{Introduction}
The standard Lambda cold dark matter ($\Lambda$CDM) model has been remarkably successful in explaining a wide range of cosmological observations with high accuracy, although several tensions have recently emerged as measurements have reached the $\sim 1\%$ precision level \citep{2021A&A...646A.140H,2022ApJ...934L...7R,2022PhRvD.105b3520A,2023PhRvD.108l3519D,2025PhRvD.112h3515A}.  In this model, structure forms hierarchically as the Universe expands \citep{1994MNRAS.271..676L}: dark matter haloes grow through gravitational mergers, with smaller haloes becoming substructures (subhaloes) within larger hosts until they are fully disrupted or merged with the central component through dynamical friction \citep{1943ApJ....97..255C}.

In parallel, a cosmological framework for galaxy formation has been developed to explain the luminous structures we observe \citep{1978MNRAS.183..341W,1991ApJ...379...52W}. Gas cools, condenses, and flows toward the centres of dark matter haloes, where star formation is triggered and galaxies form. Various sub-galactic processes—such as supernova feedback, metal enrichment, and black hole growth—proceed concurrently or subsequently. This framework has been successful in reproducing many observed galaxy properties across cosmic time when implemented through semi-analytical models or hydrodynamic simulations \citep[e.g.][]{2000MNRAS.319..168C,2005ApJ...631...21K,2015MNRAS.446..521S,2015MNRAS.451.2663H,2018MNRAS.473.4077P,2025arXiv250821126S}.

Although these physical models can reveal many details of galaxy formation, they contain numerous free parameters because of the wide range of processes involved and are computationally expensive due to the broad span of physical scales. Moreover, several key processes remain poorly understood and therefore require sub-grid treatments \citep{2015ARA&A..53...51S,2017ARA&A..55...59N}. Therefore, as a complementary approach, the galaxy–halo connection \citep{2018ARA&A..56..435W} remains a simpler and more efficient way to link the well-modelled dark matter distribution with the observed luminous universe, and to extract physically informative relations although in a highly compressed form.

Several commonly used galaxy–halo connection frameworks exist, depending on the available measurements and halo models.  
The Halo Occupation Distribution \citep[HOD;][]{1998ApJ...494....1J,2000MNRAS.318.1144P,2000MNRAS.318..203S,2002ApJ...575..587B,2005ApJ...633..791Z,2015MNRAS.454.1161Z,2021MNRAS.502.3582Y} is typically used to model the clustering and number densities of an entire galaxy sample.  
The Conditional Luminosity Function \citep[CLF;][]{2003MNRAS.339.1057Y,2012ApJ...752...41Y} models clustering and abundances as functions of galaxy luminosity.  
The Subhalo Abundance Matching \citep[SHAM;][]{2004MNRAS.353..189V,2004ApJ...609...35K,2006ApJ...647..201C,2006MNRAS.371..537W,2010ApJ...717..379B,2013ApJ...770...57B,2013MNRAS.428.3121M,2023ApJ...944..200X} requires an explicit method of identifying subhaloes
and can predict a wide range of galaxy properties. 

One of the key relations constrained by galaxy–halo connection models is the stellar mass–halo mass relation (SHMR), which encodes the star-formation efficiency (SFE) of galaxies. This relation has been constrained over a wide range of stellar and halo masses and across many redshifts \citep{2006MNRAS.371..537W,2007ApJ...667..760Z,2010MNRAS.404.1111G,2010MNRAS.402.1796W,2010ApJ...717..379B,2013ApJ...770...57B,2013MNRAS.428.3121M,2023ApJ...944..200X}. From the mass perspective, the SHMR is well constrained down to halo masses of $10^{11.0}\,\msh$ at $z\simeq0$, with the accessible mass scale increasing at higher redshifts. The limitation at $z\simeq0$ is largely set by the depth of the Sloan Digital Sky Survey \citep[SDSS;][]{2000AJ....120.1579Y} Main Sample ($r<17.7$). Probing lower-mass haloes and galaxies is important, as they provide an ideal regime for testing dark matter models \citep{2019Galax...7...81Z} and reionization physics \citep{1992MNRAS.256P..43E}.

Although substantial efforts have been made—and are ongoing—using next-generation or specially designed spectroscopic surveys \citep{2023ApJ...956....6D,2025ApJ...994..231K,2025arXiv250920458T,2025arXiv250920434T}, pushing measurements of galaxy properties in stellar-mass or luminosity-complete samples remains challenging. The presence of the local void \citep{2008ApJ...676..184T,2010Natur.465..565P,2019ApJ...872..180C,2025MNRAS.540.1635X} makes it even more difficult to constrain the SHMR at lower masses in an average environment rather than within a locally underdense region.

Given the bottleneck in spectroscopic surveys, a promising approach is to exploit photometric data. Building on \citet{2011ApJ...734...88W}, \citet[][hereafter \citetalias{2022ApJ...925...31X}]{2022ApJ...925...31X} introduced the Photometric objects Around Cosmic webs (PAC) method, which measures the excess surface density $\nwprp$ of photometric objects with desired physical properties around spectroscopic ones via angular cross-correlations. Applying PAC to the SDSS Main Sample and the DECam Legacy Survey \citep[DECaLS;][]{2019AJ....157..168D}, \citet[][hereafter \citetalias{2023ApJ...944..200X}]{2023ApJ...944..200X} measured $\nwprp$ for photometric galaxies down to stellar mass $10^{8.0}\,\ms$ at $z<0.2$, constraining the SHMR to halo mass of $\sim10^{10.0}\,\msh$ using SHAM. The PAC method was later optimized for low redshift ($z\le0.1$) in \citet[][hereafter \citetalias{2025MNRAS.540.1635X}]{2025MNRAS.540.1635X}. Applying the improved PAC to Dark Energy Survey Instrument (DESI) Year 1 Bright Galaxy Survey (Y1 BGS) Bright sample and DECaLS, \citetalias{2025MNRAS.540.1635X} measured $\nwprp$ down to a fully complete stellar mass of $10^{6.4}\,\ms$ and derived the galaxy stellar mass function (GSMF). As in \citetalias{2023ApJ...944..200X}, these measurements can be modelled with SHAM to constrain the SHMR to even lower halo masses. The resulting GSMF can also be compared with that of \citetalias{2025MNRAS.540.1635X}, providing a direct test of the galaxy-bias assumptions adopted in that work.

Therefore, in this paper we model 349 $\nwprp$ measurements across various stellar masses from DESI Y1 BGS Bright and DECaLS, combining two high-resolution $N$-body simulations with different box sizes and resolutions. We succeed in constraining the central SHMR down to halo masses of $\sim10^{8.0}\,\msh$, and identify an upturn feature at $\sim10^{10.0}\,\msh$ toward lower masses, implying an increase in SFE toward the low-mass end. This suggests the presence of a second characteristic halo mass in addition to the well-known peak around the Milky Way scale ($\sim10^{12.0}\,\msh$).  We test the robustness of this feature by allowing mass-dependent scatter and by incorporating the effects of reionization on the low-mass halo occupation fraction (HOF); the upturn persists, though with larger uncertainties in the mass-dependent scatter case. We also examine the impact of galaxy assembly bias and of assuming an incorrect cosmology, and the conclusion remains unchanged.  Finally, using the $\Delta\chi^2$ relative to the fiducial model, we find that our measurements impose $3\sigma$ and $5\sigma$ upper bounds of $10^{8.80}\,\msh$ and $10^{10.24}\,\msh$, respectively, on the minimum halo mass—indicating that dark matter haloes must exist down to at least these masses.

We describe the DESI and DECaLS data in Section~\ref{sec:data}. Section~\ref{sec:measurements} introduces the PAC method and presents the measurements. In Section~\ref{sec:sham}, we model the measurements using the fiducial SHAM model. We model or test additional effects that are not included in the fiducial model in Section~\ref{sec:extended}. A comparison with previous work and additional analyses are provided in Section~\ref{sec:compare}, and a brief summary is given in Section~\ref{sec:con}. Unless otherwise specified, the fiducial cosmology adopted throughout this paper is the Planck18 cosmology \citep{2020A&A...641A...6P}, with $\Omega_{\rm m}=0.3111$, $\Omega_{\rm b}=0.049$, $n_{\rm s}=0.9665$, $\sigma_8=0.8102$, and $H_0=67.66\,{\rm km\,s^{-1}\,Mpc^{-1}}$. We define $h\equiv H_0/100\,{\rm km\,s^{-1}\,Mpc^{-1}}$ and $h_{70}\equiv H_0/70\,{\rm km\,s^{-1}\,Mpc^{-1}}$.

\section{Observation and simulation data}\label{sec:data}
In this section, we present the observational and simulation datasets employed in this work. The observational data are identical to those used in \citetalias{2025MNRAS.540.1635X}, comprising the DESI Y1 BGS Bright spectroscopic sample together with the DECaLS photometric sample. The simulations adopted for the SHAM analysis are drawn from the Jiutian hybrid simulation suite \citep{2025SCPMA..6809511H}, where subhaloes are identified with the \textsc{HBT+} \citep{2012MNRAS.427.2437H,2018MNRAS.474..604H} and orphan galaxies are carefully tracked and validated \citep{2025JCAP...12..009X}. 

\subsection{DESI Y1 BGS}
The Dark Energy Spectroscopic Instrument (DESI) is a Stage-IV dark energy experiment designed to obtain spectra for approximately 63 million extragalactic objects over eight years \citep{2013arXiv1308.0847L, 2016arXiv161100036D, 2016arXiv161100037D, 2022AJ....164..207D}. Conducted with a multi-object fibre-fed spectrograph mounted on the 4-m Mayall Telescope at Kitt Peak National Observatory, DESI covers more than 17,000~$\rm{deg^2}$ \citep{2022AJ....164..207D}. The spectrograph operates over $3600$-$9800\,$\AA\  and can observe up to 5,000 targets simultaneously \citep{2016arXiv161100037D, 2023AJ....165....9S, 2024AJ....168...95M, 2024AJ....168..245P}. The DESI data-reduction and targeting pipelines are described in \citet{2023AJ....165..144G, 2023AJ....166..259S, 2023AJ....165...50M}. The parent imaging catalogue for DESI target selection is based on Data Release~9 of the DESI Legacy Imaging Surveys \citep{2017PASP..129f4101Z, 2019AJ....157..168D}, which combine three optical bands ($grz$) from the DECam Legacy Survey \citep[DECaLS;][]{2019AJ....157..168D}, the Dark Energy Survey \citep[DES;][]{2005astro.ph.10346T}, the Beijing-Arizona Sky Survey \citep[BASS;][]{2017PASP..129f4101Z}, and the Mayall $z$-band Legacy Survey \cite[MzLS;][]{2018PASP..130h5001Z}. Early results from DESI have already set exciting constraints on dark energy \citep{2025JCAP...07..028A,2025PhRvD.112h3515A}.

As part of its core programme, DESI carries out the Bright Galaxy Survey \citep[BGS;][]{2023AJ....165..253H}, which targets low-redshift galaxies ($z < 0.6$) observable during bright time across a $\sim14{,}000~\mathrm{deg^2}$ footprint. The BGS comprises a Bright sample ($r < 19.5$) of about 10 million galaxies and a Faint sample ($19.5 < r < 20.175$) of roughly 5 million galaxies, with detailed selection and completeness described in \citet{2023AJ....165..253H}.

In this study we employ the DESI~Y1 BGS Bright sample, now included as part of the DESI Data Release~1\footnote{\url{https://data.desi.lbl.gov/doc/releases/dr1/}} \citep{2025arXiv250314745D}, covering 5300 and 2173~$\mathrm{deg^2}$ in the Northern and Southern Galactic caps (NGC and SGC), respectively, with an average fibre-assignment completeness of~0.656. We restrict our analysis to the region overlapping the DECaLS footprint ($\mathrm{Dec}.<32^{\circ}$), resulting in a total area of 5349~$\mathrm{deg^2}$. Each galaxy is assigned a total weight
\begin{equation}
    w_{\rm tot} = w_{\rm comp}\,w_{\rm zfail}\,,
\end{equation}
where $w_{\rm comp}$ corrects for fibre-assignment incompleteness and $w_{\rm zfail}$ accounts for variations in redshift-success rate \citep{2025JCAP...01..125R}.

To compute the correlation functions, we employ the random catalogue provided by DESI \citep{2025JCAP...01..125R}, which reproduces the angular footprint and redshift distribution of the BGS sample used in this analysis.

\subsection{DECaLS}

We use the photometric catalogue from the DECaLS included in Data Release~9 of the DESI Legacy Imaging Surveys\footnote{\url{https://www.legacysurvey.org/dr9/catalogs/}} \citep{2019AJ....157..168D}. The catalogue covers approximately 9000~deg$^2$ across the Northern and Southern Galactic caps (NGC and SGC) with $\mathrm{Dec.}<32^{\circ}$, observed in the $g$, $r$, and $z$ bands to median 5$\sigma$ point-source depths of 24.9, 24.2, and 23.3, respectively. DECaLS incorporates additional imaging from DES, extending the SGC coverage by about 5000~deg$^2$.

Image processing was performed with {\textsc{Tractor}} \citep{2016ascl.soft04008L}, which models sources using parametric profiles convolved with 
exposure-specific 
point spread functions (PSFs), including  delta function, exponential, de~Vaucouleurs, and S\'ersic profiles. Best-fitting model magnitudes are adopted as default photometry and corrected for Galactic extinction following \citet{1998ApJ...500..525S}.

We restrict the analysis to regions observed in all three bands and apply bright-star and bad-pixel masks using the MASKBITS provided by the Legacy Survey. Additional cuts using the {\texttt{in\_desi}} flag ensure consistency with the DESI footprint, yielding a final coverage of 10,324~deg$^2$. Stellar contamination is removed by excluding PSF-like sources. It is possible that some residual contamination remains from non-PSF-like stellar objects, such as binaries unresolved by DECaLS \citep[][see Figure 1]{2022ApJ...939..104X}. This contamination does not significantly bias the $\nwprp$ measurements, since stars are uncorrelated with galaxies and therefore act only as a foreground noise contribution. The resulting DECaLS photometric sample contains about $5.6\times10^8$ sources. As this work focuses on the DESI~Y1 BGS Bright galaxies, we further restrict the DECaLS footprint to the region overlapping the BGS~Y1 sample, covering 5349~deg$^2$ for the cross-correlation measurements.

To compute the correlation functions, we employ the random catalogue from the DESI Legacy Imaging Surveys, which reproduces the angular footprint of the DECaLS sample used in this analysis.

\subsection{Jiutian simulation suite}

To model the measurements using SHAM, We adopt the primary runs of the Jiutian hybrid simulation suite\footnote{\url{https://jiutian.sjtu.edu.cn/}} \citep{2025SCPMA..6809511H}, developed to support extragalactic science with the China Space Station Telescope (CSST; \citealt{2019ApJ...883..203G}). The suite includes three dark matter-only $N$-body simulations with $6144^3$ particles that differ in box size $L$ and particle mass $m_{\rm p}$. The Jiutian-300, Jiutian-1G, and Jiutian-2G runs have $(L,\,m_{\rm p})=(300\,h^{-1}\mathrm{Mpc},\,1.0\times10^7\,h^{-1}\mathrm{M}_\odot)$, $(1\,h^{-1}\mathrm{Gpc},\,3.7\times10^8\,h^{-1}\mathrm{M}_\odot)$, and $(2\,h^{-1}\mathrm{Gpc},\,3.0\times10^9\,h^{-1}\mathrm{M}_\odot)$, respectively. The simulations adopt a Planck18 cosmology \citep{2020A&A...641A...6P} with $\Omega_{\rm m}=0.3111$, $\Omega_{\rm b}=0.049$, $n_{\rm s}=0.9665$, $\sigma_8=0.8102$, and $h=0.6766$. Simulations were executed with either \textsc{Gadget-3} \citep{2012MNRAS.426.2046A} or \textsc{Gadget-4} \citep{2021MNRAS.506.2871S}, and haloes are identified using the friends-of-friends algorithm \citep{1985ApJ...292..371D} with a linking length of 0.2 times the mean inter-particle separation.

Subhaloes and merger trees are generated with the time-domain subhalo finder \textsc{HBT+} \citep{2012MNRAS.427.2437H,2018MNRAS.474..604H}, which links haloes across snapshots by tracing their self-bound particles. Once accreted, a halo is tracked as a subhalo, and when its bound core becomes unresolved, its most bound particles are followed as ``orphans''. This treatment ensures continuous subhalo evolution and consistent merger histories. All subhaloes whose historical bound mass reaches at least 20 particles are tracked and stored by \textsc{HBT+}. 
We adopt a more conservative threshold of 50 particles, as detailed in Section~\ref{sec:sham}.

In this work, because our analysis aims to investigate galaxy populations across a wide range of stellar masses, and therefore halo masses, we adopt a combination of the Jiutian-300 and Jiutian-1G simulations for the SHAM modelling.

\subsection{Treatment of orphan subhaloes}

The numerical limitations of subhalo-based galaxy modelling have long been recognised \citep{2004MNRAS.352L...1G,2006MNRAS.371..537W,2018MNRAS.475.4066V,2025ApJ...981..108H,2025MNRAS.540.1107S,2025JCAP...12..009X,2025arXiv251026901C}. In simulations, subhaloes can be artificially disrupted, whereas in reality the dense central regions of dark matter haloes—and the galaxies embedded within them—are highly resilient to tidal destruction \citep{2010MNRAS.402...21N,2016ApJ...833..109S}. As a consequence, \citet{2025JCAP...12..009X} found that the subhalo abundance converges only when a subhalo contains at least $\sim5000$ particles at its peak mass. A common approach to recover the subhalo abundance is to continue tracing their most bound particles after they become unresolved as in \textsc{HBT+}. However, these particles no longer experience dynamical friction as resolved subhaloes do, and including all such “orphans” would artificially inflate the subhalo abundance, since they would persist indefinitely. It is therefore necessary to examine the merger timescales and remove subhaloes that should have already merged with their central counterparts.

We follow the same approach as \cite{2025JCAP...12..009X} for treating orphans. In that work, different merger timescale models were tested by comparing the Surviving subhalo Peak Mass Function (SPMF) of Jiutian-300 and Jiutian-1G, both with and without including orphans, and it was found that the model of \cite{2008ApJ...675.1095J} provides the best performance. To reduce dependence on any specific model, \cite{2025JCAP...12..009X} constructed the SPMFs using the resolved components from Jiutian-300 and Jiutian-1G, applying the \cite{2008ApJ...675.1095J} orphan correction only at the lowest mass end ($m_{\rm peak}<10^{10.7}\msh$), where even Jiutian-300 becomes potentially unresolved. Then, within each host halo mass and mass ratio bin, orphans are ranked by $T_{\rm ratio}=T_{\rm merger} / [T(z) - T_{\rm infall}]$ in descending order, and those with the highest values are included until the total number matches the SPMF. Here, $T(z)$ denotes the cosmic time at the redshift of interest, $T_{\rm infall}$ is the infall time of the orphan, and $T_{\rm merger}$ is the merger timescale calculated from the \cite{2008ApJ...675.1095J} model. In this way, most of the mass range depends only on the relative values of the merger timescale, while only the lowest-mass end depends on its absolute value. This approach allows accurate recovery of the subhalo abundance across a wide mass range. 

Although this method ensures that the total subhalo abundance is recovered by design, it does not guarantee that their spatial distribution within haloes is accurately reproduced. As noted above, orphan subhaloes do not experience dynamical friction and therefore are expected to follow different orbits—typically with higher energy and angular momentum—than they would if they were fully resolved. \cite{2025JCAP...12..009X} tested the recovered spatial and velocity distributions of subhaloes by comparing the corrected Jiutian-1G sample to Jiutian-300 and found that they can be recovered to within $5\%$–$10\%$ accuracy down to $0.1$–$0.2\,\mpch$ from the halo centre. Even for subhaloes with $m_{\rm peak}$ corresponding to 50--100 particles—of which 30\%--50\% are orphans—the recovery remains at a comparable level of accuracy. This level of accuracy approaches the intrinsic uncertainty in Jiutian-300 due to cosmic variance and finite-box effects. Since this is the smallest scale relevant to our study, the above treatment of orphans is sufficient for our purposes. 

\section{Measurements from the PAC method}\label{sec:measurements}
In this section, we briefly introduce the PAC method and describe its application to the DESI BGS and DECaLS datasets to obtain the $\nwprp$ measurements and their covariance matrices across different stellar mass bins. 

\subsection{Photometric objects Around Cosmic webs}
The PAC method was developed to fully exploit deep, simply selected photometric survey data \citep{2011ApJ...734...88W,2022ApJ...925...31X,2025MNRAS.540.1635X}. Its core procedure involves measuring the angular cross-correlation function (ACCF) between a spectroscopic catalogue with known redshifts and a photometric catalogue without redshift information. Based on this, the PAC method addresses two main challenges:  
\begin{enumerate}
    \item converting the measured ACCF into three-dimensional quantities that are easier to interpret and model;
    \item utilizing the implicit redshift information contained in the ACCF, since only photometric sources near spectroscopic ones at the same redshift contribute to the correlation, to infer the properties of the photometric sample.
\end{enumerate}

For the first problem, as shown in \citet{2011ApJ...734...88W} and \citetalias{2025MNRAS.540.1635X}, under the flat-sky approximation and assuming a volume-limited photometric sample ($n_2(\boldsymbol{r}_2)=\bar{n}_2$) with a spectroscopic sample located at comoving distance $r_1$, we have
\begin{align}
    \omega_{12}(\theta) &= \frac{\bar{n}_2}{\bar{S}_2} \int_0^{\infty} r_2^2 \, \xi_{12}(r_{12}) \, {\rm d}r_2\,, \\
    \Wp(r_{\rm p}) &= \int_0^{\infty} \xi_{12}(r_{12}) \, {\rm d}r_2\,, \\
    \bar{n}_2 w_{\rm p}(r_{\rm p}) &\approx \frac{\bar{S}_2}{r_1^2} \, \omega_{12}(\theta)\,, \label{eq:pac}
\end{align}
where `1' denotes the spectroscopic sample and `2' denotes the photometric sample. Here, $\omega_{12}$ is the ACCF, $\xi_{12}$ is the three-dimensional cross-correlation function, $\Wp$ is the projected $\xi_{12}$, $\bar{n}_2$ and $\bar{S}_2$ are the mean number density and mean angular surface density of the photometric sample, respectively, and $r_1$ and $r_2$ are the comoving distances to the spectroscopic and photometric sources. The separations are defined as $r_{12} = \sqrt{r_1^2 + r_2^2 - 2r_1r_2\cos{\theta}}$ and $r_{\rm p} = r_1 \sin{\theta}$. The approximation used to derive Equation~\ref{eq:pac} is  
\begin{equation}
    \int_0^{\infty} r_2^2 \, \xi_{12}(r_{12}) \, {\rm d}r_2 \approx 
    r_1^2 \int_0^{\infty} \xi_{12}(r_{12}) \, {\rm d}r_2\,,
\end{equation}
which is valid as long as $r_2^2$ varies much more slowly than $\xi_{12}(r_{12})$ along the line of sight from $r_1$ to the observer.  
\citetalias{2025MNRAS.540.1635X} (see Section~3.6) verified that this approximation holds at the percent level for the maximum scales of interest ($\sim10\,\mpch$), even at low redshifts ($z\sim0.01$–$0.02$).

Therefore, the three-dimensional physical quantity measured by the PAC method is the excess surface density distribution, $\nwp(r_{\rm p})$, of photometric sources around spectroscopic ones. Equation~\ref{eq:pac} describes the idealized case where the entire spectroscopic sample lies at a single distance $r_1$. In practice, the spectroscopic sample typically spans a redshift distribution. In this case, the angular measurements at $\theta = \arcsin(r_{\rm p}/r_1)$ from different redshifts can be stacked, provided that the evolution of $\nwp$ is negligible, allowing for a statistically consistent combination. \citetalias{2025MNRAS.540.1635X} (see Section~3.3) introduced a modified Landy–Szalay estimator \citep{1993ApJ...412...64L} to ensure unbiased stacking:
\begin{equation}
    \nwp(r_{\rm p}) =
    \frac{D_{1,{\rm w}}D_{2,{\rm w}} - D_{1,{\rm w}}R_{2,{\rm w}} - R_{1,{\rm w}}D_{2,{\rm w}} + R_{1,{\rm w}}R_{2,{\rm w}}}
    {R_1 R_2[\theta(r_1; r_{\rm p})]}\,, \label{eq:LSe}
\end{equation}
where the weighted pair counts are defined as
\begin{align}
    D_{1,{\rm w}}D_{2,{\rm w}}(r_{\rm p}) &= \bar{S}_2(r_1)\, D_2 D_1[\theta(r_1; r_{\rm p})] / r_1^2\,, \notag\\
    D_{1,{\rm w}}R_{2,{\rm w}}(r_{\rm p}) &= \bar{S}_2(r_1)\, R_2 D_1[\theta(r_1; r_{\rm p})] / r_1^2\,, \notag\\
    R_{1,{\rm w}}D_{2,{\rm w}}(r_{\rm p}) &= \bar{S}_2(r_1)\, D_2 R_1[\theta(r_1; r_{\rm p})] / r_1^2\,, \notag\\
    R_{1,{\rm w}}R_{2,{\rm w}}(r_{\rm p}) &= \bar{S}_2(r_1)\, R_2 R_1[\theta(r_1; r_{\rm p})] / r_1^2\,.
\end{align}
Here, $D_1D_2[\theta]$, $D_1R_2[\theta]$, $R_1D_2[\theta]$, and $R_1R_2[\theta]$ denote the pair counts between spec data–photo data, spec data–photo random, spec random–photo data, and spec random–photo random samples, respectively. The distance dependence of $\bar{S}_2(r_1)$ arises because different selections on the photometric catalogue are applied at different redshifts to ensure that only photometric sources with the desired physical properties contribute to the ACCF, as described below. Furthermore, the above formula does not account for the total weights $w_{\rm tot}$ from the BGS sample. These can be included by multiplying $w_{\rm tot}$ with $D_1$ and $R_1$ accordingly.

For the second problem, the PAC method addresses it by applying selections to the photometric catalogue, ensuring that only photometric sources with the desired physical properties contribute to the ACCF. As a result, $\nwprp$ measures the excess surface density of photometric sources satisfying those specific selections. 

In this study, we focus on stellar mass. For each spectroscopic source at distance $r_1$, we perform spectral energy distribution (SED) fitting for the entire photometric catalogue using multi-band photometry, \emph{assuming} that all photometric sources lie at $r_1$. We then select objects within the stellar mass bins of interest and compute the ACCF with the corresponding spectroscopic source. For photometric galaxies that are physically close to $r_1$ (and thus actually contribute to the ACCF signal), their properties are computed at approximately the correct distance, so the stellar-mass selection is meaningful. Foreground and background galaxies are also included in the SED fitting and selection, but because their assumed distance is incorrect, their inferred stellar masses are wrong. However, these galaxies are not clustered with the spectroscopic object at $r_1$ and thus only add noise rather than signal to the ACCF. This procedure introduces an additional source of uncertainty, because the photometric galaxies that truly contribute to the ACCF are not all located exactly at $r_1$, but within a finite redshift interval around it, which can lead to errors in their inferred physical properties. This approximation has also been explicitly validated in Section~3.6 of \citetalias{2025MNRAS.540.1635X}. 

The procedure is repeated for all spectroscopic sources at different $r_1$ and the results are combined using the estimator given in Equation~\ref{eq:LSe}. At each $r_1$, the value of $\bar{S}_2(r_1)$ differs because different assumed distances are used to derive the properties of the photometric catalogue, particularly for foreground and background galaxies.

In summary, the procedure for obtaining the PAC measurements is as follows:
\begin{enumerate}
    \item perform SED fitting for the entire photometric catalogue at the redshift of each spectroscopic source;
    \item select photometric sources with the desired physical properties at each redshift; 
    \item compute the ACCF and combine the results across all redshifts.
\end{enumerate} 
For further details, we refer the reader to \citetalias{2022ApJ...925...31X} and \citetalias{2025MNRAS.540.1635X}.

\begin{figure}
    \centering
    \includegraphics[width=\columnwidth]{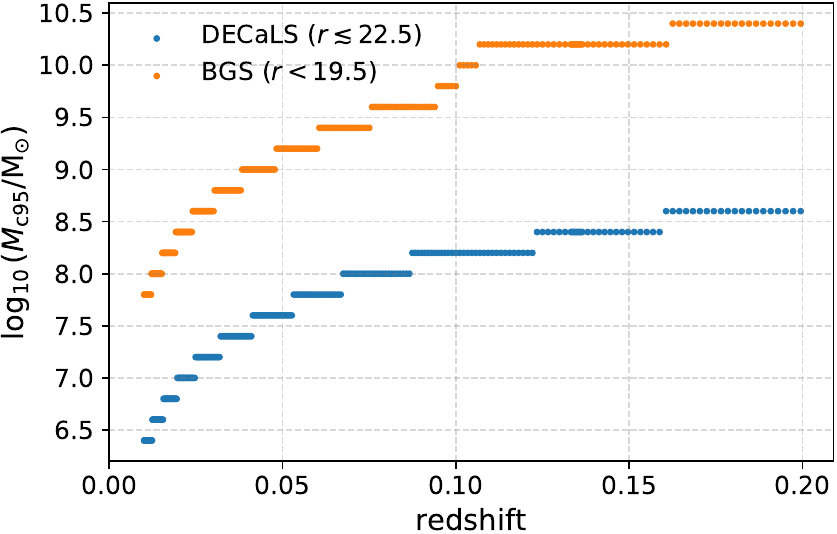}
    \caption{The 95\% completeness stellar mass limits $M_{\rm c95}(z)$ for DECaLS and BGS samples. The quantized appearance of the results arises from the comparison of stellar mass functions in bins with a width of $\Delta\log_{10}(M_{*}/\ms)=0.2$.}
    \label{fig:SM_limits}
\end{figure}

\subsection{PAC measurements from DESI BGS and DECaLS}
We follow the pipeline described in \citetalias{2025MNRAS.540.1635X} to obtain the $\nwp(r_{\rm p})$ measurements from DESI BGS and DECaLS at $z<0.2$, across various stellar mass bins for both samples. A key difference in this work is that we ensure both BGS and DECaLS samples are complete within the desired stellar mass bins to facilitate more accurate modelling, whereas \citetalias{2025MNRAS.540.1635X} only ensured completeness for DECaLS and included as much BGS data as possible to enhance the signal. In other words, \citetalias{2025MNRAS.540.1635X} prioritized reducing statistical uncertainty at the cost of larger modelling systematics, while this study emphasizes better control of systematics at the expense of signal strength.

We use the DECaLS physical property catalogue constructed in \citetalias{2025MNRAS.540.1635X}. This catalogue provides physical properties for the entire DECaLS sample, assuming redshifts over the range $0.001 < z < 1$. It was generated by performing SED fitting for the full DECaLS catalogue at 44 reference redshifts and interpolating across the entire redshift range. The interpolation accuracy for stellar mass was validated to within $\sigma[\log_{10} M_{*}/\ms] = 0.01$ using additional reference redshifts in a smaller test sample. The physical properties were derived using the SED fitting code \textsc{Cigale} \citep{2019A&A...622A.103B}, based on the $grz$-band fluxes from DECaLS. The modelling adopts the stellar population synthesis models of \citet{2003MNRAS.344.1000B}, assuming a \citet{2003PASP..115..763C} initial mass function and a delayed exponential star formation history, $\phi(t) \propto t \exp(-t/\tau)$. Three metallicities are considered, $Z/Z_{\odot} = 0.4,\ 1,\ \text{and}\ 2.5$, where $Z_{\odot}$ denotes the solar metallicity. Dust attenuation is modelled using the \citet{2000ApJ...533..682C} extinction law, with $0 < E(B-V) < 0.5$.

For the measurements in the ranges $z<0.1$ and $0.1<z<0.2$, two different colour cuts are applied to the DECaLS catalogue to exclude as many foreground and background galaxies as possible while retaining all galaxies within the target redshift ranges, thereby reducing noise in the measurements (see Section~3.4 of \citetalias{2025MNRAS.540.1635X}).

\citetalias{2025MNRAS.540.1635X} derived the 95\% completeness stellar mass limit, $M_{\rm c95}(z)$, as a function of redshift for DECaLS. This was obtained by comparing the stellar mass function of DECaLS at each assumed redshift with that of the deepest subsample, defined by regions where the $r$-band $10\sigma$ PSF depth exceeds 23.8, primarily within the DES footprint. Here, we repeat the same procedure for BGS depth by applying a cut of $r<19.5$ to the DECaLS catalogue and comparing it with the deepest sample. Figure~\ref{fig:SM_limits} shows $M_{\rm c95}(z)$ for both DECaLS and BGS. Note that the comparison is performed after applying the colour cut at each redshift range. This leads to a slightly steeper rise in $M_{\rm c95}(z)$ for BGS around $z\approx0.1$, where the colour cuts differ across the two ranges. The quantized appearance of the results in Figure~\ref{fig:SM_limits} arises from the comparison of stellar mass functions in bins with a width of $\Delta\log_{10}(M_{*}/\ms)=0.2$. We note that these limits are direct and conservative, as foreground and background galaxies with incorrectly estimated stellar masses increase the scatter in the mass-to-light ratios. The presence of unrealistically large mass-to-light values, typically from background galaxies, tends to bias the 95\% completeness limits toward higher stellar masses. For the $\nwprp$ measurement between a BGS sample of stellar mass $M_*^{\rm spec}$ and a DECaLS sample of stellar mass $M_*^{\rm photo}$, the calculation is restricted to $z < \min[z_{\rm c95}^{\rm spec}(M_*^{\rm spec}),\, z_{\rm c95}^{\rm photo}(M_*^{\rm photo})]$, where $z_{\rm c95}(M_*)$ denotes the inverse function of $M_{\rm c95}(z)$.

We obtain 349 $\nwprp$ measurements for $M_*^{\rm photo} \in [10^{6.3}, 10^{11.9}]\,{\rm M}_{\odot}$ and $M_*^{\rm spec} \in [10^{8.3}, 10^{11.7}]\,{\rm M}_{\odot}$, using a bin width of $\Delta\log_{10}(M_{*}/\ms) = 0.2$. Measurements with $M_*^{\rm spec} < 10^{9.3}\,\ms$ are used only for $M_*^{\rm photo} < 10^{7.9}\,\ms$ to reduce the total dimensionality of the data vector. The dimensionality is further reduced using principal component analysis (PCA), as described below. The measurements are performed over the radial range $0.16\,\mpch < r_{\rm p} < 16\,\mpch$ with 10 bins. Although the PAC method can reach $\rp = 0.01\,\mpch$ as in \citetalias{2025MNRAS.540.1635X}, modelling at such small scales is difficult due to the uncertain subhalo distributions in $N$-body simulations \citep{2025JCAP...12..009X}. Each $\nwprp$ measurement is obtained separately in the NGC and SGC regions and then combined according to their respective survey areas.

Photometric artifacts could in principle contaminate the PAC measurements by producing spurious DECaLS sources, as has been found around local, well-resolved galaxies at $z<0.01$ \citep{2023AJ....165...50M}. This effect has been corrected in our measurements following the procedure developed in \citetalias{2025MNRAS.540.1635X}. Specifically, we mask all DECaLS sources within $2R_{\rm e}$ of each BGS galaxy and account for the completeness of the BGS sample. This masking removes substructures and deblending artifacts that are physically associated with galaxies and can therefore contribute to the PAC correlation signal. Random artifacts that are not correlated with galaxies do not contribute to the PAC measurements. We note that contamination in the BGS sample itself does not directly bias $\bar{n}_2$, but only affect $\Wp$. It changes the effective weighting of the host galaxies of the contaminated objects; this effect is removed by applying the same weighted distributions to the random catalogue. In \citetalias{2025MNRAS.540.1635X}, we found only a 5\% difference for blue galaxies at $M_*>10^{5.0}\ms$ and a negligible difference for red galaxies. This is because most of the contamination identified in previous studies comes from well-resolved, large blue galaxies at $z<0.01$, which have been completely excluded in this study. At higher redshift, blue galaxies are not resolved as well as nearby ones. Red galaxies are also generally smoother and more uniform, with fewer substructures.

As an additional conservative test, we repeat the measurement with a cut of ${\tt FRACFLUX}<0.2$ in Appendix~\ref{sec:fracflux}. This cut removes sources whose flux may be strongly affected by nearby objects. Although it is very conservative and may also remove real galaxies, thereby changing the true $\bar{n}_2$, it provides a useful check on the possible impact of residual photometric artifacts. Figure~\ref{fig:fracflux} compares the fiducial measurements with the results after applying ${\tt FRACFLUX}<0.2$ for the lowest-mass ($M_*=10^{6.4}\ms$) red and blue DECaLS samples. For red galaxies, there is almost no difference between the two measurements. For blue galaxies, some differences appear on small scales, which may be caused either by residual contamination or by a change in the nonlinear bias of real galaxies induced by the {\tt FRACFLUX} selection. However, the two measurements are consistent on larger scales, indicating that the change in $\bar{n}_2$ is small, since low-mass galaxies should have similar $w_{\rm p}$ on linear scales. Together with the previous masking tests, we therefore conclude that our results are robust to photometric artifacts, which are expected to be more important for local blue dwarf galaxies.

The complete set of measurements is provided in Appendix~\ref{sec:fits}.

\begin{figure}
    \centering
    \includegraphics[width=\columnwidth]{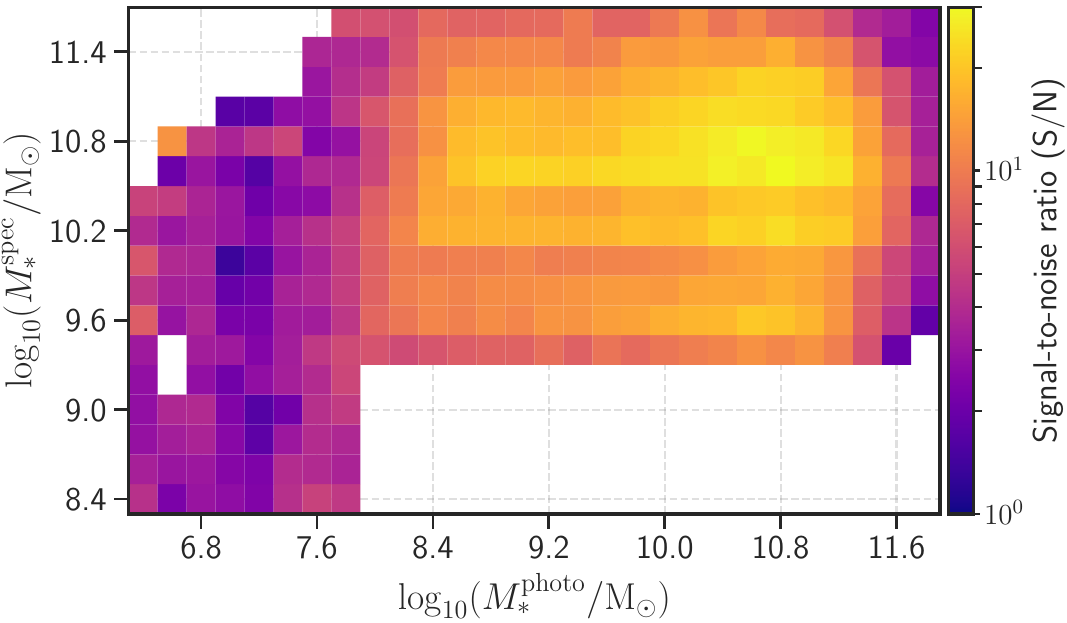}
    \caption{Total signal-to-noise ratio of each $\nwprp$ measurement in bins of $M_*^{\rm spec}$ and $M_*^{\rm photo}$.}
    \label{fig:sn_matrix}
\end{figure}

\begin{figure}
    \centering
    \includegraphics[width=\columnwidth]{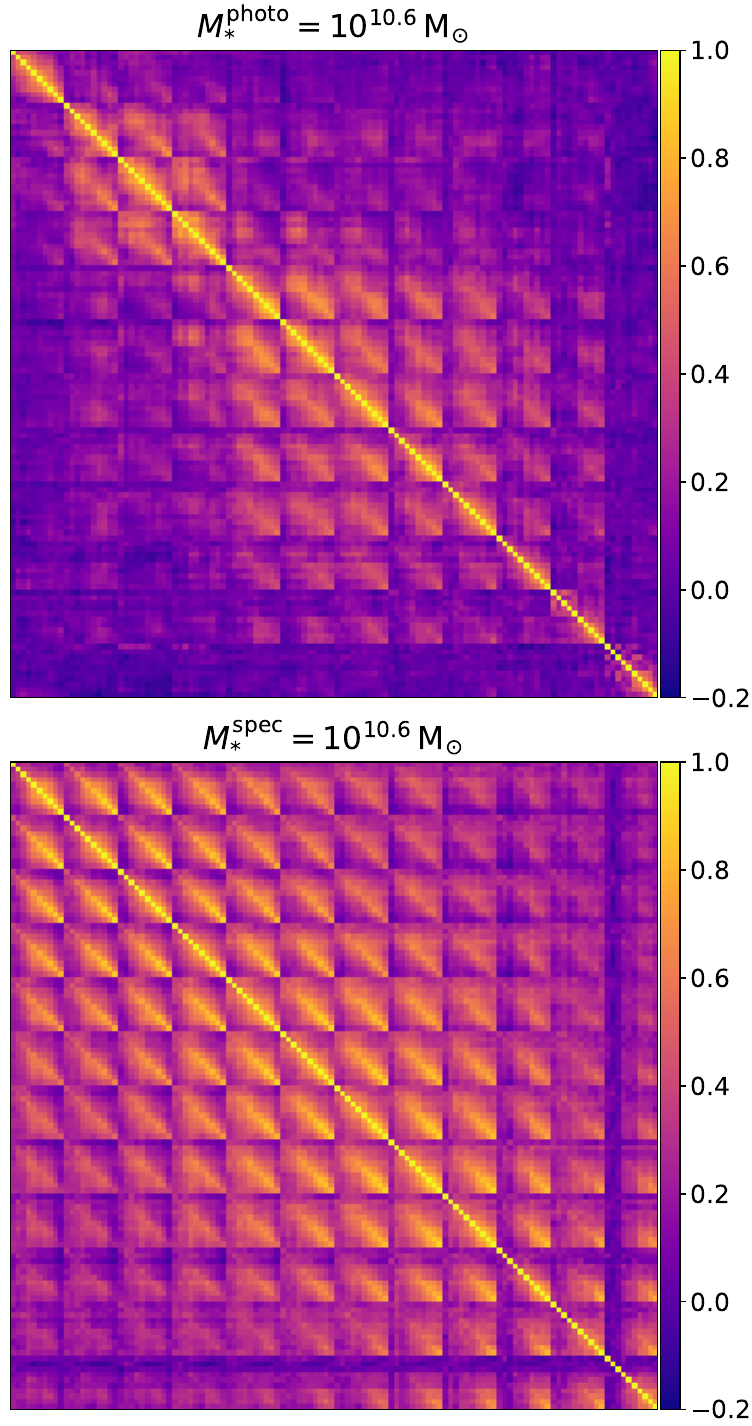}
    \caption{Normalized covariance matrices of $\nwprp$ measurements for samples with $M_*^{\rm photo}$ and $M_*^{\rm spec}$ fixed at $10^{10.6}\,\ms$, with the other sample spanning 12 stellar mass bins in the range $[10^{9.4},\,10^{11.6}]\,\ms$. Each small square represents the cross-covariance of $\nwprp$ measurements between two stellar mass bins, each containing 10 radial bins. Stellar mass increases from top to bottom and from left to right.}
    \label{fig:mass_cov}
\end{figure}

\begin{figure*}
    \centering
    \includegraphics[width=\textwidth]{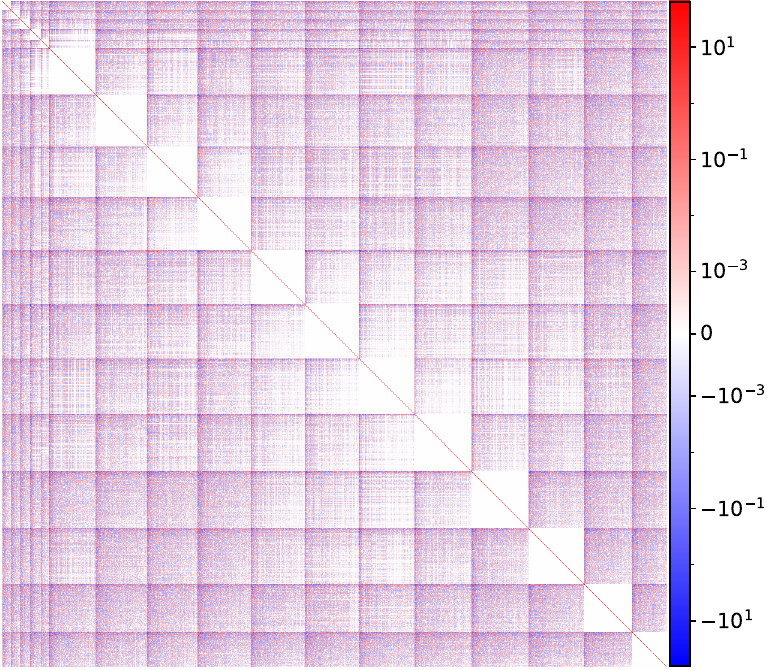}
    \caption{Estimated covariance matrix of the normalized $\nwprp$ data vector $\tilde{\boldsymbol{\mathcal{A}}}$ in the PCA-truncated space (total dimension 1454), obtained from intra-group PCA and a low-rank approximation of inter-group covariances via truncated SVD. Each rectangle represents the cross-covariance between two Groups. In total, there are 17 Groups corresponding to stacked measurements in 17 $M_*^{\rm spec}\in[10^{8.3},10^{11.7}]\,\ms$ bins. $M_*^{\rm spec}$ increases from top to bottom and from left to right.}
    \label{fig:full_cov}
\end{figure*}

\subsection{Covariance estimation, compression, and stabilization}
Because our $\nwp$ measurements extend to small $r_{\rm p}$ and are affected by complex noise from foreground and background sources at each redshift, it is difficult to derive the covariance matrix analytically or semi-analytically. Generating a large number of high-resolution mock catalogues is also impractical. Therefore, we estimate the covariance matrix directly from the data using the leave-one-out jackknife resampling method.
 
We divide the BGS sample into $N_{\rm jk}$ subsamples of equal size in the SGC and NGC regions separately, using \textsc{HEALPix} \citep{2005ApJ...622..759G} with $N_{\rm side}=512$. The $\nwp$ measurements are then obtained for each leave-one-out subsample, and the $n$th measurements from the SGC and NGC are combined to produce the final $n$th realization. We adopt $N_{\rm jk}=400$, as this configuration yields field sizes comparable to the largest scales of interest, increasing $N_{\rm jk}$ further would lead to underestimated errors on large-scale measurements. The covariance matrix for each $\nwp(r_{\rm p})$ measurement is then estimated as
\begin{equation}
    C_{\mathcal{A},ij} = \frac{N_{\rm jk}-1}{N_{\rm jk}}
    \sum_{n=1}^{N_{\rm jk}}
    \left[\mathcal{A}_n(r_{{\rm p},i}) - \bar{\mathcal{A}}(r_{{\rm p},i})\right]
    \left[\mathcal{A}_n(r_{{\rm p},j}) - \bar{\mathcal{A}}(r_{{\rm p},j})\right],
    \label{eq:cov_jk}
\end{equation}
where $\mathcal{A}\equiv\nwp$ for brevity, indices $i$ and $j$ denote the $i$th and $j$th radial bins, $\mathcal{A}_n$ is the measurement from the $n$th jackknife subsample, and $\bar{\mathcal{A}}$ is the mean of $N_{\rm jk}$ measurements. Figure~\ref{fig:sn_matrix} shows the total signal-to-noise ratio,
\begin{equation}
    S/N = \sqrt{\boldsymbol{\mathcal{A}}^\top \boldsymbol{C}_{\mathcal{A}}^{-1} \boldsymbol{\mathcal{A}}}\,,
\end{equation}
for each $\nwprp$ measurement across different $M_*^{\rm spec}$ and $M_*^{\rm photo}$ bins, providing an overview of the measurement quality.

However, different $\nwprp$ measurements are correlated because many of them share the same galaxy pairs, especially those involving the same $M_*^{\rm spec}$ and $M_*^{\rm photo}$ bins. For reliable modelling, these correlations must be properly accounted for. Although necessary, directly estimating the full covariance matrix for the 3490-dimensional (349 $\times$ 10) data vector using only $N_{\rm jk}=400$ jackknife realizations is infeasible, as the resulting covariance matrix would be ill-conditioned and non-invertible. Therefore, data compression is required, and the independent components of the data vector must be identified to make the analysis feasible.

As illustrated in Figure~\ref{fig:mass_cov}, the full covariance matrices for each $M_*^{\rm spec}$ and $M_*^{\rm photo}$ bin show that measurements sharing the same $M_*^{\rm spec}$ bin are strongly correlated, while those sharing the same $M_*^{\rm photo}$ bin are largely independent. Figure~\ref{fig:mass_cov} presents the example for $10^{10.6}\,\ms$, and we have verified that this behaviour holds across all stellar mass bins. Therefore, we compute the full covariance matrix within each $M_*^{\rm spec}$ bin (with a maximum dimension of 280), which we refer to as a “Group”, and use low-rank inter-group cross-covariance terms to capture the remaining correlations between Groups. The intra-group covariance matrices are compressed using PCA to reduce dimensionality and are stabilized for inversion using the Oracle Approximating Shrinkage \citep[OAS;][]{2010ITSP...58.5016C} estimator. The inter-group cross-covariance matrices are truncated to low rank via singular value decomposition (SVD), retaining only the leading singular components. 

As shown in Figure~\ref{fig:sn_matrix}, each $M_*^{\rm spec}$ bin contains $M_*^{\rm photo}$ bins with widely varying $S/N$, and the $S/N$ also differs substantially across radial bins. Compressing the data using the absolute covariance or the $S/N$ matrix would preferentially discard low-mass or large-scale information, which is undesirable. We therefore perform PCA and SVD on the \emph{normalized} covariance matrix, ensuring that all $M_*^{\rm photo}$ bins and all radial bins are weighted equally. In this basis, modes with larger eigenvalues reflect stronger structural correlations rather than simply higher $S/N$. This approach may remove some high–$S/N$ measurements but yields a more balanced representation of information across masses and scales. A more optimal strategy would be Fisher-informed compression \citep{1997ApJ...480...22T,2000MNRAS.317..965H}, in which compression directions are determined by the parameter derivatives of the model. However, this method requires a differentiable model with explicit parameterization and a fiducial parameter set, and has closed-form optimality only for the linear Gaussian case, where each parameter corresponds to a single compressed datum. For nonlinear problems, identifying an appropriate low-dimensional subspace remains nontrivial. For these reasons, we adopt the simpler PCA–SVD approach applied to the normalized covariance matrix, which we find performs well in practice.

Within each Group $g$, we first normalize the data matrix $\mathcal{A}_g \in \mathbb{R}^{N_{\rm jk} \times m_g}$, where $m_g$ is the dimensionality of Group $g$, and compute the full covariance:
\begin{align}
    \tilde{\mathcal{A}}_g &= \frac{\mathcal{A}_g - \bar{\mathcal{A}}_g}{\sigma_{\mathcal{A}_g}}\,, \\
    C_{gg} &= \frac{N_{\rm jk}-1}{N_{\rm jk}} \tilde{\mathcal{A}}_g^{\top} \tilde{\mathcal{A}}_g\,.
\end{align}
We then perform PCA in each Group by diagonalizing $C_{gg}$ via eigendecomposition:
\begin{align}
    C_{gg} = V_{gg} D_{gg} V_{gg}^{\top}, \qquad  
    D_{gg} = \mathrm{diag}(\lambda_1, \ldots, \lambda_{m_g})\,.
\end{align}
The top $k_g$ eigenmodes are retained such that
\begin{equation}
    \frac{\sum_{i=1}^{k_g} \lambda_i}{\sum_{i=1}^{m_g} \lambda_i} \ge \tau_g\,,
\end{equation}
where $\tau_g \in (0,1)$ is the chosen cumulative variance threshold. The truncated covariance $D_{gg} \in \mathbb{R}^{k_g \times k_g}$ and the projected measurements $Y_g \in \mathbb{R}^{N_{\rm jk} \times k_g}$ are then given by
\begin{equation}
    D_{gg} = \mathrm{diag}(\lambda_1, \ldots, \lambda_{k_g}), \qquad
    Y_g = \tilde{\mathcal{A}}_g\, V_{gg}[:,\,1:k_g]\,.
\end{equation}
To regularize the retained eigenvalues, we apply OAS shrinkage toward the identity matrix $I \in \mathbb{R}^{k_g \times k_g}$:
\begin{equation}
    \tilde{D}_{gg} = (1 - \rho_g)\, D_{gg} + \rho_g\, \mu_g\, I\,,
\end{equation}
where $\mu_g = \mathrm{tr}(D_{gg})/k_g$, and $\rho_g$ is computed analytically from $D_{gg}$ following \citet{2010ITSP...58.5016C}.

The cross-covariance between Group $g$ and Group $h$, $C_{gh} \in \mathbb{R}^{k_g \times k_h}$, is estimated as
\begin{equation}
    C_{gh} = \frac{N_{\rm jk}-1}{N_{\rm jk}}\, Y_g^{\top} Y_h\,.
\end{equation}
We perform a SVD,
\begin{equation}
    C_{gh} = U_{gh} D_{gh} V_{gh}^{\top}, \qquad 
    D_{gh} = \mathrm{diag}(s_1, \ldots, s_r)\,,
\end{equation}
where $r = \min(k_g, k_h)$, $U_{gh} \in \mathbb{R}^{k_g \times r}$, and $V_{gh} \in \mathbb{R}^{k_h \times r}$. We retain the leading $k_{gh} \le k_{\max}$ singular values contains $\tau_{gh}$ cumulative variance. Each retained mode $j$ defines a global low-rank direction linking Groups $g$ and $h$, with square-root factors
\begin{equation}
    w_{g,j} = \sqrt{s_j}\,u_{g,j}, \qquad
    w_{h,j} = \sqrt{s_j}\,v_{h,j},
\end{equation}
where $u_{g,j}$ and $v_{h,j}$ are the $j$th columns of $U_{gh}$ and $V_{gh}$, respectively.  
For each mode $j$, we construct a vector $w$ that is zero everywhere except in the components corresponding to Groups $g$ and $h$, where it contains $w_{g,j}$ and $w_{h,j}$.  
Stacking all such inter-group contributions yields
\begin{equation}
    W \in \mathbb{R}^{K \times K_{gh}}, \qquad
    K = \sum_g k_g,\qquad
    K_{gh} = \sum_{g<h} k_{gh}\,.
\end{equation}

The final covariance matrix in PCA space, $C \in \mathbb{R}^{K \times K}$, is constructed by combining the block-diagonal intra-group covariance 
$D \in \mathbb{R}^{K \times K}$ with the low-rank inter-group component 
$W \in \mathbb{R}^{K \times K_{gh}}$:
\begin{equation}
    C = D + \alpha\, (W J W^{\top}).\label{eq:cov}
\end{equation}
Here $J$ is a symmetric involutory matrix composed of disjoint $2\times2$ exchange blocks, each acting on a singular-vector pair in $W$. This structure ensures that $W J W^{\top}$ contributes only symmetric cross-group covariance terms of the form $(w_g w_h^{\top} + w_h w_g^{\top})$, while leaving all diagonal (intra-group) variances encoded in $D$ unchanged.

In the idealized limit of infinite samples and without any PCA or SVD truncation, setting $\alpha = 1$ reproduces the exact covariance expansion, and the resulting matrix is identical to the full covariance. In practice, however, the structure of Equation~\ref{eq:cov} does not guarantee that the covariance is strictly positive definite (SPD). With the limited number of jackknife samples available here, it is generally not SPD. We therefore shrink the off-diagonal blocks by a factor $\alpha \in [0,1]$ to enforce positive definiteness, where $\alpha$ is chosen as the largest value that results in an SPD covariance matrix. This reduces the inter-group correlations but ensures that the covariance is invertible. 
  
After testing different configurations, we set the hyperparameters to $\tau = 0.97$, $\tau_{gh} = 0.9$, and $k_{\max} = 8$, which reduce the total dimensionality from 3490 to 1454. This choice provides a balance between retaining information and ensuring a well-conditioned and stable covariance matrix. The estimated 1454-dimensional full covariance matrix of $\tilde{\mathcal{A}}$ in the PCA-truncated space is shown in Figure~\ref{fig:full_cov}.

\section{Modelling with Subhalo Abundance Matching}\label{sec:sham}
In this section, we model the $\nwprp$ measurements in Jiutian-1G and Jiutian-300 $N$-body simulations using the SHAM method. Through this procedure, we constrain the SHMR and infer related quantities such as the GSMF. The methodology is based on \citetalias{2023ApJ...944..200X} with several improvements.

\subsection{Tabulated correlation functions}
To model the $\nwprp$ measurements in each $M_*^{\rm spec}$ and $M_*^{\rm photo}$ bin, we compute $\n$ and $\Wp$ as weighted combinations of the halo and subhalo contributions, determined by the central and satellite SHMRs, $P_{\rm c}(M_*^{\rm c} \mid M_{\rm h}; \boldsymbol{\theta}_{\rm c})$ and $P_{\rm s}(M_*^{\rm s} \mid M_{\rm s}; \boldsymbol{\theta}_{\rm s})$. Direct pair counting scales as $O(N^2)$, and repeating it for every sampled set of parameters $\boldsymbol{\theta}_c$ and $\boldsymbol{\theta}_s$ would render the parameter inference computationally infeasible. Therefore, we adopt the tabulated correlation function approach \citep{2016MNRAS.458.4015Z,2022ApJ...928...10G}, in which tables of halo–halo ($\Wp^{\rm hh}$), halo–subhalo ($\Wp^{\rm hs}$), and subhalo–subhalo ($\Wp^{\rm ss}$) projected correlation functions are precomputed in fine mass bins, together with the corresponding halo and subhalo number densities $\bar{n}_{\rm h}$ and $\bar{n}_{\rm s}$ in those bins. For $N_{\rm h}$ halo mass bins, $N_{\rm s}$ subhalo mass bins, and $N_{\rm r}$ radial bins, the table dimensions are
$
\Wp^{\rm hh} \in \mathbb{R}^{N_{\rm h} \times N_{\rm h} \times N_{\rm r}}$, $
\Wp^{\rm hs} \in \mathbb{R}^{N_{\rm h} \times N_{\rm s} \times N_{\rm r}}$,
$\Wp^{\rm ss} \in \mathbb{R}^{N_{\rm s} \times N_{\rm s} \times N_{\rm r}}$,
where $\Wp^{\rm hh}$ and $\Wp^{\rm ss}$ are symmetric with respect to their mass-bin indices, while $\Wp^{\rm hs}$ is not.
 
With these tables, the model $\Wp(r_{\rm p}; M_*^{\rm spec}, M_*^{\rm photo})$ in each pair of stellar-mass bins can be computed efficiently as
{\small
\begin{align}
    &\n(M_{*}^{\rm spec})\n(M_{*}^{\rm photo})\Wp(\rp;M_*^{\rm spec},M_*^{\rm photo})\notag\\
    &=\sum_i^{N_{\rm h}}\sum_j^{N_{\rm h}}\bar{n}_{{\rm h},i}\bar{n}_{{\rm h},j}P(M_{*}^{\rm spec}|M_{{\rm h},i})P(M_{*}^{\rm photo}|M_{{\rm h},j})\Wp^{\rm hh}(\rp;M_{{\rm h},i},M_{{\rm h},j})\notag\\
    &+\sum_i^{N_{\rm h}}\sum_j^{N_{\rm s}}\bar{n}_{{\rm h},i}\bar{n}_{{\rm s},j}[P(M_{*}^{\rm spec}|M_{{\rm h},i})P(M_{*}^{\rm photo}|M_{{\rm s},j})\notag\\
    &+P(M_{*}^{\rm photo}|M_{{\rm h},i})P(M_{*}^{\rm spec}|M_{{\rm s},j})]\Wp^{\rm hs}(\rp;M_{{\rm h},i},M_{{\rm s},j})\notag\\
    &+\sum_i^{N_{\rm s}}\sum_j^{N_{\rm s}}\bar{n}_{{\rm s},i}\bar{n}_{{\rm s},j}P(M_{*}^{\rm spec}|M_{{\rm s},i})P(M_{*}^{\rm photo}|M_{{\rm s},j})\Wp^{\rm ss}(\rp;M_{{\rm s},i},M_{{\rm s},j})\,,\label{eq:wp}
\end{align}}
where $i$ and $j$ denote the indices of the halo and subhalo mass bins, and
\begin{align}
    P(M_{*}|M_{{\rm h},i})&=\int_{M_*^{\rm min}}^{M_*^{\rm max}}P_{\rm c}(M_*^{\rm c} \mid M_{{\rm h},i}; \boldsymbol{\theta}_{\rm c}){\rm d}M_*^{\rm c}\,\\
    P(M_{*}|M_{{\rm s},i})&=\int_{M_*^{\rm min}}^{M_*^{\rm max}}P_{\rm s}(M_*^{\rm s} \mid M_{{\rm s},i}; \boldsymbol{\theta}_{\rm s}){\rm d}M_*^{\rm s}\,,\\
    \n(M_*)&=\sum_i^{N_{\rm h}}\bar{n}_{{\rm h},i}P(M_{*}|M_{{\rm h},i})+\sum_i^{N_{\rm s}}\bar{n}_{{\rm s},i}P(M_{*}|M_{{\rm s},i})\,.
\end{align}
Here $M_*\in[M_*^{\rm min},M_*^{\rm max}]$. 

We construct these tables using the Jiutian-1G and Jiutian-300 simulations at a snapshot around $z=0.1$. We adopt the virial mass $M_{\rm vir}$ as the halo mass, defined in a redshift dependent manner following the spherical collapse overdensity criterion of \citet{1998ApJ...495...80B}. For subhalos, we use the peak bound mass $m_{\rm peak}$, i.e., the maximum bound mass attained over the subhalo’s history. To fully utilize the larger volume of Jiutian-1G, we impose 50-particle thresholds for $M_{\rm vir}$ and $m_{\rm peak}$, above which the halo and subhalo statistics are considered reliable. Therefore, if both mass bins involved in a given $\Wp$ entry exceed $10^{10.3}\,\msh$, we compute it from Jiutian-1G; otherwise, we use Jiutian-300. The lowest mass bin satisfying the 50-particle threshold in Jiutian-300 is $10^{8.7}\,\msh$. We adopt very fine mass bins of width $\Delta\log_{10}(M_{\rm h}/\ms)=0.01$, which is sufficient for our purposes. This yields 670 bins across the range $[10^{8.7},10^{15.4}]\,\ms$.

The low halo-mass limit of $10^{8.7}\,\msh$ is only marginally sufficient for modelling the lowest stellar-mass bin at $10^{6.4}\,\ms$. To prevent this limit from introducing modelling systematics, we apply a controlled extrapolation of the halo terms in $\Wp$ that depend on halos and on $\bar{n}_{\rm h}$. No extrapolation is applied to the subhalo terms, since satellite galaxies inhabit more massive subhalos, as shown below; thus the $10^{8.7}\,\msh$ limit is adequate. 

The halo abundance $\bar{n}_{\rm h}$ is extrapolated using an analytic halo mass function (HMF). This is relatively robust because the low-mass end of the HMF follows a well-established power-law form. We adopt a modified \citet{2016MNRAS.456.2486D} model, which matches the Jiutian simulations best at high masses. For $\Wp^{\rm hh}$ and $\Wp^{\rm hs}$, we extrapolate to lower halo masses under the assumption that halo bias varies only weakly at the low-mass end. Accordingly, the low-mass entries in these tables are constructed using measurements from halos in the range $[10^{8.7},10^{9.5}]\,\msh$. This assumption is validated by comparing $\Wp^{\rm hh}$ and $\Wp^{\rm hs}$ within this mass range, where we find only mild variation that is well below the statistical uncertainties of the low-mass measurements. A detailed validation of the extrapolation is provided in Appendix~\ref{sec:extra}.

The tables are extrapolated down to a halo mass of $10^{7.0}\,\msh$. The resulting dimensions are
$(840,840,10)$ for $\Wp^{\rm hh}$, $(840,670,10)$ for $\Wp^{\rm hs}$, and $(670,670,10)$ for $\Wp^{\rm ss}$. 
These tables are applicable only to models in which the (sub)halo bias depends solely on (sub)halo mass. 
Incorporating additional dependencies related to halo assembly bias \citep{2005MNRAS.363L..66G,2006ApJ...652...71W,2007ApJ...657..664J} would require extra dimensions in the tables, which would increase their size substantially and may necessitate further compression.

In summary, the Jiutian-1G tables are used for $M_{\rm h}>10^{10.3}\msh$, while the Jiutian-300 tables are used for $10^{8.7}\msh<M_{\rm h}<10^{10.3}\msh$. At $M_{\rm h}<10^{8.7}\msh$, we use an analytic HMF together with extrapolated $\Wp$ tables validated in Appendix~\ref{sec:extra}. This extrapolation affects only the modelling of the two lowest photometric stellar-mass bin, $M_*^{\rm photo}=10^{6.4}\ms$ and $10^{6.6}\ms$, and does not alter the main results of this paper.

\begin{figure}
    \centering
    \includegraphics[width=\columnwidth]{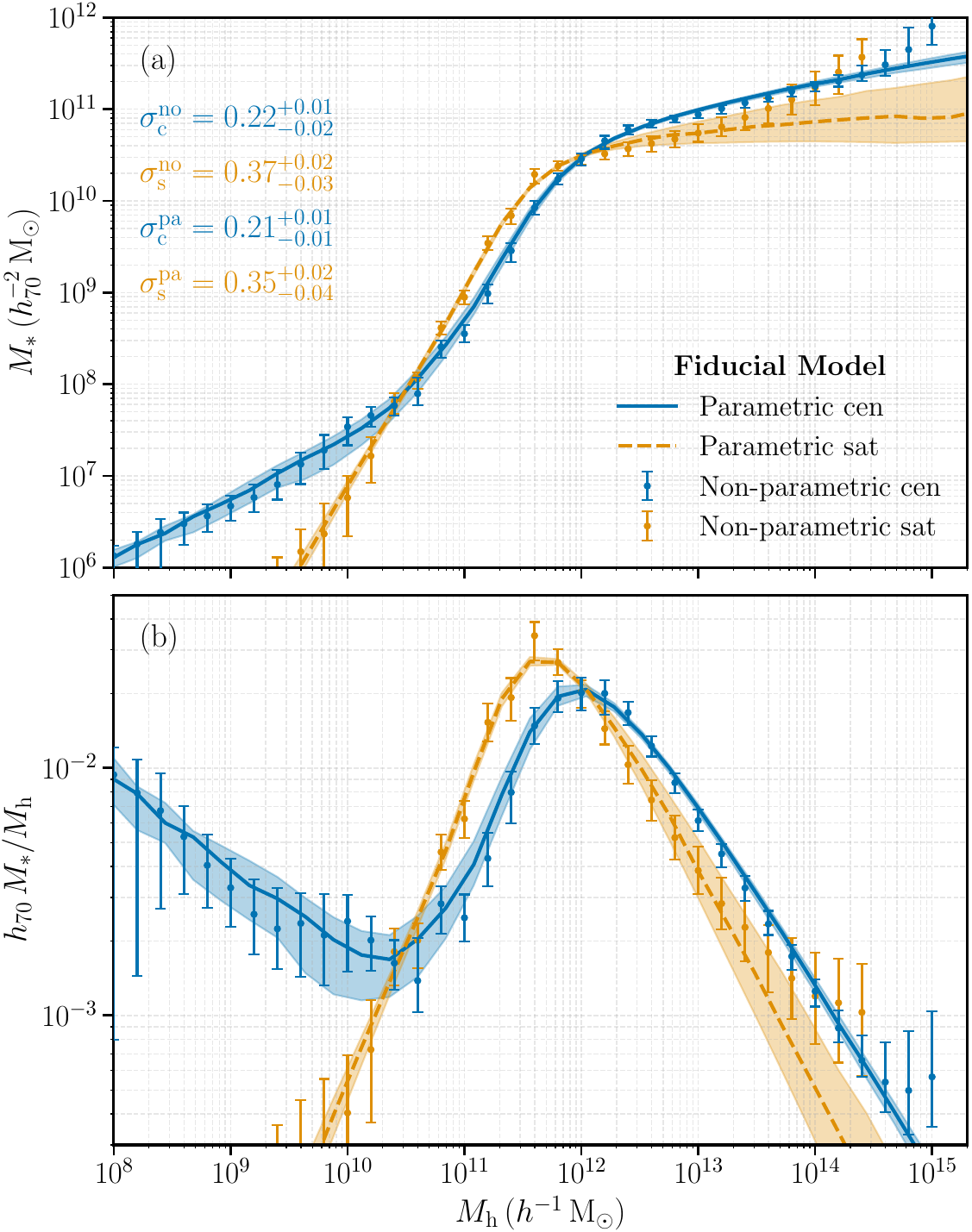}
    \caption{Mean stellar–halo mass relations (top) and stellar-to-halo mass ratios (bottom) for central (blue) and satellite (orange) galaxies constrained by the fiducial models. Results from both the non-parametric and parametric models are shown. Dots with error bars and curves with shaded regions denote the maximum a posterior (MAP) estimates and the corresponding $1\sigma$ intervals. The scatters of the relations, along with their uncertainties, are also shown in the top panel.
}
    \label{fig:fiducial_shmr}
\end{figure}

\begin{figure}
    \centering
    \includegraphics[width=\columnwidth]{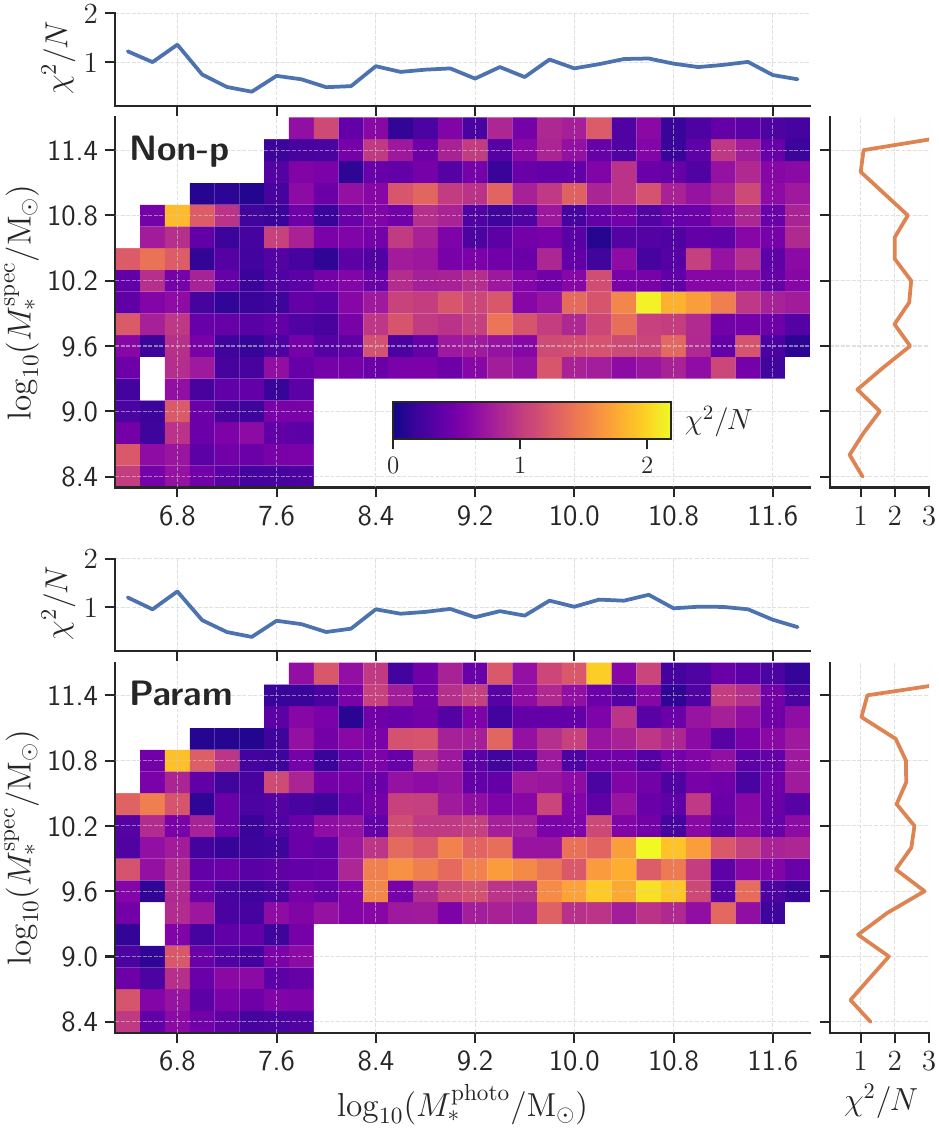}
    \caption{Reduced $\chi^2$ values of the fits to each $\nwprp$ measurement from the fiducial model. Results for the non-parametric model (top) and the parametric model (bottom) are shown. The top and right marginal panels in each plot display the reduced $\chi^2$ values aggregated over all $\nwprp$ measurements associated with each $M_*^{\rm photo}$ and $M_*^{\rm spec}$ bin.}
    \label{fig:fiducial_chi2}
\end{figure}

\begin{figure*}
    \centering
    \includegraphics[width=\textwidth]{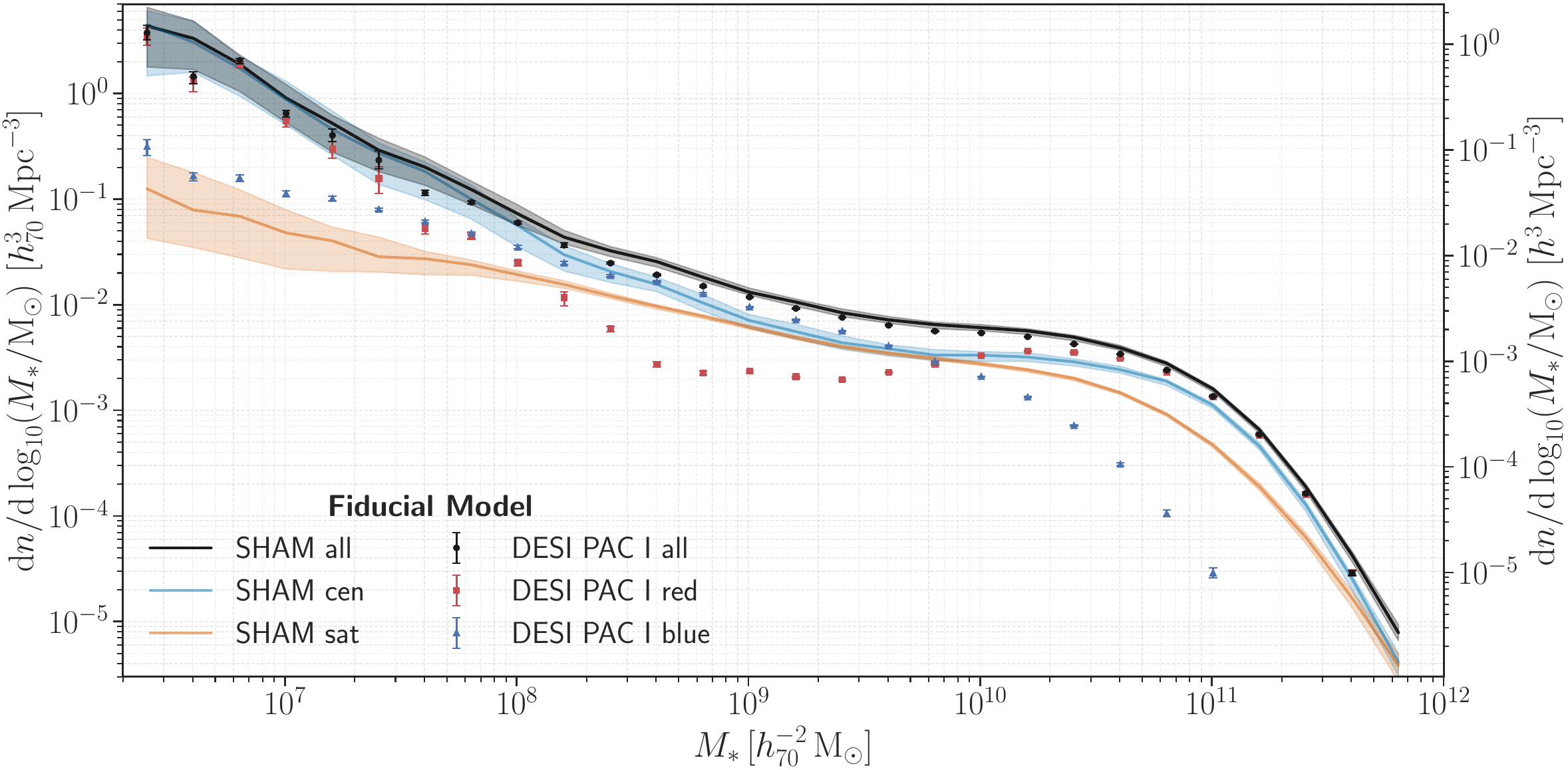}
    \caption{GSMFs derived from the fiducial non-parametric SHAM model. Results for central galaxies, satellite galaxies, and the total population are shown. For comparison, the GSMFs of the total, red, and blue populations from \citetalias{2025MNRAS.540.1635X} are also included.}
    \label{fig:fiducial_smf}
\end{figure*}

\subsection{SHMR-based SHAM model}
To model the $\nwprp$ measurements, we use the SHAM method, which assumes that more massive galaxies reside in more massive (sub)haloes with some level of scatter. Our SHAM model is constructed from the central and satellite SHMRs, $P_{\rm c}(M_* \mid M_{\rm h}; \boldsymbol{\theta}_{\rm c})$ and $P_{\rm s}(M_* \mid M_{\rm s}; \boldsymbol{\theta}_{\rm s})$, which are typically modelled as log-normal distributions around their mean relations: 
\begin{align}
    &P_{\rm c}(\log_{10}M_* \mid M_{{\rm h}}; \boldsymbol{\theta}_{\rm c})=\frac{1}{\sqrt{2\pi}\sigma_{\rm c}(M_{\rm h};\boldsymbol{\theta}_{\rm c}^{\sigma})}\times\notag\\
    &\exp{\left[-\frac{\log_{10}M_*-\left\langle\log_{10} (M_*|M_{\rm h};\boldsymbol{\theta}_{\rm c}^{\rm m})\right\rangle}{2\sigma_{\rm c}^2(M_{\rm h};\boldsymbol{\theta}_{\rm c}^{\sigma})}\right]}\,,\notag\\
    &P_{\rm s}(\log_{10}M_* \mid M_{{\rm s}}; \boldsymbol{\theta}_{\rm s})=\frac{1}{\sqrt{2\pi}\sigma_{\rm s}(M_{\rm s};\boldsymbol{\theta}_{\rm s}^{\sigma})}\times\notag\\
    &\exp{\left[-\frac{\log_{10}M_*-\left\langle\log_{10} (M_*|M_{\rm s};\boldsymbol{\theta}_{\rm s}^{\rm m})\right\rangle}{2\sigma_{\rm s}^2(M_{\rm s};\boldsymbol{\theta}_{\rm s}^{\sigma})}\right]}\,,\label{eq:shmr}
\end{align}
where $\boldsymbol{\theta}_{\rm c}^{\sigma}$ and $\boldsymbol{\theta}_{\rm s}^{\sigma}$ specify the scatter, and $\boldsymbol{\theta}_{\rm c}^{\rm m}$ and $\boldsymbol{\theta}_{\rm s}^{\rm m}$ describe the parameters of the mean relations. 

In \citetalias{2023ApJ...944..200X}, we found that assuming central and satellite galaxies share the same SHMR ($P_{\rm c}=P_{\rm s}$) provides a good fit to the $\nwprp$ measurements for SDSS galaxies with $M_*^{\rm spec}\in[10^{10.3},10^{11.3}]\,\ms$ and for DECaLS galaxies with $M_*^{\rm photo}\in[10^{7.9},10^{11.7}]\,\ms$. However, we find that this assumption fails for $\nwprp$ from the lower-mass DESI galaxies and low-mass DECaLS galaxies considered here, which were not explored in \citetalias{2023ApJ...944..200X}. Consequently, we model $P_{\rm c}$ and $P_{\rm s}$ separately in this work. We further find that the SHMR inferred in \citetalias{2023ApJ...944..200X} effectively corresponds to a combination of the true $P_{\rm c}$ at the high-mass end and $P_{\rm s}$ at the low-mass end. The necessity of treating $P_{\rm c}$ and $P_{\rm s}$ independently has also been noted in previous studies \citep{2016MNRAS.459.3040G, 2019MNRAS.488.3143B}.

There are several possible parametrizations for $\langle M_* \mid M_{\rm h}\rangle$ and $\langle M_* \mid M_{\rm s}\rangle$, such as the double–power-law form or the modified form from \citet{2013ApJ...770...57B}. However, given the large amount of information contained in hundreds of $\nwprp$ measurements, and the fact that explicit parametrizations may introduce additional systematics, we choose to adopt a non-parametric model with only a monotonicity prior. We specify 43 anchor points in $M_*$ for $M_{\rm vir}\in[10^{7.0},10^{15.4}]\,\msh$, evenly spaced with $\Delta\log_{10}(M_{\rm vir}/\msh)=0.2$, and 34 anchor points in $M_*$ for $m_{\rm peak}\in[10^{8.8},10^{15.4}]\,\msh$. The relations $\langle M_* \mid M_{\rm h}\rangle$ and $\langle M_* \mid M_{\rm s}\rangle$ over the full mass ranges are then obtained by linearly interpolating between these anchor points.
The lowest and highest halo masses may not host galaxies within the stellar mass range of interest, and the corresponding parameters cannot be constrained. Nevertheless, we retain all anchor points in the model, as we do not impose any prior on where the effective limits should lie.

For the scatter, \citetalias{2023ApJ...944..200X} found it to be nearly constant over $M_{\rm vir}\in[10^{10.0},10^{15.0}]\,\msh$. Although it is unclear whether this remains true when $P_{\rm c}$ and $P_{\rm s}$ are modelled separately or when extending to lower masses, we adopt a constant scatter as a first step and introduce two parameters, $\sigma_{\rm c}$ and $\sigma_{\rm s}$, to describe the scatter for haloes and subhaloes, respectively. This choice is motivated by the fact that the model already contains many parameters for the mean relations and that there is a potentially strong degeneracy between the low-mass SHMR slope and the scatter, so increasing the number of parameters in $\boldsymbol{\theta}_{\rm c}^{\sigma}$ and $\boldsymbol{\theta}_{\rm s}^{\sigma}$ at this stage may not be helpful. We will return to this point after the fiducial model is constrained.

The effect of reionization should also be taken into account when incorporating low-mass data, as it implies that low-mass (sub)haloes are not guaranteed to host a galaxy \citep{1992MNRAS.256P..43E}. This means that the normalizations of $P_{\rm c}$ and $P_{\rm s}$ are no longer unity at low masses and should be modified according to the halo occupation fraction (HOF). However, the HOF depends on the details of gas cooling and the physics of reionization, and several models with different predictions exist \citep{2020MNRAS.498.4887B,2023MNRAS.524.2290N,2025ApJ...983L..23N}. Therefore, we first adopt a fiducial model in which haloes are fully occupied, and then test different HOF models to examine how the results change.
  
There are two additional effects that may influence our results. The first is the so-called galaxy assembly bias or secondary bias \citep{2007MNRAS.374.1303C}, which remains a subject of active debate in HOD-related studies \citep{2014MNRAS.443.3044Z,2021MNRAS.502.3582Y}. In the context of our SHAM model, galaxy assembly bias implies that the scatter in the SHMR may not be random but may depend on secondary halo properties. For example, at fixed halo mass, haloes hosting higher-mass galaxies may form earlier \citep{2021NatAs...5.1069C} and therefore exhibit higher or lower halo bias than the average \citep{2007ApJ...657..664J}. Consequently, at fixed stellar mass, galaxies in lower-mass host haloes would preferentially reside in earlier-formed haloes and should follow the $\Wp$ of these haloes rather than the average, while galaxies in higher-mass host haloes would follow the $\Wp$ of later-formed haloes. Thus, the potential impact of galaxy assembly bias on our model is a differential effect across stellar-mass bins, depending on the competing changes in $\Wp$ for galaxies residing in earlier- and later-forming haloes, and remains a source of concern. The second effect is the use of an incorrect cosmology in the modelling; in our case, the Jiutian simulations assume the Planck18 cosmology. Our measurements are of $\nwprp$, which combines both abundance and clustering. It is therefore possible that, even if the Planck18-based model provides an apparently good fit to all $\nwprp$ measurements, the underlying $\bar{n}$ and $\Wp$ may both be incorrect in compensating ways. These two effects will be examined later.

\begin{figure*}
    \centering
    \includegraphics[width=\textwidth]{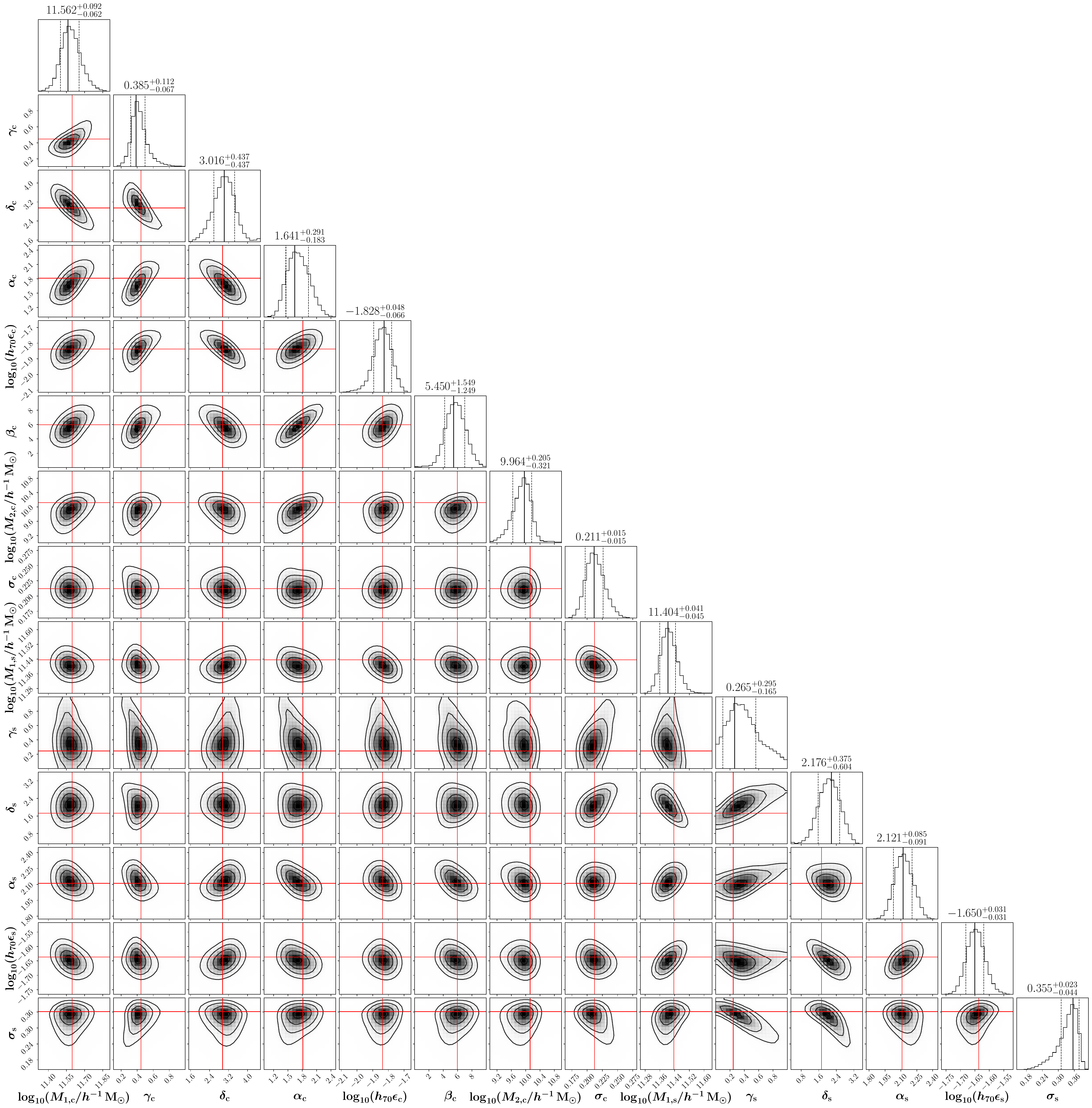}
    \caption{Posterior distributions of the parameters for the parametric fiducial SHMR model based on Equation~\ref{eq:bwcmod}. The contours show the joint distributions for each parameter pair, with levels corresponding to the 39.3\%, 86.5\%, and 98.9\% confidence intervals. The MAP estimates and their associated $1\sigma$ marginalized uncertainties are shown above each column. The maximum-posterior points are marked by the intersection of the two red lines.}
    \label{fig:mcmc}
\end{figure*}

\subsection{Constraining the fiducial non-parametric model}
We first constrain a fiducial model that adopts separate non-parametric mean SHMRs for central and satellite galaxies and a constant scatter, without including the HOF and assuming that (sub)haloes are fully occupied. This serves as a baseline for incorporating additional effects and extending the model to more complex forms.

To obtain the model prediction, once we specify $\boldsymbol{\theta}_{\rm c}^{\rm m}$ (43 parameters), $\boldsymbol{\theta}_{\rm s}^{\rm m}$ (34 parameters), and the scatter parameters $\boldsymbol{\theta}_{\rm c}^{\sigma}$ and $\boldsymbol{\theta}_{\rm s}^{\sigma}$ (each of dimension 1), we compute $P_{\rm c}$ and $P_{\rm s}$ using Equation~\ref{eq:shmr}. With the $\Wp$ and (sub)halo abundance tables, we then predict $\nwprp$ in the corresponding stellar-mass bins using Equation~\ref{eq:wp}. The model predictions are normalized in the same manner as the measurements and projected into the same PCA-truncated space. We define the $\chi^2$ as
\begin{equation}
    \chi^2 =
    (\tilde{\boldsymbol{\mathcal{A}}} - \tilde{\boldsymbol{\mathcal{A}}}_{\rm mod})^\top
    \boldsymbol{C}_{\tilde{\mathcal{A}}}^{-1}
    (\tilde{\boldsymbol{\mathcal{A}}} - \tilde{\boldsymbol{\mathcal{A}}}_{\rm mod})\,,
\end{equation}
where the covariance matrix $\boldsymbol{C}_{\tilde{\mathcal{A}}}$ is shown in Figure~\ref{fig:full_cov}. Then, we adopt a multivariate $t$-likelihood \citep{2016MNRAS.456L.132S} whose effective degrees of freedom are set by the  number of jackknife samples $N_{\mathrm{jk}}$:
\begin{equation}
\ln\mathcal{L}_{t}(d)
= -\frac{1}{2}\,N_{\mathrm{jk}}\,
  \ln\!\left(1 + \frac{\chi^2}{N_{\mathrm{jk}}-1}\right),
\end{equation}
which reduces to the Gaussian likelihood in the limit $N_{\mathrm{jk}}\rightarrow\infty$. The $t$-form provides a well-defined likelihood even when the covariance is estimated from a finite number of jackknife realizations, and naturally regularizes the impact of small-sample noise in the inverse covariance. We adopt wide uniform priors of 
$\boldsymbol{\theta}_{\rm c}^{\rm m},\,\boldsymbol{\theta}_{\rm s}^{\rm m} \in [10^{5.0},10^{13.0}]\,\ms$ and $\boldsymbol{\theta}_{\rm c}^{\sigma},\,\boldsymbol{\theta}_{\rm s}^{\sigma} \in [0,1]$, together with a penalty term that enforces the mean SHMRs to be monotonically increasing.
    
Since we have a high-dimensional parameter space with $p=79$, we use the Hamiltonian Monte Carlo (HMC) sampler \citep{1987PhLB..195..216D} to efficiently explore the posterior. HMC is a gradient-based method that simulates Hamiltonian dynamics and is well suited for sampling high-dimensional spaces with strong parameter correlations. In practice, we perform the sampling using the No-U-Turn Sampler \citep[NUTS;][]{2011arXiv1111.4246H} implemented in the \textsc{NumPyro} package \citep{2019arXiv191211554P}, an adaptive variant of HMC that automatically tunes the trajectory length and step size to avoid inefficient random-walk behaviour. All model components are written in fully differentiable form and compiled with \textsc{JAX} just-in-time (JIT), so that gradients and likelihood evaluations are executed as optimized XLA kernels, substantially accelerating each NUTS trajectory while keeping the entire inference pipeline end-to-end differentiable. Here and throughout, our fiducial sampling setup uses 36 chains, 1000 warm-up steps, and 5000 retained posterior samples per chain.

The constrained SHMRs from the non-parametric fiducial model are shown in panel~(a) of Figure~\ref{fig:fiducial_shmr}, with the corresponding stellar-to-halo mass ratios displayed in panel~(b). The model achieves reasonably good constraints over $M_{\rm vir}\in[10^{8.4},10^{15.0}]\,\msh$ and $m_{\rm peak}\in[10^{10.0},10^{14.4}]\,\msh$, with particularly high accuracy in the intermediate mass range. To assess the goodness of fit, we show in the top panel of Figure~\ref{fig:fiducial_chi2} the reduced $\chi^2$ for each $\nwprp$ measurement, without applying PCA or SVD truncations. We also present the reduced $\chi^2$ values obtained by combining all measurements within each $M_*^{\rm spec}$ and $M_*^{\rm photo}$ bin, using the full covariance matrices for these measurements. The corresponding model fits are listed in Appendix~\ref{sec:fits}. Overall, the fits across all mass bins are reasonably good, with reduced $\chi^2$ values lying in the range $0.5$–$2$. We observe that the reduced $\chi^2$ values are systematically larger when grouping measurements by $M_*^{\rm spec}$ ($\sim 2$) than by $M_*^{\rm photo}$ ($\sim 1$). This difference may arise because $\n$ varies much more strongly with $M_*$ than $\Wp$, making it easier for the $\n$ measurements to exhibit relatively larger residuals compared to $\Wp$. We also note that the reduced $\chi^2$ values fall below unity for $M_*^{\rm photo}<10^{8.4}\,\ms$. This is likely due to the presence of long-tailed noise distributions in the low-$S/N$ measurements. Assuming Gaussian uncertainties in these bins may therefore lead to an overestimation of the errors. In addition, jackknife resampling may also overestimate the uncertainties \citep{2009MNRAS.396...19N}.

To diagnose the convergence of the HMC sampling, in Figure~\ref{fig:diagnostics}, we show the rank-normalized split $\hat{R}$, bulk effective sample size $\mathrm{ESS}_{\rm bulk}$, and tail effective sample size $\mathrm{ESS}_{\rm tail}$ for the mean $M_*$ and scatter parameters at each $M_{\rm h}$. Over $10^{9.0}\msh<M_{\rm vir}<10^{14.4}\msh$ and $10^{10.0}\msh<m_{\rm peak}<10^{14.0}\msh$ for central and satellite galaxies, respectively, $\hat{R}<1.01$ indicates good mixing among the chains, while $\mathrm{ESS}>400$ indicates a sufficient number of effective samples. At lower and higher masses, $\hat{R}$ gradually increases because the measurements provide weaker constraints.

From Figure~\ref{fig:fiducial_shmr}, we find that haloes and subhaloes do exhibit distinct SHMRs. For $M_{\rm h}$ above the ratio peak around $10^{12.0}\,\msh$, haloes appear to host more massive galaxies than subhaloes, although the relatively large uncertainties for the subhalo measurements prevent a firm conclusion. In contrast, for $10^{10.4}\,\msh < M_{\rm h} < 10^{12.0}\,\msh$, haloes tend to host less massive galaxies than subhaloes. These trends for satellites and centrals are consistent with \citet[][see their Figure~10]{2019MNRAS.488.3143B}. According to \citet{2019MNRAS.488.3143B}, in low-mass (sub)haloes, satellite quenching timescales after infall are long, allowing satellites to continue growing while $m_{\rm peak}$ remains fixed, leading to a higher SHMR. In contrast, in high-mass (sub)haloes, galaxy growth occurs mainly through mergers, a process that is suppressed for satellites, causing a lower SHMR. 

Moreover, we find an interesting feature: the central SHMR exhibits a clear change in slope around $M_{\rm h}\simeq10^{10}\,\msh$, becoming shallower toward lower masses. This leads to an increase in the ratio at the low-mass end, as shown in panel~(b), a trend not reported in previous studies. This suggests that star formation in small haloes may be more efficient than previously thought. In principle, a similar upturn might also appear in the satellite SHMR at sufficiently low $m_{\rm peak}$. However, the satellite SHMR is poorly constrained for $m_{\rm peak}<10^{10.5}\,\msh$, as indicated by the large uncertainties; the residual constraint in this regime is likely dominated by the imposed monotonic prior. This is because the satellite constraints mainly arise from the one-halo term of the $\nwprp$ measurements. Many low-mass satellites reside in low-mass centrals whose virial radii are smaller than our minimum scale cut of $0.16\,\mpch$. Improving the constraints on the satellite SHMR will require a more accurate treatment of orphan subhaloes in order to fully exploit $\nwprp$ measurements on smaller scales. Meanwhile, another DESI study based on the group catalogue \citep{2021ApJ...909..143Y} and the conditional stellar mass function of BGS also finds a similar upturn feature in the SHMR at low masses \citep{2026Inpre}.

For the scatters, we find that under the constant-scatter assumption the model requires a larger value for subhaloes, $\sigma_{\rm s}=0.37^{+0.02}_{-0.03}$, than for haloes, $\sigma_{\rm c}=0.22^{+0.01}_{-0.02}$.

Once the SHMR model is constrained, the GSMF can be derived. The GSMF is an important quantity on its own, and comparing it with existing measurements helps assess potential modelling biases. In Figure~\ref{fig:fiducial_smf}, we show the GSMF derived from the fiducial non-parametric SHMR model, along with the separate contributions from central and satellite galaxies. The results are compared with those from \citetalias{2025MNRAS.540.1635X}, which agree well with the DESI BGS measurements obtained using the $V_{\rm max}$ method \citep{2025arXiv251101803M} down to $10^{9.0}\,h_{70}^{-2}\,\ms$, and with COSMOS-Web results \citep{2025A&A...695A..20S} at slightly higher redshift down to $10^{7.6}\,h_{70}^{-2}\,\ms$. We find that our GSMF for the full galaxy population is consistent with that of \citetalias{2025MNRAS.540.1635X} within $2\sigma$ for most stellar-mass bins. However, our results are systematically higher by about $\sim15\%$, which becomes non-negligible when all mass bins are considered together. Given that the high-mass GSMF in \citetalias{2025MNRAS.540.1635X} is relatively model independent and agrees well with the BGS $V_{\rm max}$ measurements, this offset is more likely to originate from modelling biases in our analysis. We will examine this in later sections. Examining centrals and satellites separately, we find that central galaxies dominate at fixed stellar mass, with satellites becoming comparable only within $M_*\in[10^{9.0},10^{10.0}]\,h_{70}^{-2}\,\ms$. This behaviour is driven primarily by the relative abundances of haloes and subhaloes at fixed (sub)halo mass, and is further influenced by differences in their SHMRs. We also show the GSMFs for red and blue galaxies from \citetalias{2025MNRAS.540.1635X}. An interesting result from \citetalias{2025MNRAS.540.1635X} is that red dwarfs dominate the low-mass end. Combined with our findings, this implies that most red dwarfs are central galaxies.

\begin{figure}
    \centering
    \includegraphics[width=\columnwidth]{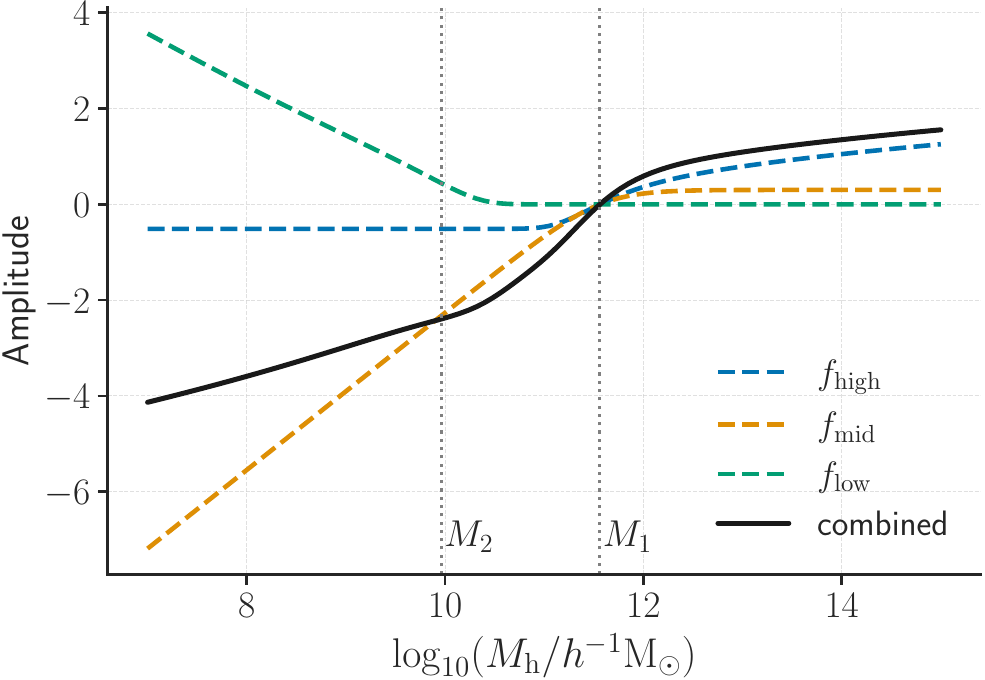}
    \caption{Illustration of the contributions of $f_{\rm low}$, $f_{\rm mid}$, and $f_{\rm high}$ to the fiducial parametric central SHMR. The curves are shown using the MAP parameter values.}
    \label{fig:bwc_decom}
\end{figure}

\begin{figure}
    \centering
    \includegraphics[width=\columnwidth]{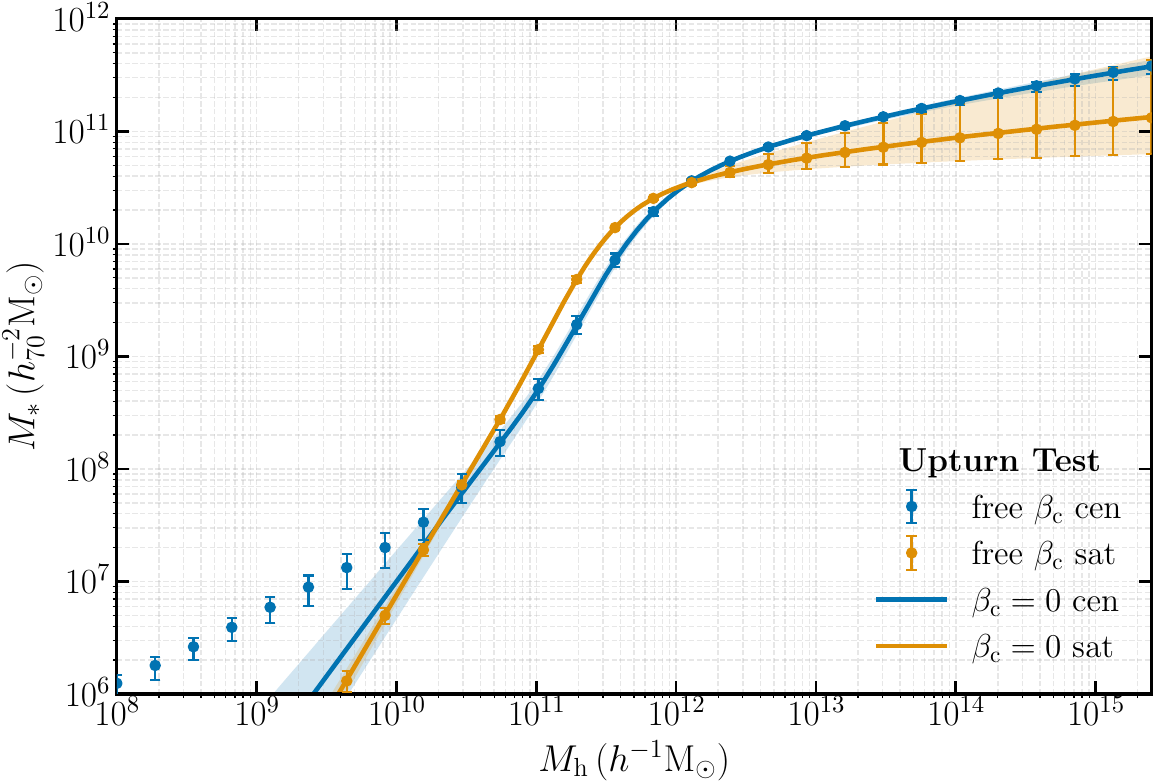}
    \caption{Comparison of the stellar-to-halo mass relations inferred with the fiducial parametric SHAM model when the low-mass upturn parameter is allowed to vary and when it is fixed to $\beta_{\rm c}=0$ \citep{2013ApJ...770...57B}. Points with error bars show the MAP and $1\sigma$ intervals for the free-$\beta_{\rm c}$ model, while solid curves and shaded regions show the MAP and $1\sigma$ intervals for the $\beta_{\rm c}=0$ model. Blue and orange denote the central- and satellite-galaxy SHMRs, respectively.}
    \label{fig:fiducial_upturn_test}
\end{figure}

\begin{figure}
    \centering
    \includegraphics[width=\columnwidth]{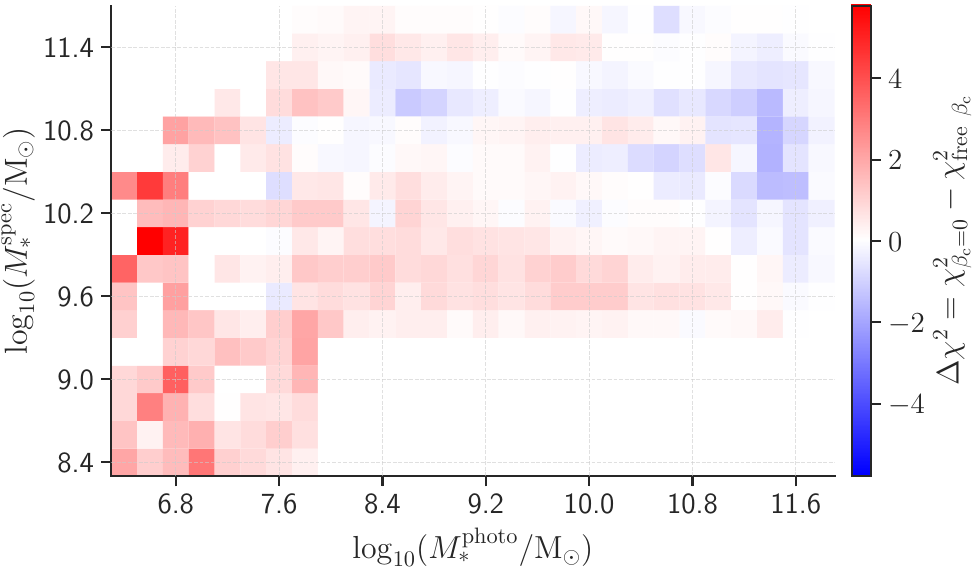}
    \caption{Difference in $\chi^2$ between the parametric SHMR models with a low-mass upturn (free $\beta_{\rm c}$) and without an upturn ($\beta_{\rm c}=0$) for the $\nwprp$ measurements in each pair of stellar-mass bins.}
    \label{fig:dchi2_fiducial_noupturn}
\end{figure}

\begin{figure}
    \centering
    \includegraphics[width=\columnwidth]{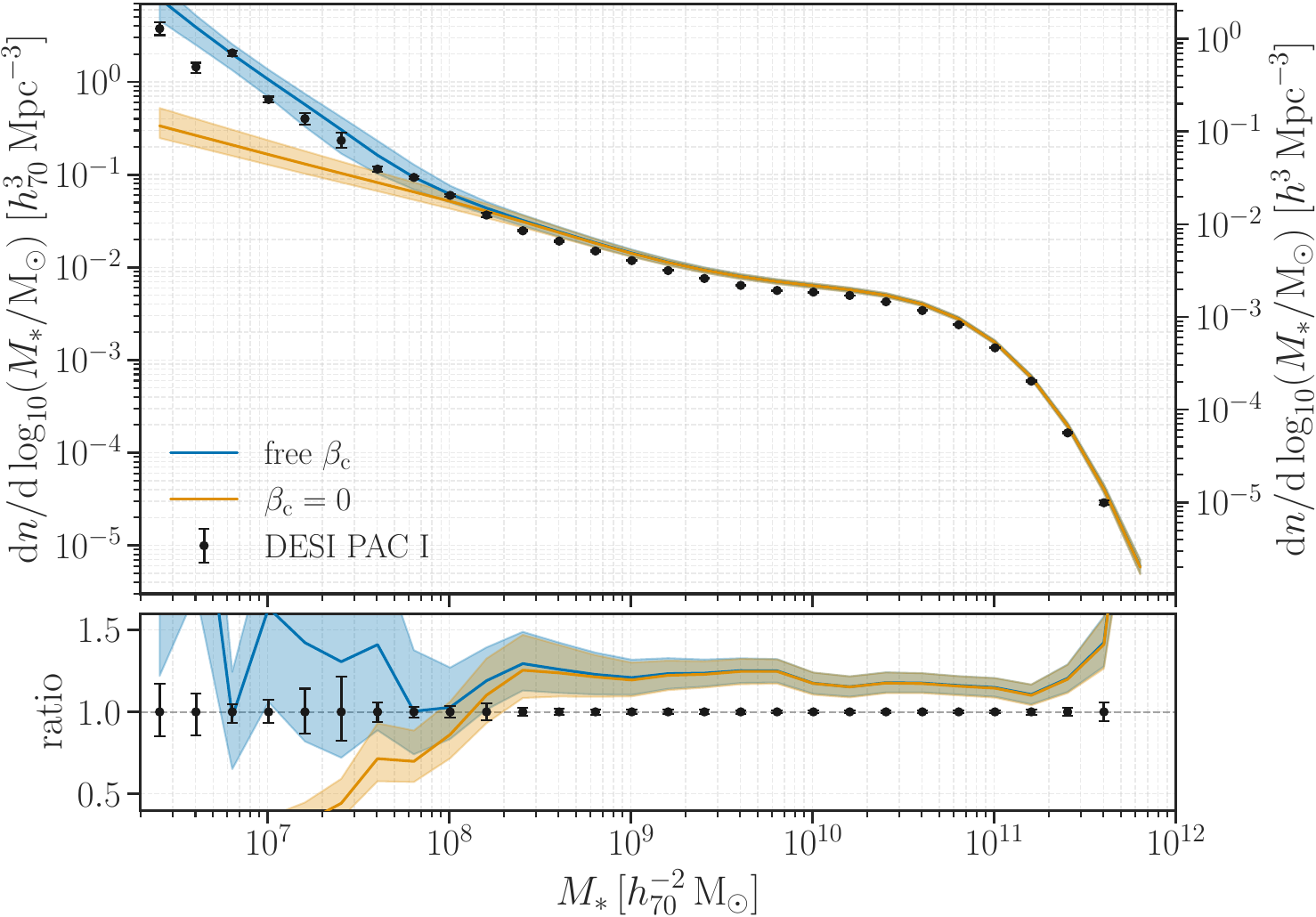}
    \caption{GSMFs derived from the parametric SHMR models with a low-mass upturn (free $\beta_{\rm c}$) and without an upturn ($\beta_{\rm c}=0$). For comparison, the total GSMF from \citetalias{2025MNRAS.540.1635X} is also shown.}
    \label{fig:fiducial_smf_noupturn}
\end{figure}

\subsection{A parametric model}

Although flexible, the non-parametric model may be susceptible to overfitting. Therefore, motivated by its results, we also explore a parametric formulation. We follow \citet{2013ApJ...770...57B}, who introduced a mean SHMR parametrization capable of capturing the upturn in the GSMF near $M_*\simeq10^{9.5}\,\ms$. For the mass range relevant here, an additional upturn around $M_*\simeq10^{7.5}\,\ms$ must also be described. We tested the \citet{2013ApJ...770...57B} form and found that it does not adequately capture the data and requires further modification. Therefore, we introduce additional terms beyond the \citet{2013ApJ...770...57B} formulation:
\begin{align}
    \log_{10}(M_*/h^{-2}_{70}\,\ms)
    &= \log_{10}\!\left([\epsilon h_{70}] [M_1/\msh]\right)
       + \log_{10}(h_{70}/h) \notag\\
    &\quad + f_{\rm high}\!\left(\log_{10}\!\left[\frac{M_{\rm h}}{M_1}\right]\right) - f_{\rm high}(0) \notag\\
    &\quad + f_{\rm mid}\!\left(\log_{10}\!\left[\frac{M_{\rm h}}{M_1}\right]\right) - f_{\rm mid}(0) \\
    &\quad + f_{\rm low}\!\left(\log_{10}\!\left[\frac{M_{\rm h}}{M_2}\right]\right)
         - f_{\rm low}\!\left(\log_{10}\!\left[\frac{M_1}{M_2}\right]\right)\,, \notag\label{eq:bwcmod}
\end{align}
where
\begin{align}
    f_{\rm high}(x;\delta,\gamma)
        &= \delta\,\frac{\left(\log_{10}[1+\exp (x)]\right)^{\gamma}}{1+\exp(10^{-x})}\,, \notag\\[3pt]
    f_{\rm mid}(x;\alpha)
        &= -\log_{10}\!\left(10^{-\alpha x}+1\right)\,, \notag\\[3pt]
    f_{\rm low}(x;\beta)
        &= \beta\,\frac{\log_{10}[1+\exp(-x)]}{1+\exp(10^{x})}\,.
\end{align}
Here, $M_1 > M_2$ are two characteristic halo masses at which the SHMR slope changes appreciably. Note that the parameter $M_1$ in the \citet{2013ApJ...770...57B} formulation does not coincide exactly with the location of the peak in $M_*/M_{\rm h}$. In our case, the peak occurs at a value roughly 0.4\,dex higher than $M_1$. The functions $f_{\rm low}$, $f_{\rm mid}$, and $f_{\rm high}$ primarily regulate the behaviour of the SHMR in the regimes $M_{\rm h}<M_2$, $M_2 < M_{\rm h} < M_1$, and  $M_{\rm h}>M_1$, respectively, though each contributes across a broader range. In the limit $M_{\rm h}\ll M_2$, the asymptotic slope approaches $\alpha - \beta/(2\ln 10)$, while for $M_{\rm h}\gg M_1$ the relation becomes a sub–power law with index $\gamma$. The intermediate regime involves a non-trivial combination of all terms. The normalization is chosen such that $\epsilon$ corresponds to the star-formation efficiency at $M_1$, although in practice only the combination $\epsilon h_{70}$ is constrained. The additional term $\log_{10}(h_{70}/h)=0.155$ accounts for the change in mass units. Our model reduces to that of \citet{2013ApJ...770...57B} when the $f_{\rm low}$ term is omitted or, equivalently, when $\beta=0$. 

We refit the measurements using a parametric SHMR model. For central galaxies, we adopt the mean relation in Equation~\ref{eq:bwcmod}. For satellite galaxies, we use the \citet{2013ApJ...770...57B} form with $\beta_{\rm s}=0$. Each relation is assigned a constant scatter across the full (sub)halo mass range. In total, the model includes 14 parameters—8 for centrals and 6 for satellites—significantly fewer than the 77 anchors plus 2 scatters in the non-parametric model. Parameter sampling is performed using the NUTS algorithm implemented in \textsc{NumPyro}. We find good convergence, with a maximum $\hat{R}$ of 1.0087 and minimum $\mathrm{ESS}_{\rm bulk}$ and $\mathrm{ESS}_{\rm tail}$ values of 4444.4 and 1792.4, respectively.

The posterior distributions of the parameters are shown in Figure~\ref{fig:mcmc}, generated with \textsc{corner} \citep{2016JOSS....1...24F}. The resulting SHMRs are presented in Figure~\ref{fig:fiducial_shmr}, and the corresponding $\chi^2/N$ values for each mass bin are displayed in the bottom panel of Figure~\ref{fig:fiducial_chi2}. From Figure~\ref{fig:fiducial_shmr}, the parametric model agrees well with the non-parametric result, except at the highest halo masses, where the parametric SHMR is slightly lower. This deviation remains within $2\sigma$, and both observational and simulation uncertainties are substantial in this regime. In addition, the monotonic prior in the non-parametric model may dominate the posterior in these regions where the uncertainties are large and cause the increase in the non-parametric model. From Figure~\ref{fig:fiducial_chi2}, the parametric model shows a reduced $\chi^2$ only slightly higher than that of the non-parametric model, as expected. This indicates that the chosen parametrization provides an adequate description of the SHMR. 

To illustrate the contributions of $f_{\rm low}$, $f_{\rm mid}$, and $f_{\rm high}$ to the central SHMR, we show their decomposition in Figure~\ref{fig:bwc_decom}, using the MAP parameter values from Figure~\ref{fig:mcmc}. The central SHMR closely follows the shape of $f_{\rm high}$ for $M_{\rm h}>M_1$, is governed by $f_{\rm mid}$ for $M_2<M_{\rm h}<M_1$, and is further modified by $f_{\rm low}$ at $M_{\rm h}<M_2$.

The posterior constraint, $\beta_{\rm c}=5.450^{+1.549}_{-1.249}$, shown in Figure~\ref{fig:mcmc}, provides evidence for a low-mass upturn in the central SHMR. To further quantify and understand this feature, we refit the $\nwprp$ measurements using a parametric model with $\beta_{\rm c}=0$, corresponding to the \citet{2013ApJ...770...57B} model. Figure~\ref{fig:fiducial_upturn_test} compares the SHMRs inferred from the models with free $\beta_{\rm c}$ and with $\beta_{\rm c}=0$. The differences in $\chi^2$ for the $\nwprp$ measurements in each pair of stellar-mass bins are shown in Figure~\ref{fig:dchi2_fiducial_noupturn}, while the corresponding GSMFs are compared in Figure~\ref{fig:fiducial_smf_noupturn}. The inferred SHMRs differ primarily in the central-galaxy relation at $M_{\rm h}<10^{10}\msh$, where the low-mass upturn appears. Consequently, the $\beta_{\rm c}=0$ model generally yields larger $\chi^2$ values than the fiducial model, particularly for measurements with $M_*^{\rm photo}<10^{8.0}\ms$. The upturn is driven by the rapid rise of the GSMF toward low stellar masses below $M_*<10^{8.0}\ms$, which produces a corresponding increase in the amplitude of the low-mass $\nwprp$ measurements. The $\beta_{\rm c}=0$ model fails to reproduce this steep low-mass rise in the GSMF. The overall difference in goodness of fit between the $\beta_{\rm c}=0$ and free-$\beta_{\rm c}$ models is $\Delta\chi^2=17.86$. The corresponding difference in the Bayesian information criterion is
\begin{equation}
    \Delta{\rm BIC}=2\Delta\ln\mathcal{L}-\Delta k\ln N=3.30,
\end{equation}
where $\Delta k=2$ is the difference in the number of free parameters ($\beta_\mathrm{c}$ and $M_\mathrm{2,c}$) and $N=1454$ is the dimension of the data vector. This indicates moderate evidence for a non-zero $\beta_{\rm c}$. More precise $\nwprp$ measurements at $M_*<10^{8.0}\ms$ will be important for strengthening the evidence for the low-mass upturn.

\begin{figure}
    \centering
    \includegraphics[width=\columnwidth]{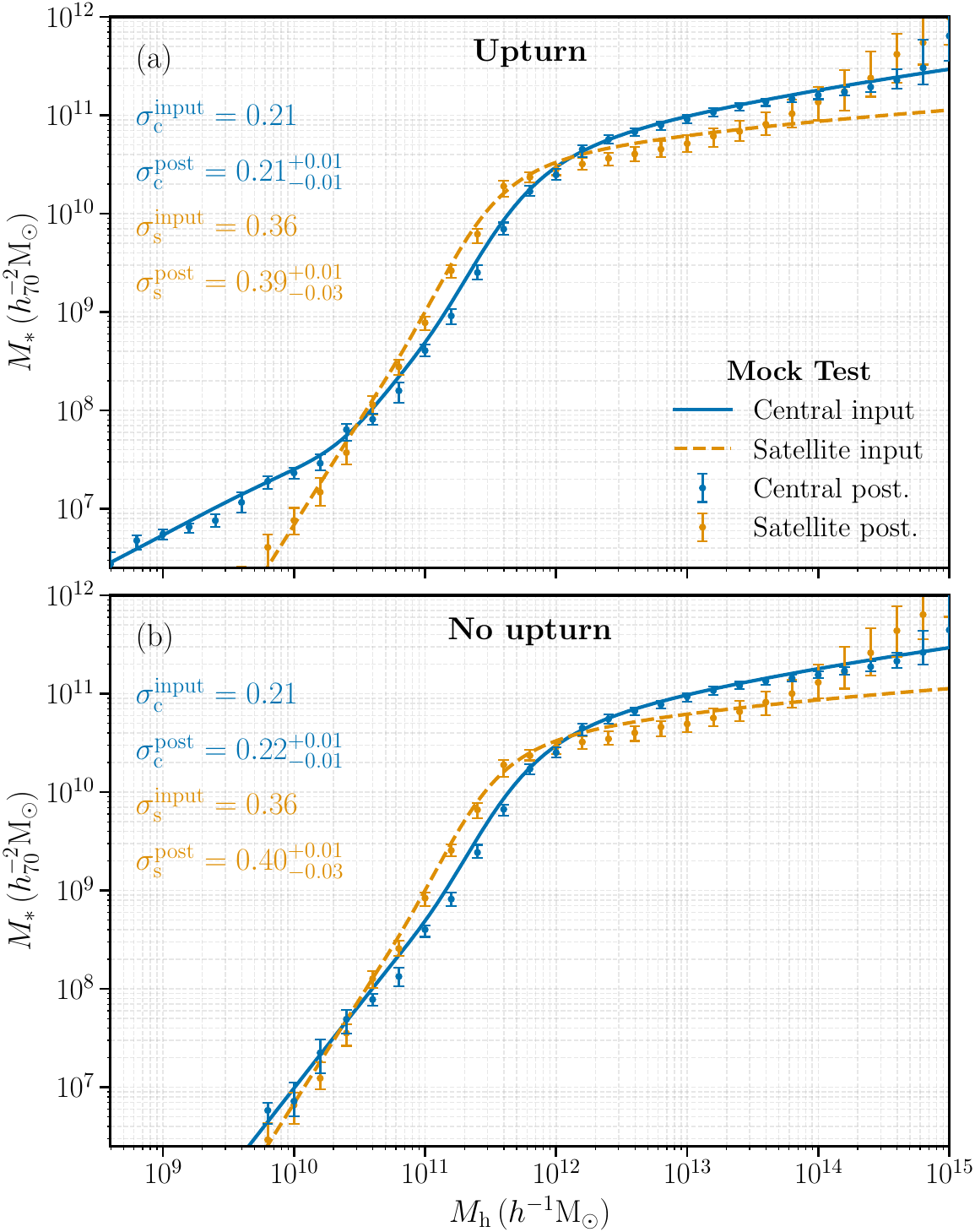}
    \caption{Mock-recovery tests of the SHAM model with (a) and without (b) a low-mass upturn in the central-galaxy stellar-to-halo mass relation. Solid and dashed curves show the input relations for central and satellite galaxies, respectively, while points with error bars show the MAP and $1\sigma$ intervals recovered from the mock measurements. Blue and orange denote central and satellite galaxies. The input and recovered scatters, $\sigma_{\rm c}$ and $\sigma_{\rm s}$, are listed in each panel.}
    \label{fig:mock}
\end{figure}

\begin{figure}
    \centering
    \includegraphics[width=\columnwidth]{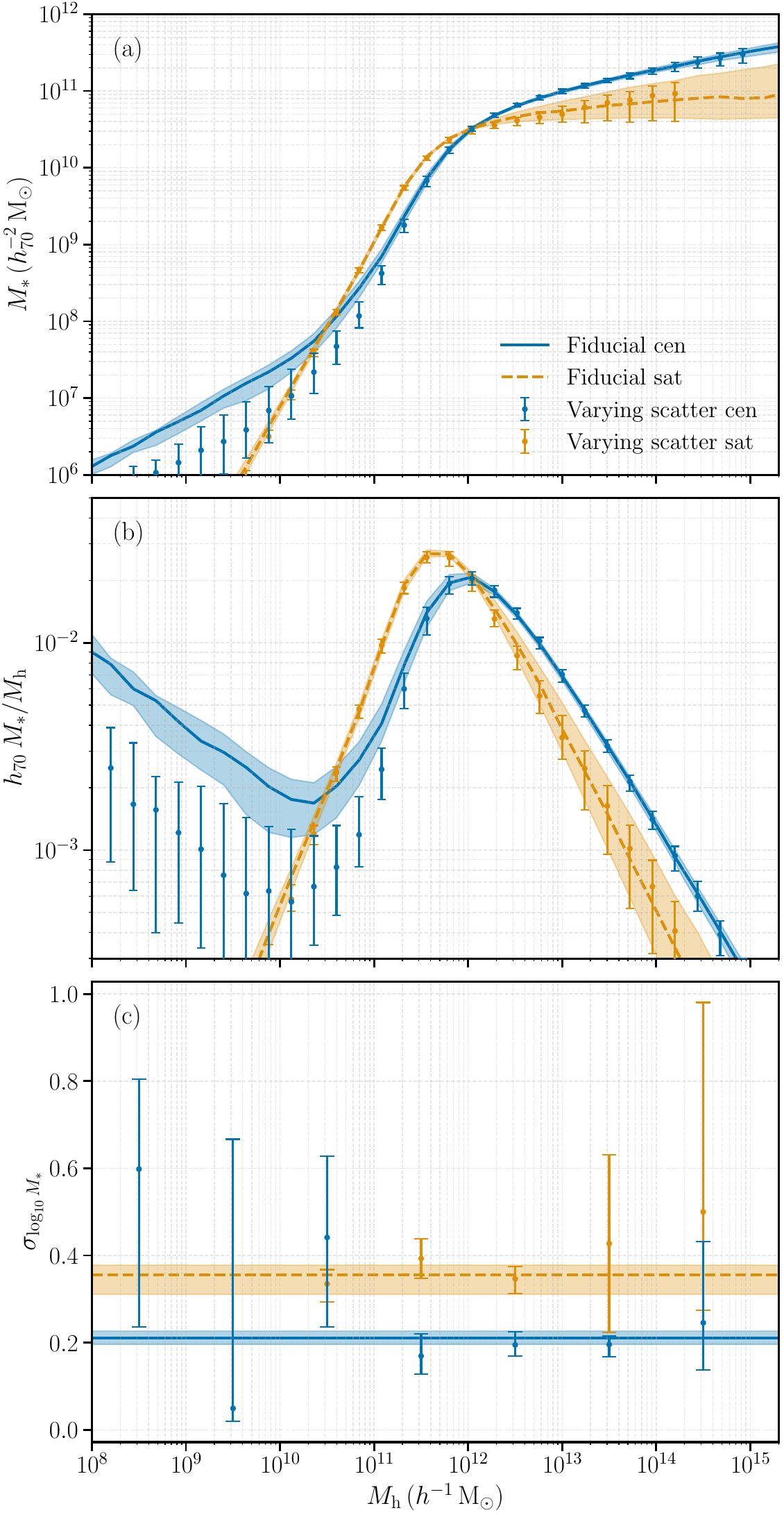}
    \caption{Same as Figure~\ref{fig:fiducial_shmr}, but comparing the fiducial parametric SHAM model with constant scatter to the parametric model with mass-dependent scatter in the SHMRs. Solid and dashed curves with shaded regions show the fiducial central- and satellite-galaxy results, respectively, while points with error bars show the corresponding results for the mass-dependent-scatter model. Panel~(c) shows the inferred scatter as a function of halo mass.}
    \label{fig:vary_shmr}
\end{figure}

\begin{figure}
    \centering
    \includegraphics[width=\columnwidth]{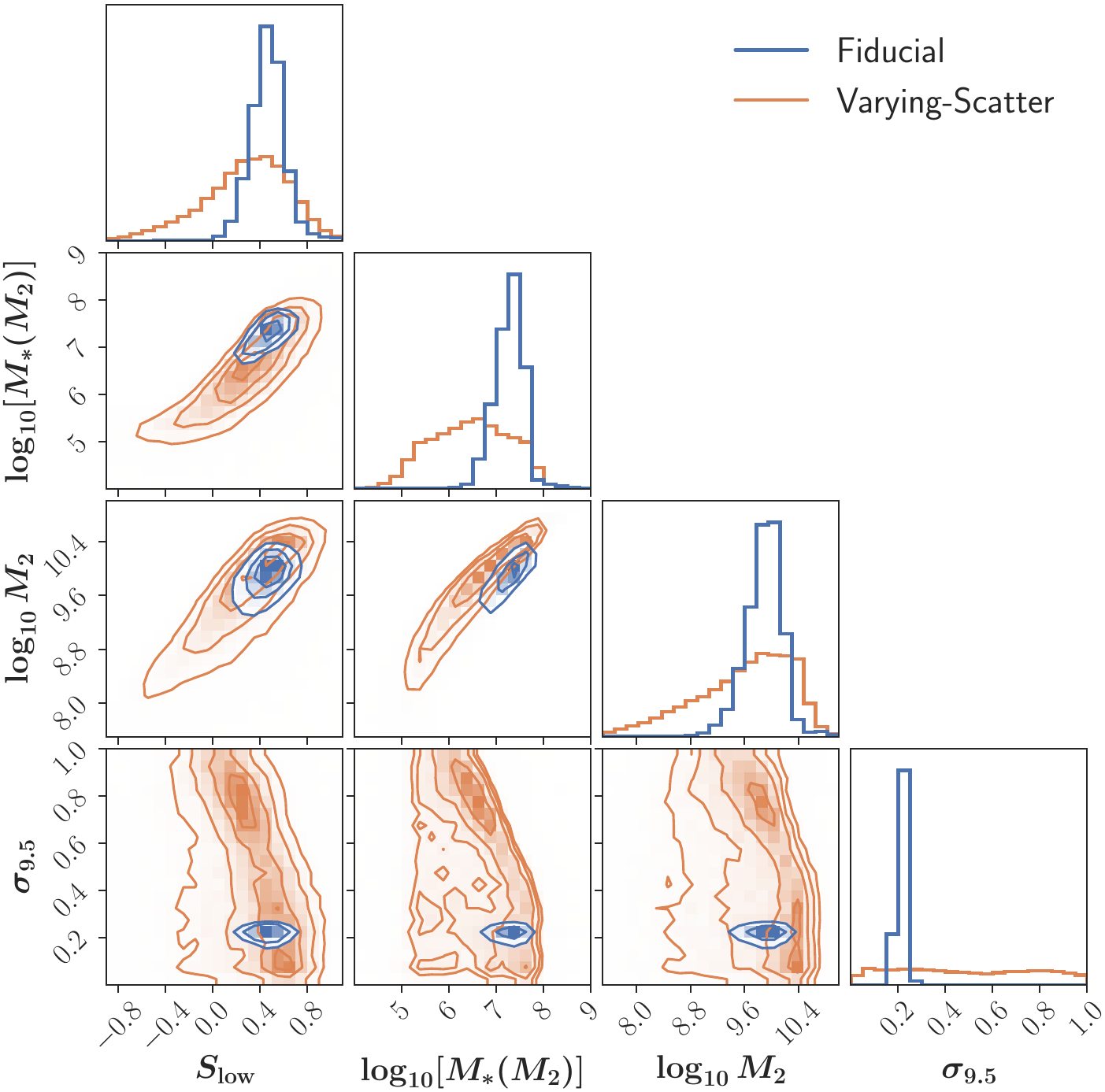}
    \caption{Posterior distributions of key parameters that determine the central SHMR at the low-mass end ($M_{\rm h}<M_2$), including the low-mass slope $S_{\rm low}=\alpha-\beta/(2\ln 10)$, the characteristic halo mass $M_2$ and its corresponding mean stellar mass $M_*(M_2)$, and the stellar-mass scatter $\sigma_{9.5}$ evaluated over $M_{\rm vir}\in[10^{9.0},10^{10.0}]\,\msh$. Results from the fiducial and varying-scatter parametric models are compared.}
    \label{fig:mcmc_vary}
\end{figure}

\begin{figure}
    \centering
    \includegraphics[width=\columnwidth]{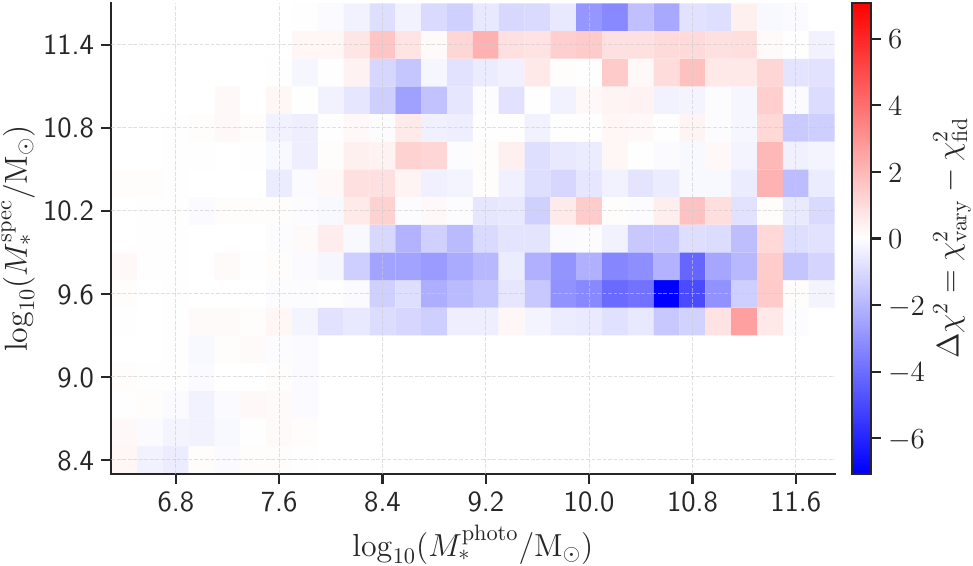}
    \caption{Difference in $\chi^2$ ($\Delta \chi^2 =\chi^2_{\rm vary}-\chi^2_{\rm fid}$) between the constant-scatter and varying–scatter parametric SHMR model for the $\nwprp$ measurements in each stellar-mass bin.}
    \label{fig:vary_chi2}
\end{figure}

\begin{figure}
    \centering
    \includegraphics[width=\columnwidth]{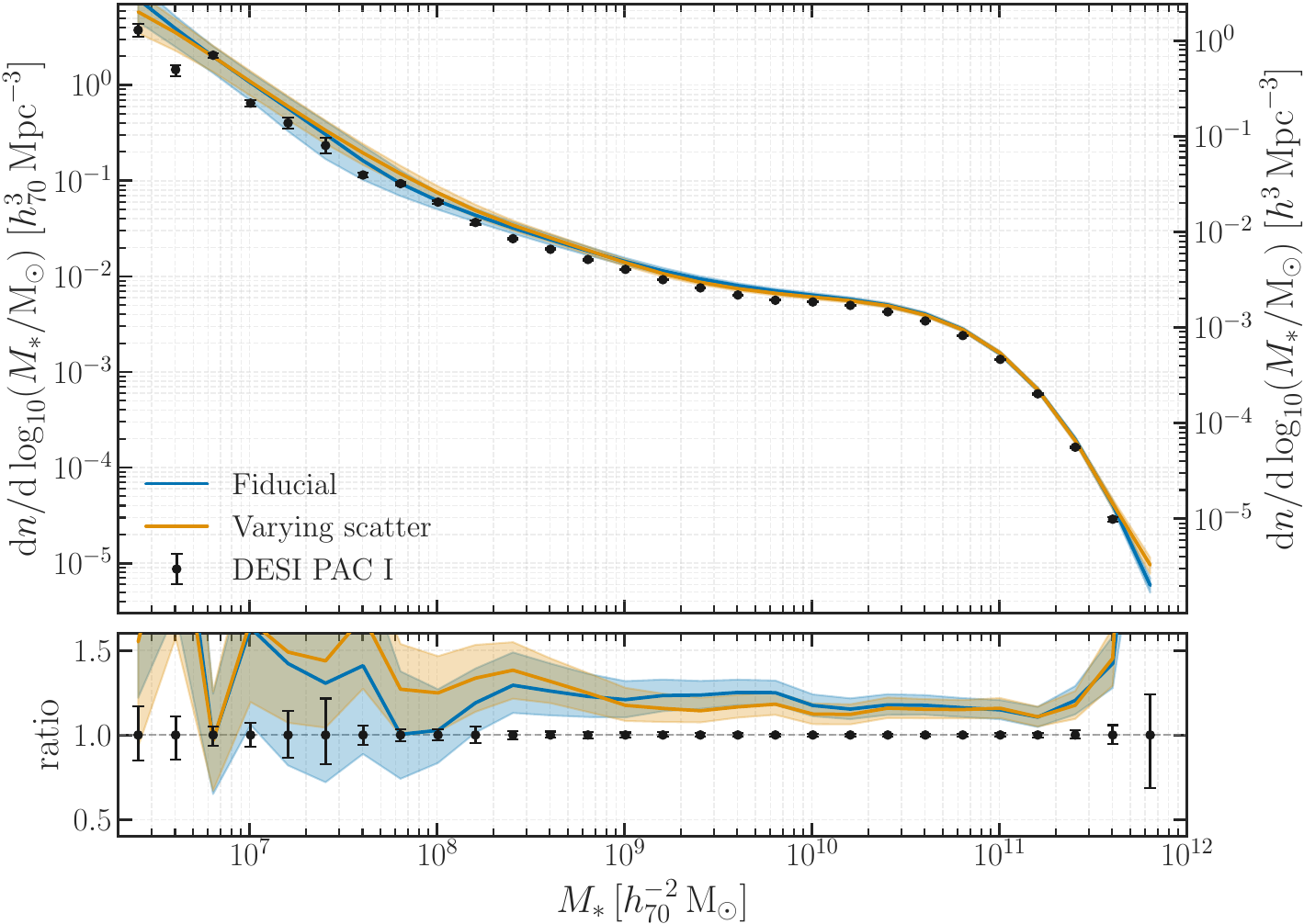}
    \caption{GSMFs derived from the fiducial and varying-scatter parametric models. For comparison, the total GSMF from \citetalias{2025MNRAS.540.1635X} is also shown.}
    \label{fig:vary_smf}
\end{figure}

\subsection{Mock Validation of the PAC and SHAM Pipelines}

Given the complexity of the PAC measurement and SHAM modelling pipelines, we perform an end-to-end light-cone mock test to validate whether the input SHMRs can be recovered without bias. Since constructing a full DECaLS light-cone mock down to $r\lesssim 22.5$ is currently unrealistic, we instead generate a pseudo-mock using several approximations while retaining all the key ingredients that affect the PAC measurements.

We construct the light cone using the Jiutian-300 simulation. First, we replicate the periodic simulation box at $z=0.1$ in a $3\times3\times3$ configuration. We then select a random observer position and orientation within the central box and calculate the angular coordinates (RA and Dec.) and redshift-space comoving distances of the (sub)haloes. The redshift-space comoving distances are subsequently converted into redshifts. We retain and save a full-sky light cone of (sub)haloes over $0<z<0.25$. In this construction, we neglect the redshift evolution of the (sub)halo population over this redshift range.

We then assign stellar masses to the (sub)haloes using the best-fitting fiducial parametric SHMRs. To test whether the inferred upturn in the central-galaxy SHMR can be reliably recovered, we construct a second stellar-mass assignment using a central-galaxy SHMR without an upturn, obtained by setting $\beta_{\rm c}=0$, while keeping the other parameters unchanged. To reproduce the depths of the BGS and DECaLS samples, we select mock galaxies above the corresponding stellar-mass completeness limits, $M_{\rm c95}(z)$, shown in Figure~\ref{fig:SM_limits}, consistent with the selections used in this study. Next, we mimic the angular selection of the BGS sample. Instead of running a full fibre-assignment pipeline, we adopt a simpler approach based on a HEALPix map with $N_{\rm side}=512$. We retain only mock galaxies located in pixels that contain observed BGS galaxies. Within each retained pixel, mock BGS galaxies are randomly down-sampled according to the mean spectroscopic completeness of that pixel and are assigned the corresponding completeness weights. In this approximation, we neglect fibre-collision effects among BGS galaxies, which have a negligible impact on the PAC measurements. Meanwhile, we generate two random catalogues that reproduce the angular and redshift distributions of the mock BGS and DECaLS galaxy samples, respectively.

Using the mock BGS and DECaLS galaxy samples, we reproduce the mock $\nwprp$ measurements with a pipeline similar to that described in Section~\ref{sec:measurements}. We use mock BGS galaxies over $0.01<z<0.2$ and mock DECaLS galaxies over $0<z<0.25$, matching the redshift ranges adopted for the observational measurements. For each mock BGS galaxy, we reassign stellar masses to the full mock DECaLS catalogue according to
\begin{equation}
    M_*^{\prime}
    =
    M_*
    \frac{D_{L,\mathrm{BGS}}^2}
         {D_{L,\mathrm{DECaLS}}^2},
\end{equation}
where $D_{L,\mathrm{BGS}}$ is the luminosity distance of the BGS galaxy and $D_{L,\mathrm{DECaLS}}$ is the true luminosity distance of the DECaLS galaxy. This transformation mimics the observational procedure in which the stellar masses of DECaLS galaxies are estimated by assuming that they lie at the redshift of the corresponding BGS galaxy. The simple luminosity-distance scaling neglects differences in the $k$-correction. This approximation should be adequate for DECaLS galaxies physically associated with the BGS galaxy, which dominate the clustering signal, because their redshifts are similar. It mainly affects uncorrelated foreground and background galaxies, whose contribution is corrected for in the subsequent measurement procedure.

Because the mock DECaLS light cone extends only to $z=0.25$, higher-redshift background galaxies are absent from the ACCF measurement. As a result, the mock $\nwprp$ measurements would contain fewer uncorrelated pairs and could have a different S/N from the observations. To obtain more realistic measurements, we add an additional Poisson-distributed random pair count in each angular-separation bin. The expected number of added pairs is determined by the difference between the observed and mock angular number densities, $\bar{S}_2^{\rm obs}-\bar{S}_2^{\rm mock}$, with the effective area of each $\theta$ bin inferred from the corresponding random pair counts. These additional pairs represent the uncorrelated high-redshift DECaLS background that is absent from the mock light cone. Accordingly, in the estimator given by Equation~\ref{eq:LSe}, we replace $\bar{S}_2^{\rm mock}$ with $\bar{S}_2^{\rm obs}$.

Finally, we obtain the PAC measurements and their covariance matrices for $M_*^{\rm photo}>10^{6.7}\ms$, using the same $M_*^{\rm spec}$ bins as in the observations. The mass resolution of Jiutian-300 is insufficient to model the two lowest $M_*^{\rm photo}$ bins. We then fit the mock $\nwprp$ measurements with the fiducial non-parametric SHAM model for both the upturn and no-upturn mocks. The input and recovered posterior SHMRs are compared in Figure~\ref{fig:mock}. For all resolved (sub)halo-mass bins, the input SHMRs are recovered within $2\sigma$. The model also correctly recovers the presence of the low-mass upturn in the upturn mock and its absence in the no-upturn mock. The only noticeable deviation occurs at the high-mass end of the satellite SHMR, where the non-parametric model exhibits an offset similar to that found in the observational fit. This supports our interpretation that the measurements provide only weak constraints in this regime, so the inferred SHMR is largely controlled by the monotonicity prior. Overall, the upturn mock demonstrates that the pipeline can recover an injected low-mass upturn, while the no-upturn mock provides a null test showing that the fitting procedure does not generate a spurious upturn.

It is also worth noting that the mock constraints on the central-galaxy SHMR at the low-mass end are slightly tighter than those obtained from the observations. This may indicate that the mock remains somewhat optimistic. For example, uncertainties in the stellar-mass estimates are not included. Nevertheless, the mock tests provide a strong validation of our measurement and modelling pipeline.

\begin{figure}
    \centering
    \includegraphics[width=\columnwidth]{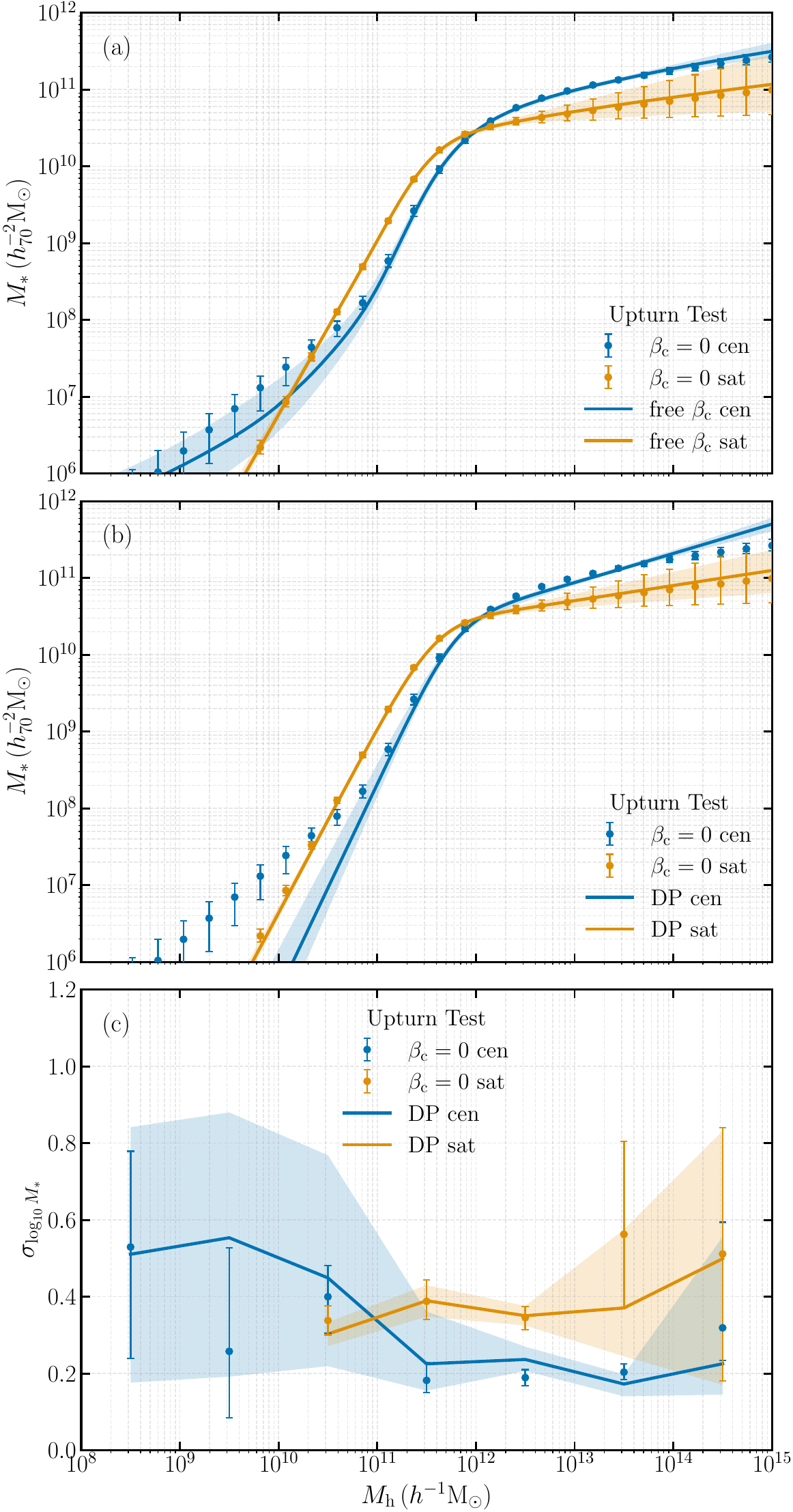}
    \caption{Low-mass upturn test for the parametric SHAM models with mass-dependent scatter. Panel~(a) compares the model with free $\beta_{\rm c}$ to that with $\beta_{\rm c}=0$, while panel~(b) compares the $\beta_{\rm c}=0$ model to the double-power-law model. Panel~(c) shows the corresponding mass-dependent scatters for the latter two models. Points with error bars show the $\beta_{\rm c}=0$ results, while solid curves with shaded regions show the free-$\beta_{\rm c}$ results in panel~(a) and the double-power-law results in panels~(b) and~(c).}
    \label{fig:vary_upturn_test}
\end{figure}

\begin{figure}
    \centering
    \includegraphics[width=\linewidth]{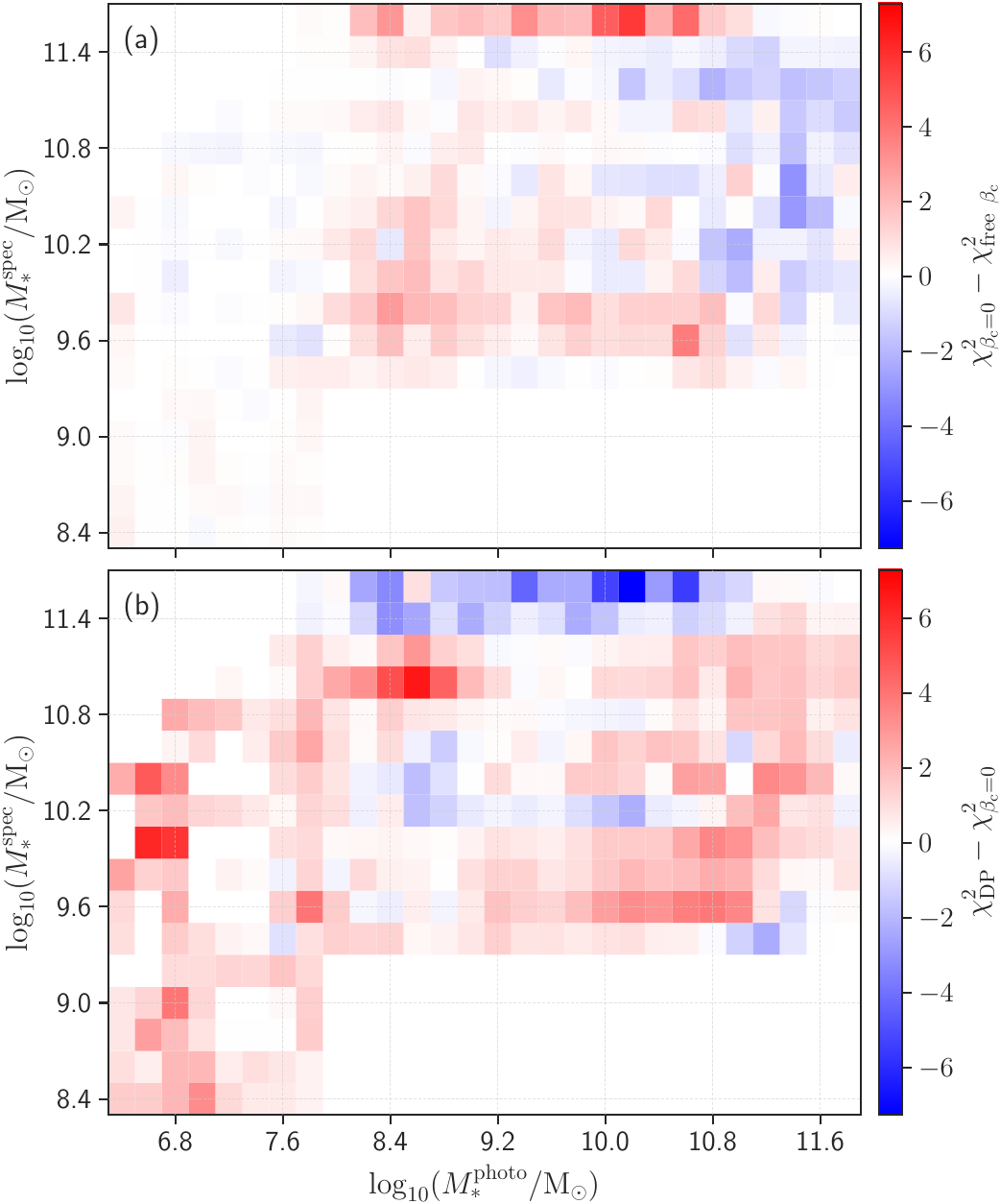}
    \caption{Differences in model $\chi^2$ to the PAC measurements between the parametric SHAM models with mass-dependent scatter. Panel~(a) shows $\Delta\chi^2=\chi^2_{\beta_{\rm c}=0}-\chi^2_{\mathrm{free}\ \beta_{\rm c}}$, such that positive values favour the model with free $\beta_{\rm c}$. Panel~(b) shows $\Delta\chi^2=\chi^2_{\rm DP}-\chi^2_{\beta_{\rm c}=0}$, such that positive values favour the $\beta_{\rm c}=0$ model and negative values favour the double-power-law model. Each cell corresponds to one combination of spectroscopic and photometric stellar-mass bins.}
    \label{fig:dchi2_vary_noupturn}
\end{figure}

\section{Extended SHAM models and tests}\label{sec:extended}
In this section, we model or test additional effects that may influence the galaxy–halo connection but are not included in the fiducial model.

\subsection{Effects of Mass-dependent scatter}
Building on the fiducial model, we now examine whether the scatters in the SHMRs can vary with halo mass and whether such variation can be constrained using only the $\nwprp$ measurements.

Here, we focus only on the parametric model, as allowing additional freedom in the scatter of the non-parametric model makes convergence difficult to achieve. We introduce eight values of $\sigma_{\rm c}(M_{\rm vir})$ for haloes over $M_{\rm vir}\in[10^{8.0},10^{15.0}]\,\msh$ and six values of $\sigma_{\rm s}(m_{\rm peak})$ for subhaloes over $m_{\rm peak}\in[10^{10.0},10^{15.0}]\,\msh$. We rerun the fits and show the resulting mean SHMRs (panel~(a)) and the scatters (panel~(c)) as functions of (sub)halo mass in Figure~\ref{fig:vary_shmr}. Results from the fiducial models are also shown for comparison. The varying-scatter model also shows reasonably good convergence, with a maximum $\hat{R}$ of 1.0136 and minimum $\mathrm{ESS}_{\rm bulk}$ and $\mathrm{ESS}_{\rm tail}$ values of 2948.4 and 1718.0, respectively. Only two parameters have $\hat{R}>1.01$.

From Figure~\ref{fig:vary_shmr}(c), we find that the scatters are reasonably well constrained and remain nearly unchanged relative to the fiducial case over $M_{\rm vir}\in[10^{11.0},10^{15.0}]\,\msh$ and $m_{\rm peak}\in[10^{10.0},10^{13.0}]\,\msh$, consistent with the conclusions of \citetalias{2023ApJ...944..200X}. At $M_{\rm h}>10^{14.0}\,\msh$, the scatters are poorly constrained due to limited signal in the measurements, and at $M_{\rm h}<10^{10.0}\,\msh$ they are also poorly constrained because of degeneracies with the mean relation. Allowing mass-dependent scatter also leads to larger uncertainties in the inferred mean SHMRs. 

To better illustrate the degeneracy at the low-mass end, we show in Figure~\ref{fig:mcmc_vary} the posterior distributions of four key parameters that govern the central SHMR in this regime from the parametric model. These include the low-mass slope of the mean relation, $S_{\rm low}=\alpha-\beta/(2\ln 10)$, the characteristic halo mass $M_2$ and its corresponding mean stellar mass $M_*(M_2)$, and the scatter $\sigma_{9.5}$ evaluated over $M_{\rm vir}\in[10^{9.0},10^{10.0}]\,\msh$. A clear bimodality is present in the posterior of $\sigma_{9.5}$: one mode prefers a smaller $\sigma_{9.5}$, a steeper slope, and a larger turnover mass, while the other shows the opposite trend. In the fiducial model, the scatter is tightly constrained by the high-mass measurements, and the prior of a constant scatter removes the degeneracy. As a result, the low-mass behaviour is more tightly constrained and falls into the branch with smaller $\sigma_{9.5}$. However, in principle this degeneracy is difficult to break using only the $\nwprp$ measurements. As discussed in Appendix~\ref{sec:extra}, the halo $\Wp$ varies very little at the low-mass end, leaving the abundance as the primary source of information. Consequently, many different models can produce nearly identical galaxy abundances at low masses, as illustrated in Figure~\ref{fig:mcmc_vary}. Additional low-mass observables that are more sensitive to halo mass, such as gravitational lensing \citep{2025arXiv250920458T,2025arXiv250920434T} or higher-order extensions of the PAC method, are required to break this degeneracy.
 
To illustrate the improvement obtained by introducing additional $\sigma_{\rm c}$ and $\sigma_{\rm s}$ parameters, Figure~\ref{fig:vary_chi2} shows the $\Delta\chi^2$ values relative to the fiducial model for the parametric case. The mass-dependent-scatter model yields slightly lower $\chi^2$ values in many stellar-mass bins, with a total improvement of $\Delta\chi^2=-12.7$.

The GSMF derived from the varying-scatter parametric model is shown in Figure~\ref{fig:vary_smf}, together with the results from the fiducial model and from \citetalias{2025MNRAS.540.1635X}. The varying-scatter model agrees well with the fiducial model across the full stellar-mass range. Allowing a mass-dependent scatter does not resolve the $\sim15\%$ systematic offset at the high-mass end. At the low-mass end, although the SHMR is allowed to vary substantially, the posterior GSMF remains as tightly constrained as in the fiducial model, confirming that different SHMR forms can yield similar galaxy abundances.

Similar to the test performed for the fiducial model, we also examine the low-mass upturn in the varying-scatter model, particularly to determine whether this feature persists given the degeneracy between the low-mass slope and the scatter. In Figure~\ref{fig:vary_upturn_test}(a), we compare the constrained varying-scatter model with $\beta_{\rm c}=0$ to the corresponding model with free $\beta_{\rm c}$, while Figure~\ref{fig:dchi2_vary_noupturn}(a) shows the resulting $\Delta\chi^2$. We find, however, that the varying-scatter model with $\beta_{\rm c}=0$ can still exhibit an upturn at $M_{\rm vir}<10^{10.0}\msh$. This occurs because the $\beta_{\rm c}=0$ functional form can itself produce an upturn at the low-mass end. In the fiducial model, its curvature is instead used to describe the weaker feature at $10^{10.0}\msh<M_{\rm vir}<10^{12.0}\msh$. Once the scatter is allowed to vary with halo mass, this intermediate-mass curvature can be absorbed by the mass dependence of the scatter, allowing the curvature in the mean relation to reproduce the upturn at the lowest masses. Relative to the model with free $\beta_{\rm c}$, the $\beta_{\rm c}=0$ model yields only $\Delta\chi^2=5.18$ and $\Delta{\rm BIC}=-9.39$. Thus, the modest improvement in $\chi^2$ does not justify the additional freedom in $\beta_{\rm c}$, and the varying-scatter model with $\beta_{\rm c}=0$ already provides a good fit to the measurements. This does not, however, imply that a low-mass upturn is unnecessary, because the $\beta_{\rm c}=0$ varying-scatter model can itself reproduce this feature.

To further examine the low-mass upturn, we fit the $\nwprp$ measurements using a simpler double-power-law (DP) model for the mean relation:
\begin{equation}
    M_{*}=
    \frac{2k}
    {(M_{\rm h}/M_0)^{-\alpha}+(M_{\rm h}/M_0)^{-\beta}}\,.
\end{equation}
By construction, this model does not exhibit a low-mass upturn, although it is not guaranteed to reproduce exactly the same relation as the $\beta_{\rm c}=0$ model at other halo masses. In Figure~\ref{fig:vary_shmr}(b), we compare the mean relations inferred from the varying-scatter DP model and the $\beta_{\rm c}=0$ model. As designed, the DP model shows no low-mass upturn. The corresponding $\Delta\chi^2$ distribution is shown in Figure~\ref{fig:dchi2_vary_noupturn}. The DP model yields larger $\chi^2$ values at $M_*^{\rm photo}<10^{8.0}\ms$, similar to the comparison between the fiducial models with free $\beta_{\rm c}$ and $\beta_{\rm c}=0$. The total difference is $\Delta\chi^2=25.6$, with $\Delta{\rm BIC}=11.03$, indicating a strong preference for the varying-scatter model with $\beta_{\rm c}=0$. Although the difference may also receive contributions from other stellar-mass ranges, the distribution of $\Delta\chi^2$ in Figure~\ref{fig:dchi2_vary_noupturn}(b) supports the presence of a low-mass upturn in the central-galaxy SHMR even when the scatter is allowed to vary with halo mass.

\begin{figure}
    \centering
    \includegraphics[width=\columnwidth]{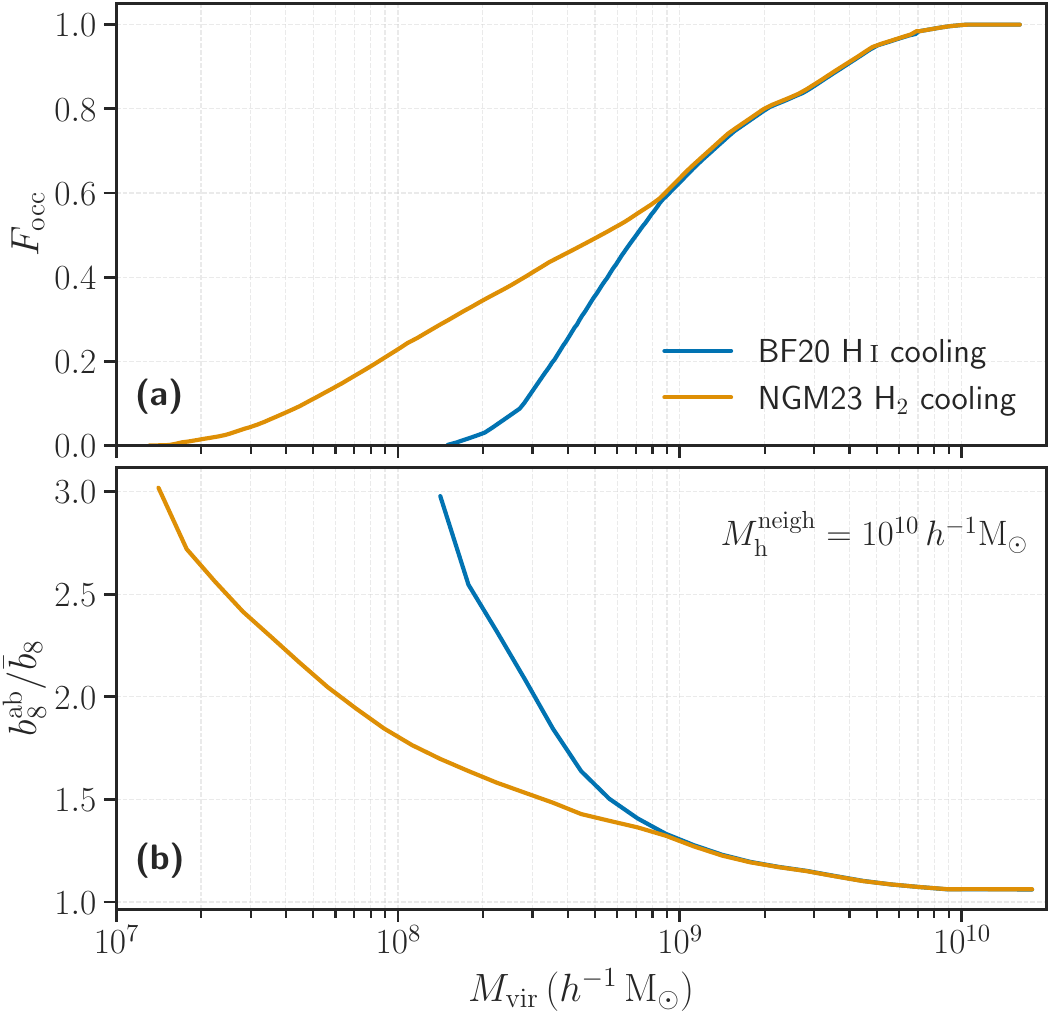}
    \caption{(a) Halo occupation fraction at $z=0$ from \citet{2025ApJ...983L..23N}, based on the \citetalias{2020MNRAS.498.4887B} H\,\textsc{i} and \citetalias{2023MNRAS.524.2290N} H$_2$ gas-cooling models for the onset of galaxy formation. (b) Difference between the mean $\Wp$ of all haloes and the mean $\Wp$ of the earliest formed fraction $F_{\rm occ}$ of haloes, evaluated at $\rp=8\,\mpch$. The $\Wp$ values are computed using neighbouring haloes and subhaloes of mass $10^{10.0}\,\msh$.
}
    \label{fig:HOF}
\end{figure}

\begin{figure}
    \centering
    \includegraphics[width=\columnwidth]{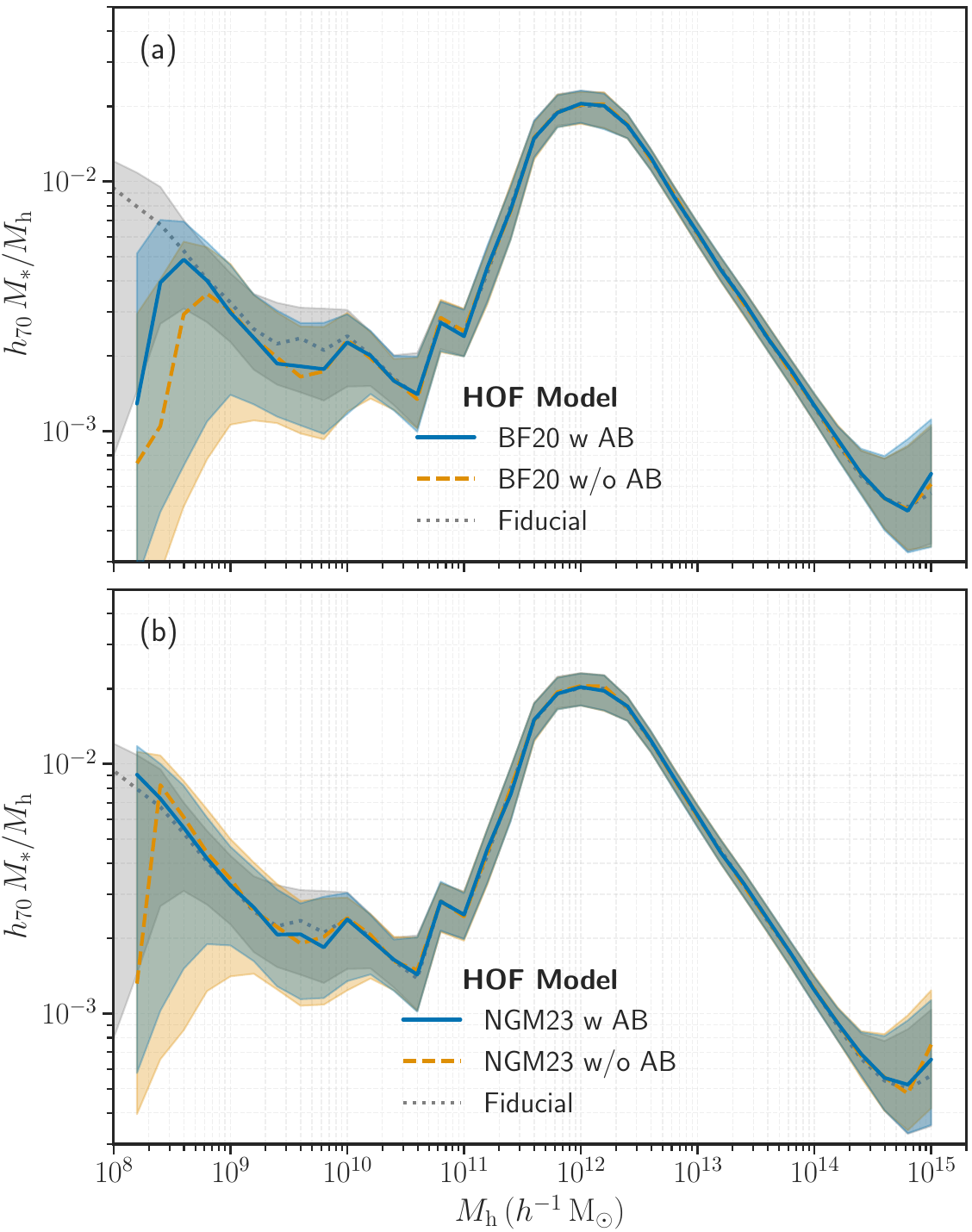}
    \caption{Mean central SHMRs from the non-parametric models incorporating different HOF prescriptions. Results obtained with and without accounting for assembly bias are shown. The fiducial non-parametric model is included for comparison.}
    \label{fig:HOF_shmr}
\end{figure}

\begin{figure}
    \centering
    \includegraphics[width=\columnwidth]{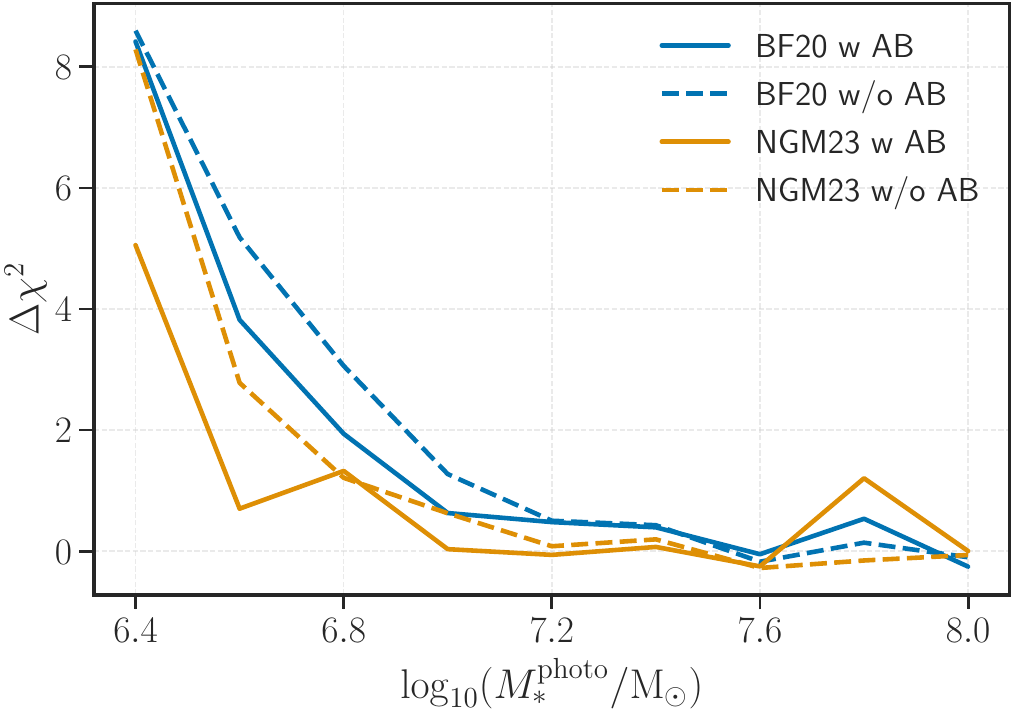}
    \caption{Total $\Delta\chi^2$ for the $\nwprp$ measurements in each $M_{*}^{\rm photo}$ bin, comparing various HOF models to the non-parametric fiducial model.}
    \label{fig:HOF_chi2}
\end{figure}

\begin{figure}
    \centering
    \includegraphics[width=\columnwidth]{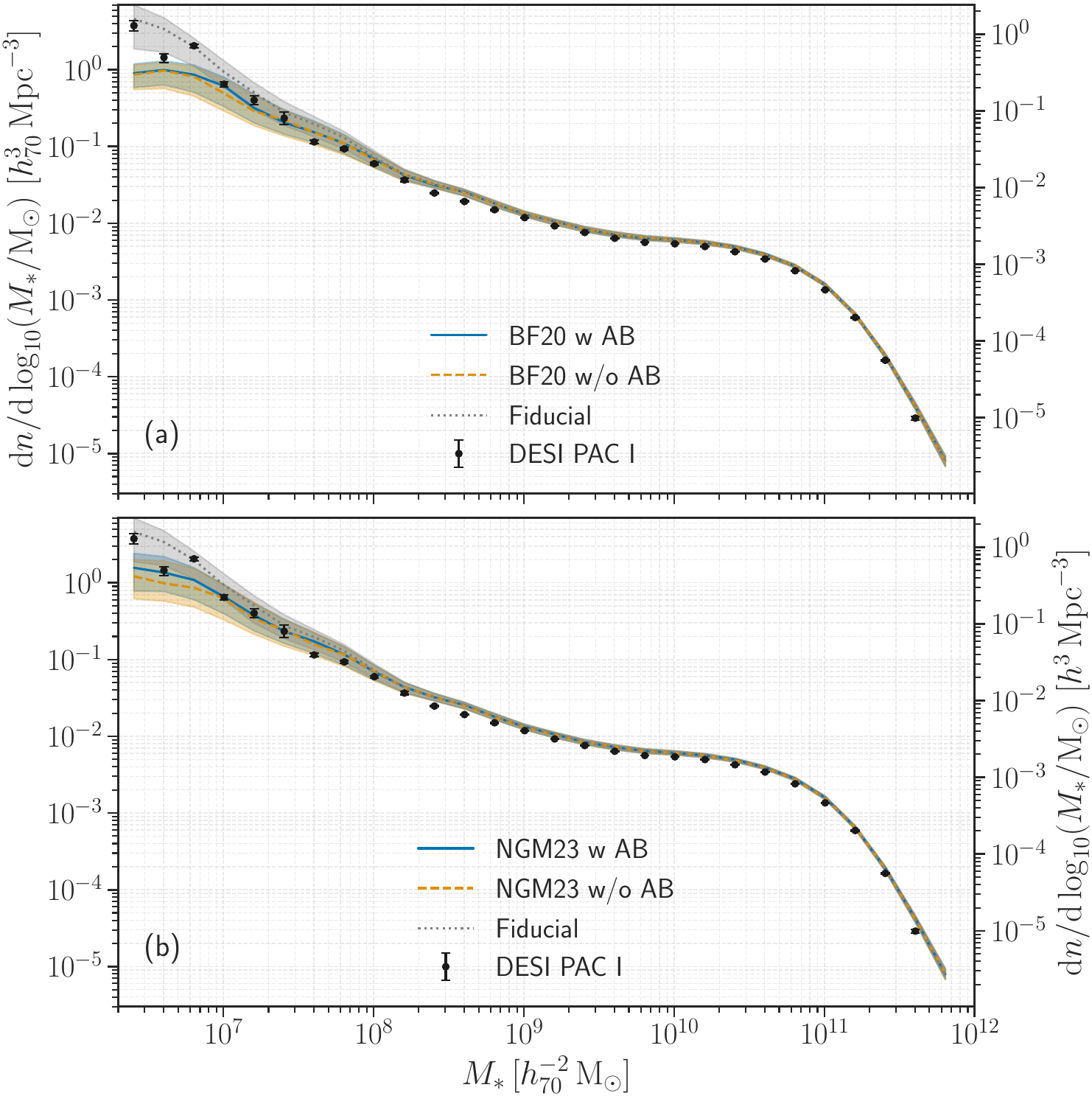}
    \caption{GSMFs derived from the non-parametric SHAM models incorporating different HOF prescriptions. Results obtained with and without accounting for assembly bias are shown. The fiducial non-parametric model and the GSMF from \citetalias{2025MNRAS.540.1635X} are included for comparison.}
    \label{fig:HOF_smf}
\end{figure}

\begin{figure*}
    \centering
    \includegraphics[width=\textwidth]{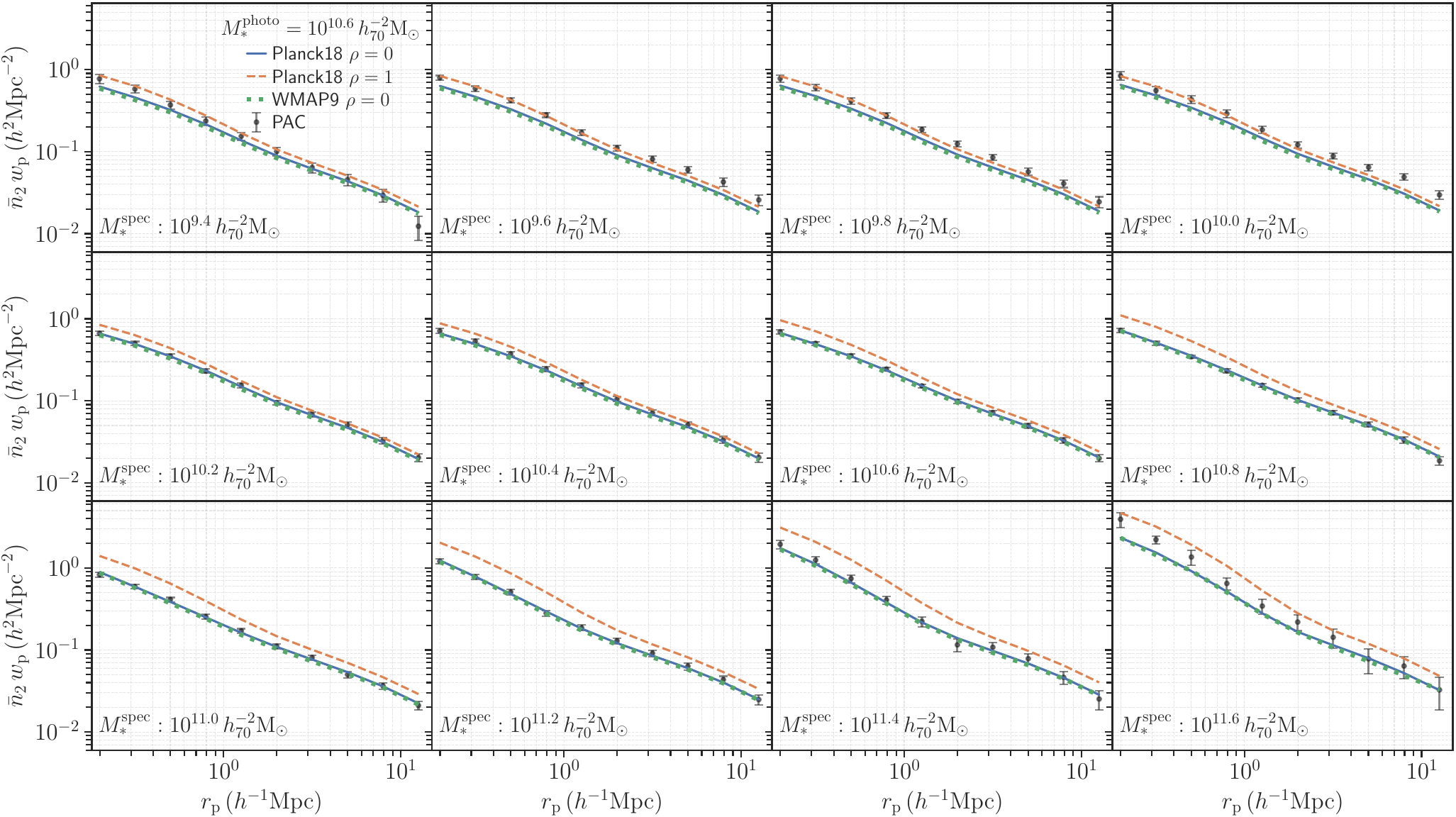}
    \caption{Comparison of $\nwprp$ predictions for $M_{*}^{\rm photo}=10^{10.6}\,h^{-2}_{70}\,\ms$ and various $M_{*}^{\rm spec}$ bins, generated from mocks with different cosmologies and galaxy assembly bias assumptions. Shown are predictions from Planck18 cosmology with no assembly bias ($\rho=0$) and with maximal assembly bias ($\rho=1$) using Jiutian-1G, as well as WMAP9-like cosmology with no assembly bias using \textsc{CosmicGrowth}. All mocks adopt the same SHAM prescription using the fiducial parametric MAP SHMR. Observation data are also shown for comparison.
}
    \label{fig:test_sys_nwp}
\end{figure*}

\begin{figure*}
    \centering
    \includegraphics[width=\textwidth]{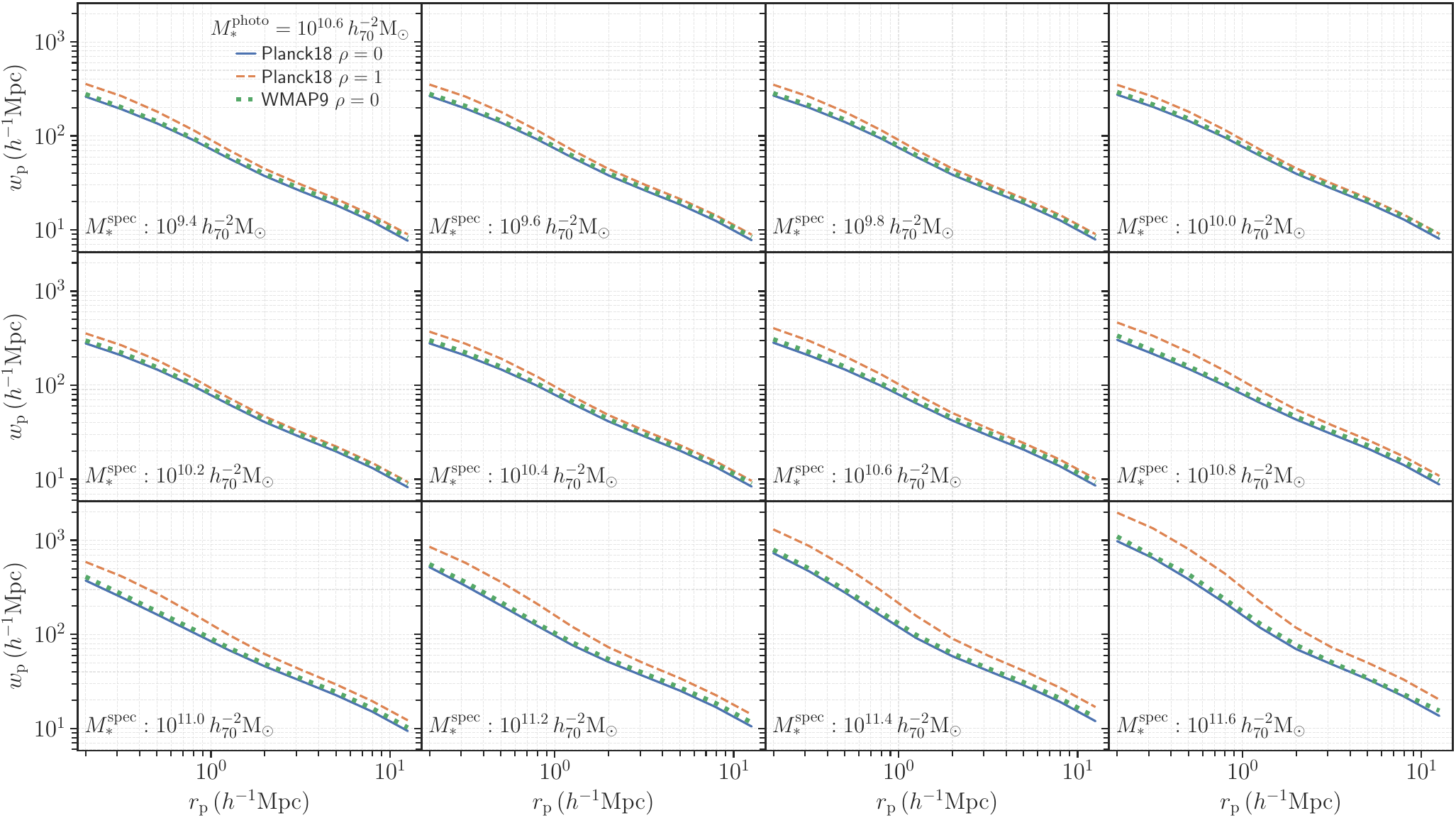}
    \caption{Same as Figure~\ref{fig:test_sys_nwp}, but showing the comparison of the predicted $\Wp(\rp)$.}
    \label{fig:test_sys_wp}
\end{figure*}

\subsection{Effects of reionization on halo occupation fraction}
Another factor that likely affects our results at the low-mass end is the impact of reionization. After reionization, the characteristic halo mass above which gas can cool increases substantially. Consequently, many haloes relevant to our analysis may fail to form a galaxy. In this situation, the forms of Equation~\ref{eq:shmr} for $P_{\rm c}$ and $P_{\rm s}$ may no longer be appropriate, since their normalizations are unity and therefore assume that every (sub)halo hosts a galaxy.
 
The simplest correction is to introduce a halo occupation fraction (HOF) $F_{\rm occ}$ into the SHMR model, such that $P'_{\rm c}=F_{\rm occ}(M_{\rm h},z)\,P_{\rm c}$ and $P'_{\rm s}=F_{\rm occ}(m_{\rm peak},z_{\rm peak})\,P_{\rm s}$. This formulation implies a bimodal stellar-mass distribution at fixed halo mass—a combination of a log-normal component and a $\delta$-function at $M_{*}=0$. Although this is somewhat idealized, as it effectively assumes that reionization either completely suppresses galaxy formation or has no effect, the true forms of $P'_{\rm c}$ and $P'_{\rm s}$ under reionization are still uncertain. The HOF approach therefore serves as a reasonable first step.

The HOF is determined by the galaxy-formation threshold at each redshift and by the subsequent halo assembly history \citep{2002ApJ...568...52W,2003MNRAS.339...12Z,2009ApJ...707..354Z} up to the epoch of interest. The formation threshold itself is set by gas-cooling physics prior to reionization \citep{2020MNRAS.498.4887B,2023MNRAS.524.2290N} and by the reionization process \citep{1992MNRAS.256P..43E}. Because the HOF depends on several poorly constrained ingredients and our low-mass $\nwprp$ measurements have limited $S/N$, we adopt two representative HOF models from \citet{2025ApJ...983L..23N}: the H\,\textsc{i}-cooling and H$_2$-cooling models. These are based on the H\,\textsc{i}-dominated cooling model of \citet[][hereafter \citetalias{2020MNRAS.498.4887B}]{2020MNRAS.498.4887B} and the H$_2$ cooling model of \citet[][hereafter \citetalias{2023MNRAS.524.2290N}]{2023MNRAS.524.2290N} before reionization, a detailed reionization prescription from \citetalias{2023MNRAS.524.2290N}, and a \citet{2002ApJ...568...52W} assembly-history model.

The resulting HOFs at $z=0$ are shown in Figure~\ref{fig:HOF}(a). Both models begin to depart from unity at $M_{\rm vir}\simeq10^{10}\,\msh$, decreasing toward lower masses. They agree down to $M_{\rm vir}\simeq10^{9}\,\msh$, after which the H\,\textsc{i} model reaches zero at $\sim10^{8}\,\msh$, whereas the H$_2$ model remains non-zero down to $\sim10^{7}\,\msh$. This behaviour reflects the fact that, over $10^{9}$–$10^{10}\,\msh$, reionization physics dominates and is identical in both models, while below $10^{9}\,\msh$ pre-reionization cooling becomes the controlling factor. Allowing H$_2$ cooling enables gas to cool in substantially lower-mass haloes, leading to a higher HOF in the \citetalias{2023MNRAS.524.2290N} model. We apply the HOFs to our SHMR models to obtain $P'_{\rm c}$ for haloes. For subhaloes, reionization has no appreciable impact over the mass range relevant to our analysis, so $P_{\rm s}$ is left unchanged. 

After applying the HOF, assembly-bias effects may become more significant and warrant greater concern. For mass bins with $F_{\rm occ}=1$, assembly bias manifests as a differential effect: at fixed stellar mass, galaxies in lower-mass host haloes preferentially reside in earlier-formed haloes and should follow the $\Wp$ of those haloes rather than the average, while galaxies in higher-mass host haloes tend to follow the $\Wp$ of later-formed haloes. In contrast, when $F_{\rm occ}<1$, all galaxies necessarily reside in the subset of early-forming haloes capable of forming galaxies before reionization. In this regime, assembly bias becomes a net effect and can be substantially stronger.

Therefore, for each HOF model, we consider two limiting cases: (i) galaxies randomly occupy haloes, and (ii) galaxies occupy the fraction $F_{\rm occ}$ of earliest-forming haloes. These correspond to the cases with no assembly-bias effect and with the maximal assembly-bias effect, respectively. They are identical for haloes with $F_{\rm occ}=1$. We define the formation redshift $z_{\rm form}$ as the redshift at which a halo first assembles half of its present mass. To implement the second case, we construct new $\Wp^{\rm hh}$ and $\Wp^{\rm hs}$ tables in which $\Wp$ is computed using, at each halo mass, the $F_{\rm occ}$ subset of haloes with the highest $z_{\rm form}$ rather than the full sample.

However, this introduces an additional complication for the extrapolation of the halo tables below $M_{\rm vir}<10^{8.7}\,\msh$, since the same procedure must now be performed for the earliest-forming halo subsets. Fortunately, a reasonable extrapolation is still possible: we find that haloes with $M_{\rm vir}\in[10^{8.7},10^{9.9}]\,\msh$ selected at the same percentile of $z_{\rm form}$ have $\Wp$ values that agree to within $\sim 10\%$, even though $\Wp$ can vary by nearly a factor of three across different $z_{\rm form}$ percentiles. A detailed validation is provided in Appendix~\ref{sec:extra}. Following the previous strategy, we therefore construct the extrapolated $\Wp^{\rm hh}$ and $\Wp^{\rm hs}$ tables for $M_{\rm vir}\in[10^{7.0},10^{8.7}]\,\msh$ using haloes in $[10^{8.7},10^{9.5}]\,\msh$. For each mass bin, haloes are selected according to the corresponding $F_{\rm occ}(M_{\rm vir})$ fraction with the highest $z_{\rm form}$. Since the \citetalias{2020MNRAS.498.4887B} and \citetalias{2023MNRAS.524.2290N} cooling models predict different $F_{\rm occ}$, we generate separate $\Wp^{\rm hh}$ and $\Wp^{\rm hs}$ tables for each. In Figure~\ref{fig:HOF}(b), we show the ratio of $\Wp$ at $\rp=8\,\mpch$ between the two occupancy cases, computed by cross-correlating with neighbouring (sub)haloes within $M_{*}^{\rm neigh}\in[10^{9.75},10^{10.25}]\,\ms$. The ratio is close to 1 at the high-mass end and rises toward lower halo masses, reaching values of about 3 in the most extreme cases.

We then refit the $\nwprp$ measurements using four non-parametric SHMR models, corresponding to the two HOF prescriptions, each evaluated with and without assembly bias. The resulting central SHMRs are compared to the fiducial model in Figure~\ref{fig:HOF_shmr}; all appear broadly consistent within the uncertainties. However, agreement in the inferred SHMR does not imply that all models reproduce the measurements equally well. Figure~\ref{fig:HOF_chi2} shows the total $\Delta\chi^2$ of each model relative to the fiducial case for every $M_{*}^{\rm photo}$ bin. All HOF models yield poorer fits at $M_{*}^{\rm photo}<10^{7.0}\,\ms$, with $\Delta\chi^2$ increasing toward lower masses. The \citetalias{2023MNRAS.524.2290N} model performs better than the \citetalias{2020MNRAS.498.4887B} model, and including assembly bias improves the fit in both cases. This trend may suggest that the HOF is underestimated in these models, leaving too few haloes to match the observed number densities at low stellar masses. Given the relatively low quality of the measurements at $M_{*}^{\rm photo}<10^{7.0}\,\ms$, the differences are not highly significant. The \citetalias{2023MNRAS.524.2290N} model with assembly bias provides a fit comparable to the fiducial case, while the remaining models show discrepancies at the $\sim2$–$3\sigma$ level. This may indicate that H$_2$ cooling deserves more consideration, though improved measurements are needed to confirm this. Furthermore, reionization physics—fixed in the adopted HOF models—also affects the HOF; its variation should be explored before reaching firm conclusions.

In Figure~\ref{fig:HOF_smf}, we show the GSMFs derived from the HOF models and compare them with the fiducial results. All models agree well for $M_{*}>10^{7.0}\,h^{-2}_{70}\ms$. At $M_{*}<10^{7.0}\,h^{-2}_{70}\ms$, the \citetalias{2020MNRAS.498.4887B} model yields the lowest GSMF, the \citetalias{2023MNRAS.524.2290N} model predicts higher values, and the fiducial model gives the highest among the three. The \citetalias{2025MNRAS.540.1635X} GSMF is shown for reference, though it is not guaranteed to be accurate at these masses, as it relies on a constant-bias assumption at the low-mass end and does not include the assembly-bias effects implied by a non-unity HOF. Considering the $\Delta\chi^2$ trends in Figure~\ref{fig:HOF_chi2} and the expectation that some form of HOF must be present, the true GSMF at the faint end likely lies between the fiducial model and the \citetalias{2023MNRAS.524.2290N} prediction.

The results obtained with the \citetalias{2020MNRAS.498.4887B} HOF model also provide a conservative test of omitting the halo-catalogue extrapolation at $M_{\rm h}<10^{8.7}\msh$, because this model suppresses the contribution of low-mass haloes more strongly than simply excluding haloes below this mass threshold. Therefore, omitting haloes with $M_{\rm h}<10^{8.7}\msh$ should have a smaller effect on the inferred results than adopting the \citetalias{2020MNRAS.498.4887B} HOF model.

\begin{figure}
    \centering
    \includegraphics[width=\columnwidth]{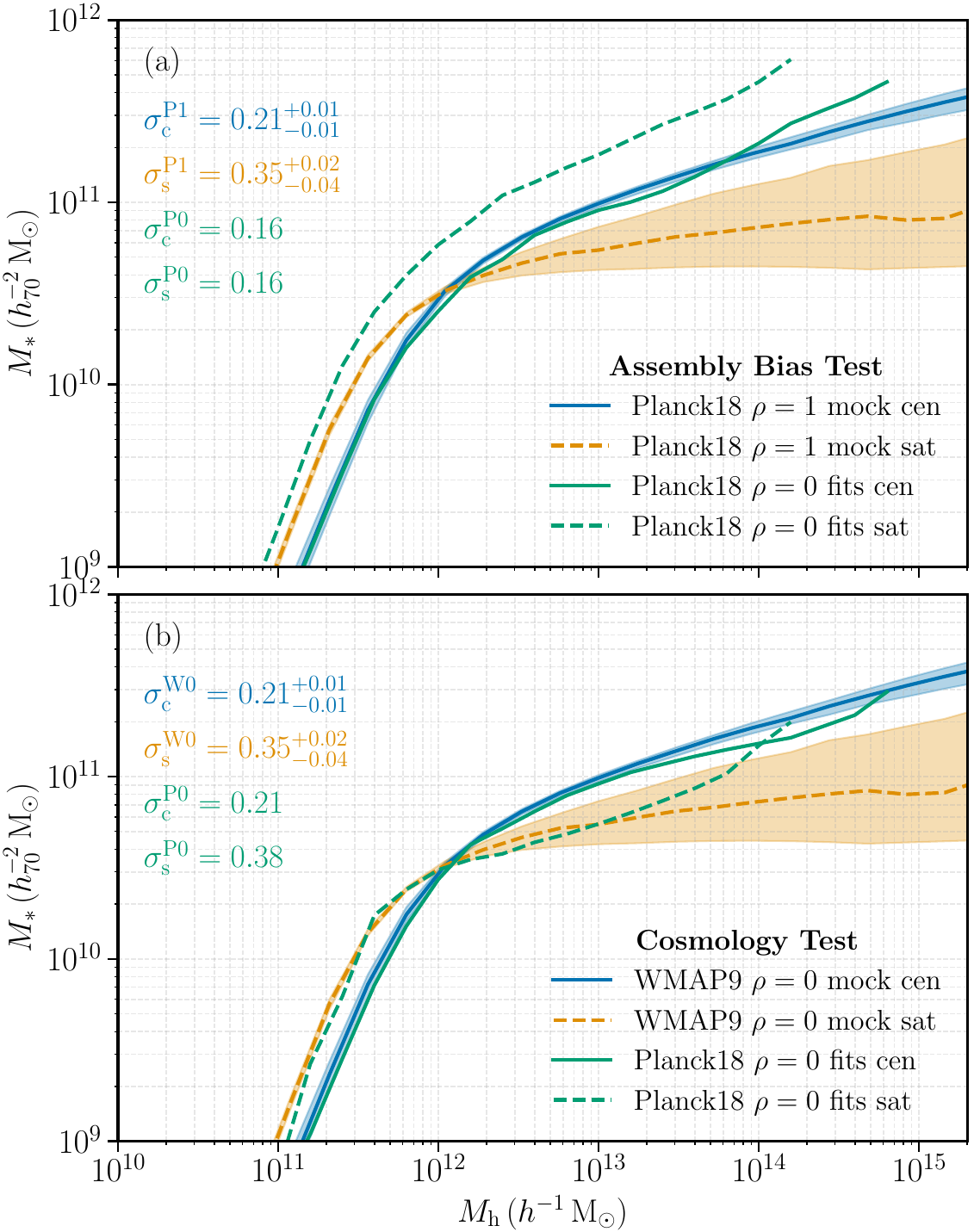}
    \caption{Tests of how galaxy assembly bias and assumed cosmology affect SHMR inference using mocks. (a) MAP SHMRs (green) obtained by fitting Jiutian-1G mocks with maximal assembly bias ($\rho=1$) using non-parametric models without assembly bias ($\rho=0$). (b) MAP SHMRs (green) obtained by fitting a \textsc{CosmicGrowth} mock with WMAP9-like cosmology using Jiutian-1G (Planck18 cosmology) non-parametric models.}
    \label{fig:AB_cosmo_SHMR}
\end{figure}

\begin{figure*}
    \centering
    \includegraphics[width=\textwidth]{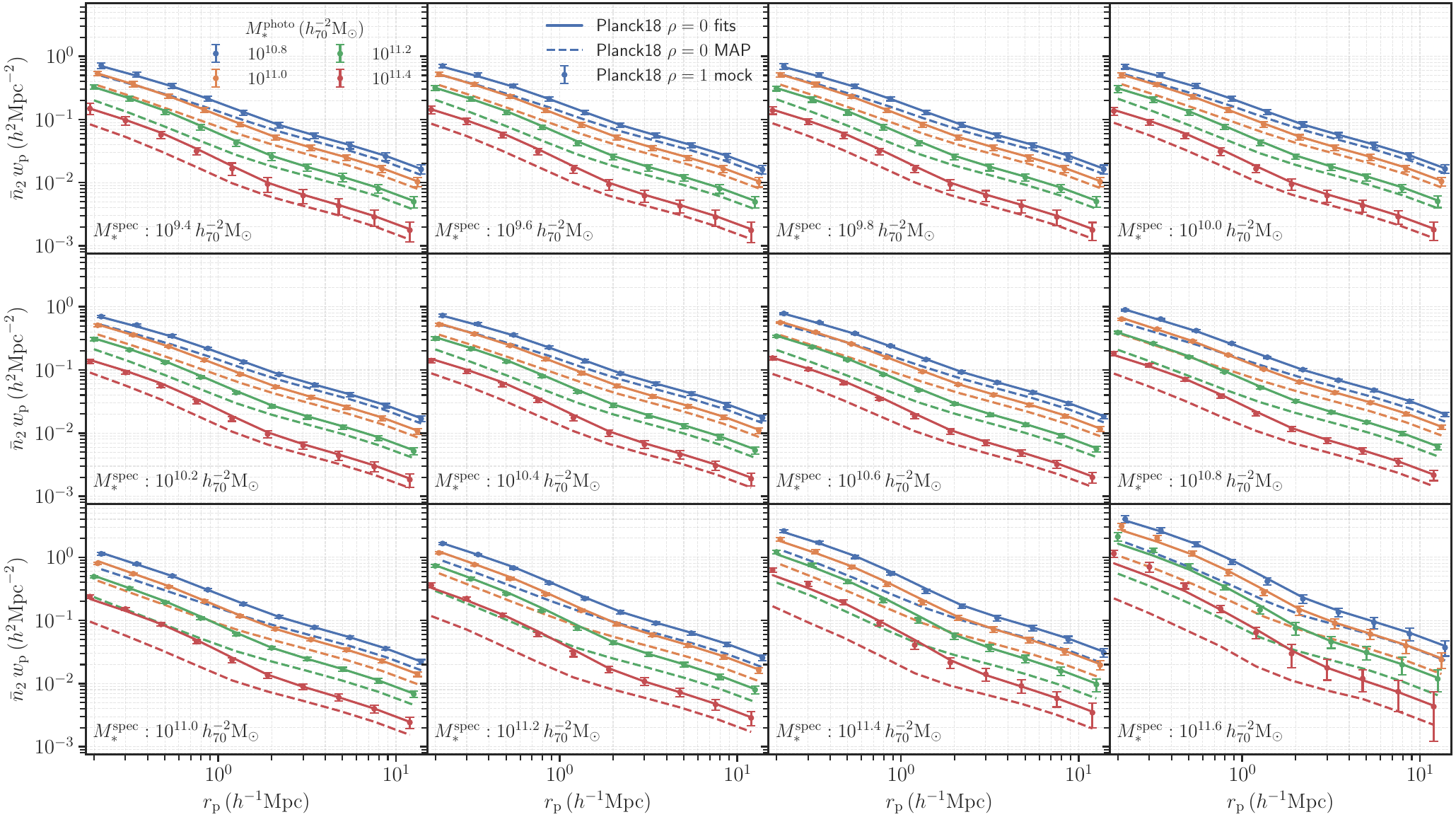}
    \caption{Examples of fits to $\nwprp$ measurements from the Jiutian-1G mock with maximal assembly bias ($\rho=1$; points with error bars), using a $\rho=0$ SHAM model (solid lines), shown for several stellar-mass bins. For comparison, $\nwprp$ measurements from the $\rho=0$ mocks are also shown (dashed lines). Both $\rho=0$ and $\rho=1$ mocks are generated using the MAP fiducial parametric SHMRs.
}
    \label{fig:AB_nwp_fitting}
\end{figure*}

\subsection{Effects of galaxy assembly bias}
We next consider galaxy assembly bias in a more general context, even for the case with $F_{\rm occ}=1$, to test whether it could account for the $\sim15\%$ systematic offset observed in the GSMF. We examine this possibility within our SHAM framework, where several studies have explored related directions \citep{2021MNRAS.504.5205C,2022MNRAS.511.1789Z}.

Under our SHMR-based SHAM framework, galaxy assembly bias arises when the scatter in the SHMR is not random but depends on secondary halo properties, for example through a conditional relation such as $P(M_* \mid M_{\rm h}, z_{\rm form})$. Other halo properties (e.g., concentration) could play a similar role. In a more general setting, $P(M_* \mid M_{\rm h})$ need not follow a lognormal distribution. Because halo assembly bias implies that the halo correlation function $w_{\rm p}(r_{\rm p}; M_{\rm h}, z_{\rm form})$ also depends on $z_{\rm form}$, the galaxy $w_{\rm p}$, a weighted average of halo $w_{\rm p}$ according to $P(M_* \mid M_{\rm h}, z_{\rm form})$, will in general depend on this conditional SHMR. This is how galaxy assembly bias enters our modelling.

With a log-normal form for $P(M_* \mid M_{\rm h}; \sigma)$ obtained after integrating over $z_{\rm form}$, and with $P(z_{\rm form}\mid M_{\rm h})$ measured from simulations, recovering the conditional relation $P(M_* \mid M_{\rm h}, z_{\rm form})$ requires specifying the joint distribution $P(M_*, z_{\rm form}\mid M_{\rm h})$ at fixed halo mass:
\begin{align}
    P(M_*, z_{\rm form}\mid M_{\rm h}) &= P(M_* \mid M_{\rm h}, z_{\rm form})\,P(z_{\rm form}\mid M_{\rm h})\,, \\
    P(M_* \mid M_{\rm h}) &= \int P(M_* \mid M_{\rm h}, z_{\rm form})\,P(z_{\rm form}\mid M_{\rm h})\,{\rm d}z_{\rm form}\,.\notag
\end{align}
Given only the two marginal distributions $P(M_* \mid M_{\rm h})$ and $P(z_{\rm form}\mid M_{\rm h})$, the joint distribution $P(M_*, z_{\rm form}\mid M_{\rm h})$—and therefore the conditional relation $P(M_* \mid M_{\rm h}, z_{\rm form})$—is not uniquely determined. Additional parametrization and constraints on $P(M_*, z_{\rm form}\mid M_{\rm h})$ are therefore required.

The simplest choice for $P(M_*, z_{\rm form}\mid M_{\rm h})$ is a bivariate Gaussian parametrized by a correlation coefficient $\rho$.  
For a constant-scatter model $P(M_* \mid M_{\rm h}; \sigma)$, we first normalize $\log_{10}M_{*}$ and $z_{\rm form}$:
\begin{align}
    x &= \frac{\log_{10} M_{*} - \left\langle \log_{10} M_{*}\mid M_{\rm h} \right\rangle}{\sigma}\,, \notag\\
    y &= \frac{z_{\rm form} - \bar{z}_{\rm form}(M_{\rm h})}{s_z(M_{\rm h})}\,,
\end{align}
where $s_z(M_{\rm h})$ is the standard deviation of $z_{\rm form}$ at fixed halo mass.  
This implicitly assumes $P(z_{\rm form}\mid M_{\rm h})$ is approximately Gaussian; if not, $z_{\rm form}$ may first be Gaussianized. We then assume $(x,y)$ follows a bivariate normal distribution at fixed $M_{\rm h}$:
\begin{equation}
\begin{pmatrix}
x \\
y
\end{pmatrix}
\Big|\,M_{\rm h}
\sim
\mathcal{N}\!\left(
\begin{pmatrix}
0 \\
0
\end{pmatrix},
\;
\begin{pmatrix}
1 & \rho \\
\rho & 1
\end{pmatrix}
\right).
\end{equation}
Here $\rho$ is taken to be constant across halo mass, although a mass-dependent $\rho(M_{\rm h})$ could also be considered. Then, $P(M_* \mid M_{\rm h}, z_{\rm form})$ can be written as
\begin{align}
    &P\bigl(\log_{10} M_* \mid M_{\rm h}, z_{\mathrm{form}}\bigr)
=
\frac{1}{\sqrt{2\pi \left(1-\rho^2\right)\sigma^2}}\times\notag\\
&\exp\left[
 -\frac{
   \left(\log_{10} M_* - \left\langle \log_{10} M_{*}\mid M_{\rm h} \right\rangle - \rho\,\sigma\,y\right)^2
 }{
   2\left(1-\rho^2\right)\sigma^2
 }
\right]\,.
\end{align}

In this formulation, a single additional parameter, $\rho$, controls the correlation between $M_*$ and $z_{\rm form}$. In practice, however, implementing this model within our HMC framework would require extending the $\Wp$ tables with an additional dimension for $z_{\rm form}$, which would dramatically increase both memory usage and computational cost. A dedicated investigation would be needed to explore possible compression schemes for the $\Wp$ tables before such a model becomes tractable. Therefore, in this work we do not attempt to constrain $\rho$, but instead test the limiting case $\rho=1$, where $M_*$ and $z_{\rm form}$ are in one-to-one correspondence at fixed $M_{\rm h}$—i.e., galaxies in earlier-formed haloes have larger stellar masses \citep{2021NatAs...5.1069C}. In this limit, $P(\log_{10} M_* \mid M_{\rm h}, z_{\rm form})$ collapses to a $\delta$-function,
\begin{align}
&P(\log_{10} M_* \mid M_{\rm h}, z_{\rm form})
=\notag\\
&\delta\!\left[
\log_{10} M_* -
\left\langle \log_{10} M_* \mid M_{\rm h} \right\rangle
- \sigma\,
\Phi^{-1}\!\bigl(F_z(z_{\rm form}\mid M_{\rm h})\bigr)
\right],
\end{align}
where $F_z$ is the cumulative distribution (CDF) of $z_{\rm form}$ and $\Phi^{-1}$ is the inverse CDF of the standard normal distribution. The corresponding stellar mass then has the explicit form
\begin{align}
\log_{10} M_*(M_{\rm h}, z_{\rm form})
=
\left\langle \log_{10} M_* \mid M_{\rm h} \right\rangle
+ \sigma\,
\Phi^{-1}\!\bigl(F_z(z_{\rm form}\mid M_{\rm h})\bigr).
\end{align}

\begin{figure*}
    \centering
    \includegraphics[width=\textwidth]{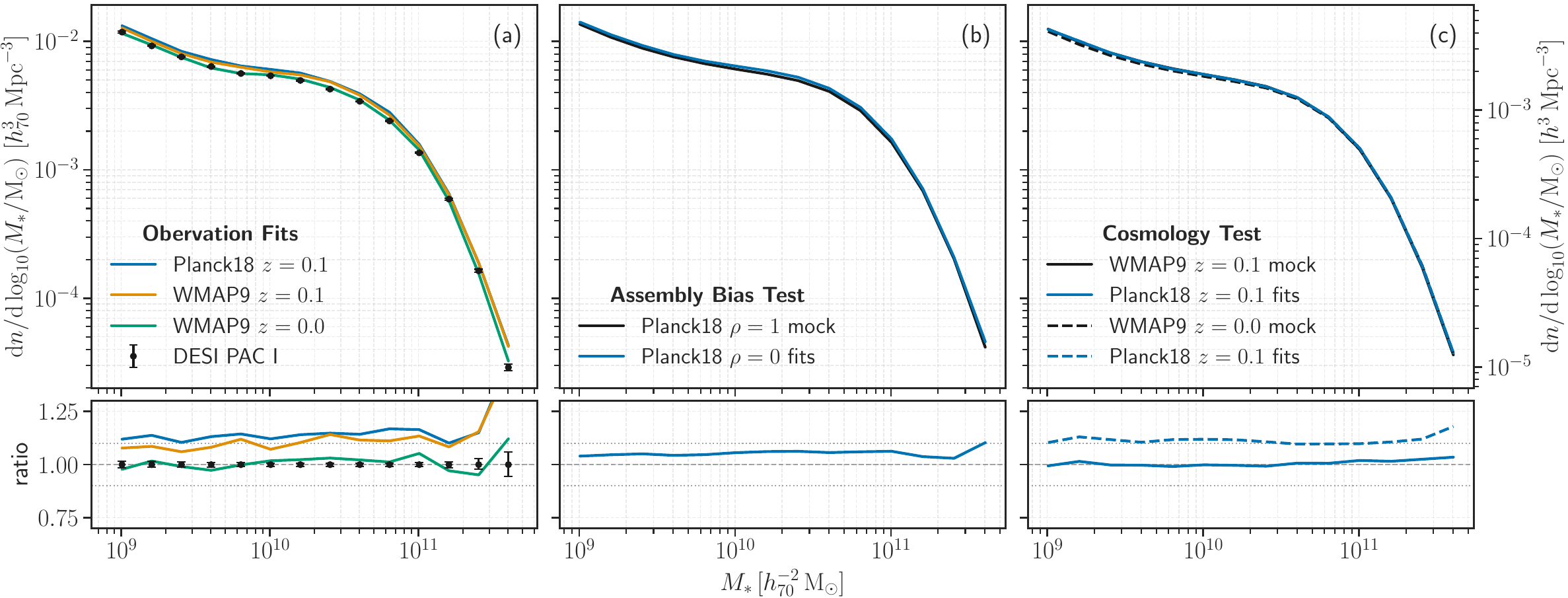}
    \caption{Tests of how assembly bias and the assumed cosmology affect the derived GSMFs.  
(a) Comparison of GSMFs from the MAP fiducial models in Jiutian\textendash1G (Planck18 cosmology) and \textsc{CosmicGrowth} (WMAP9-like cosmology) with those from \citetalias{2025MNRAS.540.1635X}. We also include results obtained using the WMAP9-like cosmology at $z=0$, which, when interpreted at $z=0.1$, effectively mimic a cosmology with $\Omega_{\rm m}=0.216$ and $\sigma_8=0.87$.  
(b) Comparison of GSMFs in the Jiutian\textendash1G mock with maximal assembly bias ($\rho=1$) to the GSMFs inferred from fitting its $\nwprp$ using non-parametric SHMR models without assembly bias ($\rho=0$).  
(c) Comparison of GSMFs in the CosmosGrowth mock to the GSMFs inferred from fitting its $\nwprp$ using Jiutian\textendash1G non-parametric SHMR models.
Dotted horizontal lines in the lower panels indicate the $\pm 10\%$ region.}
    \label{fig:AB_cosmo_SMF}
\end{figure*}

In Figures~\ref{fig:test_sys_nwp} and \ref{fig:test_sys_wp}, we compare $\nwprp$ and $\Wp(\rp)$ for $M_{*}^{\rm photo}=10^{10.6}\,h^{-2}_{70}\ms$ between the $\rho=0$ and $\rho=1$ cases in Jiutian-1G, using the SHAM model with the fiducial parametric MAP SHMR. For subhaloes, $z_{\rm form}$ is defined as the redshift at which they reach half of their $m_{\rm peak}$. The $M_*$–$z_{\rm form}$ correlation is applied separately to haloes and subhaloes.
Since $P(M_{*}\mid M_{\rm h})$ is identical in the two cases, the resulting GSMF is unchanged, and the ratios of $\nwprp$ and $\Wp(\rp)$ are likewise expected to be the same. We find that the $\rho=1$ case produces a larger $\Wp(\rp)$ than the $\rho=0$ case, with the difference increasing toward higher $M_*^{\rm spec}$ and reaching a factor of $\sim1.5$ at $10^{11.6}\,h^{-2}_{70}\,\ms$. For larger $M_*^{\rm photo}$, the discrepancy can grow further. This behaviour arises from the combined effects of the mass dependence of (sub)halo assembly bias and the slope of the mean SHMR. Although the $\rho=1$ case represents an extreme scenario, the result indicates that galaxy assembly bias can influence the modelling of our measurements.

To assess how the presence of galaxy assembly bias affects SHMR modelling and the inferred GSMF, we fit the $\rho=1$ mocks using the $\rho=0$ non-parametric SHMR model. For simplicity, we perform this test only for $M_{*}^{\rm spec}\geq10^{9.4}\,h^{-2}_{70}\ms$ and $M{*}^{\rm photo}\geq10^{9.0}\,h^{-2}_{70}\ms$ in Jiutian-1G. To mimic the relative weighting of data across stellar-mass bins and scales in the real analysis, we use the observational covariance matrix in the fitting. Since the mock $\nwprp$ measurements from Jiutian-1G have much smaller statistical uncertainties, this procedure yields very small $\chi^2$ values but does not bias the MAP parameters. A more realistic mock-based test would require adding noise to the mock $\nwprp$ according to the observational covariance, which we do not perform here. Instead, in the comparisons below we display the posterior distributions from the observational analysis around the true parameters; the MAP uncertainties should be similar if the mocks were perturbed using the same covariance. This approach provides a reasonable indication of the significance of the systematic effects.

The MAP SHMR obtained by fitting the $\rho=1$ mock with the $\rho=0$ model is compared to the true SHMR in Figure~\ref{fig:AB_cosmo_SHMR}(a), and examples of the corresponding $\nwprp$ fits are shown in Figure~\ref{fig:AB_nwp_fitting}. We find that the $\rho=0$ model can reproduce the $\nwprp$ measurements of the $\rho=1$ mock extremely well. However, the inferred SHMR is biased: it requires a much higher mean satellite SHMR and smaller scatters for both centrals and satellites, while the mean central SHMR remains similar to the truth. In Figure~\ref{fig:AB_cosmo_SMF}(b), we show the GSMF derived from the fitted model and compare it to the true GSMF. Even in the extreme case of $\rho=1$, ignoring galaxy assembly bias leads to only a $\sim5\%$ overestimation of the GSMF when fitting $\nwprp$, and thus a corresponding $\sim5\%$ underestimation of $\Wp(\rp)$. This is insufficient to explain the $\sim15\%$ systematic offset in the GSMF relative to the model-independent measurements shown again in Figure~\ref{fig:AB_cosmo_SMF}(a). In other words, within Planck18 cosmology we are unable to find a SHAM model that simultaneously fits both $\n$ and $\Wp(\rp)$, indicating a degree of internal tension.

\begin{figure}
    \centering
    \includegraphics[width=\columnwidth]{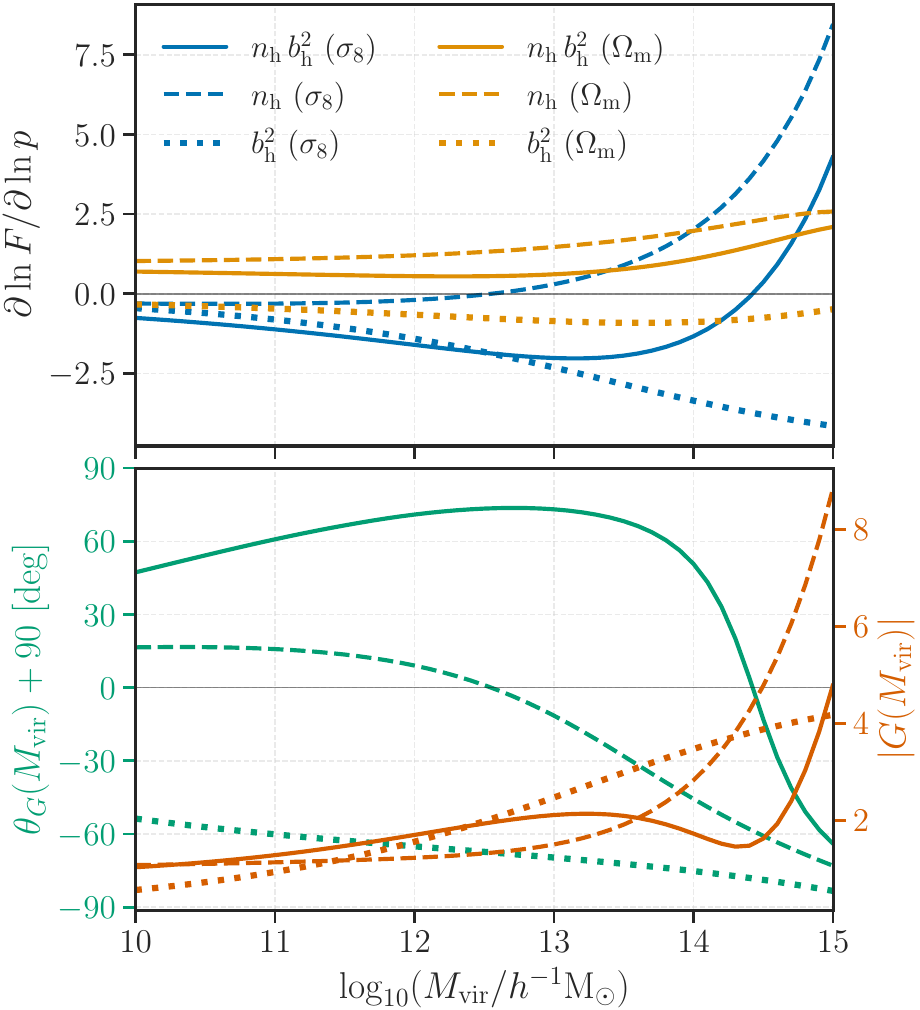}
    \caption{Dependence of $n_{\mathrm h} b_{\mathrm h}^2$, $n_{\mathrm h}$, and $b_{\mathrm h}^2$ on $\sigma_8$ and $\Omega_{\mathrm m}$ around the Planck18 cosmology at $z=0.1$, shown as a function of $M_{\rm vir}$. The top panel shows the fractional derivatives with respect to each parameter separately. The bottom panel shows the gradient amplitude $|G|$ and the degenerate direction $\theta_G+90^\circ$ in the joint dependence, where $x=\tan(\theta_G+90^\circ)$ represents the effective power‐law index in the combined dependence $\sigma_8\,\Omega_{\rm m}^{x}$.}
    \label{fig:cosmo_depend}
\end{figure}

\begin{figure}
    \centering
    \includegraphics[width=\columnwidth]{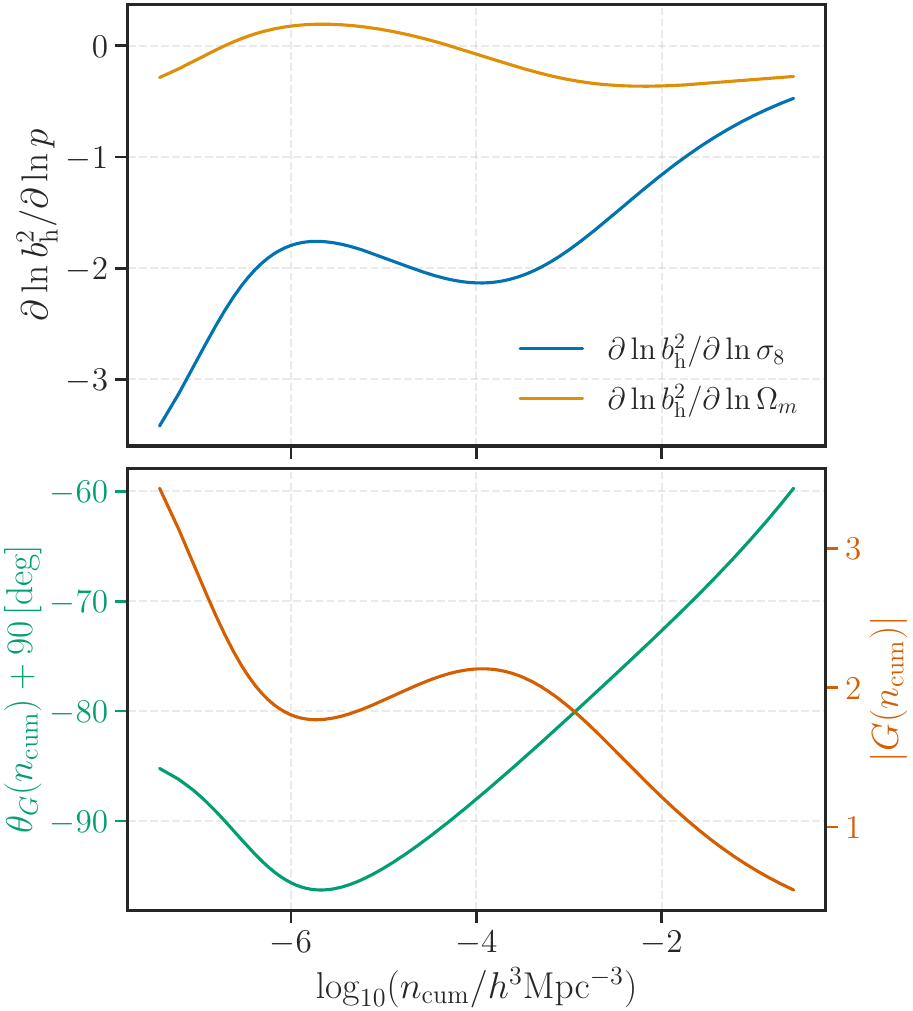}
    \caption{Similar to Figure~\ref{fig:cosmo_depend}, but for $b_{\rm h}^2$ as a function of the cumulative number density $n_{\rm cum}$.}
    \label{fig:cosmo_depend_ncum}
\end{figure}

\begin{figure}
    \centering
    \includegraphics[width=\columnwidth]{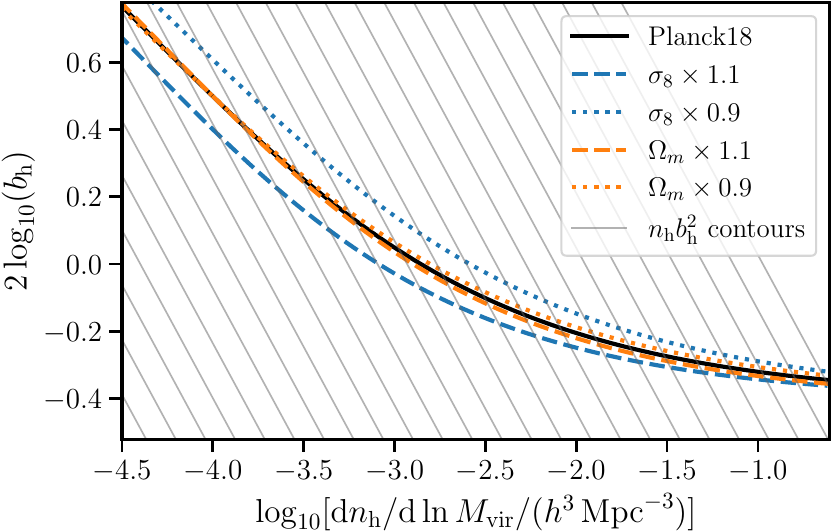}
    \caption{$n_{\rm h}$–$b_{\rm h}^2$ relation for cosmologies with $\sigma_8$ and $\Omega_{\rm m}$ varied around the Planck18 baseline. Grey lines indicate contours of constant $n_{\rm h}b_{\rm h}^2$.}
    \label{fig:cosmo_depend_nwp}
\end{figure}

\begin{figure}
    \centering
    \includegraphics[width=\columnwidth]{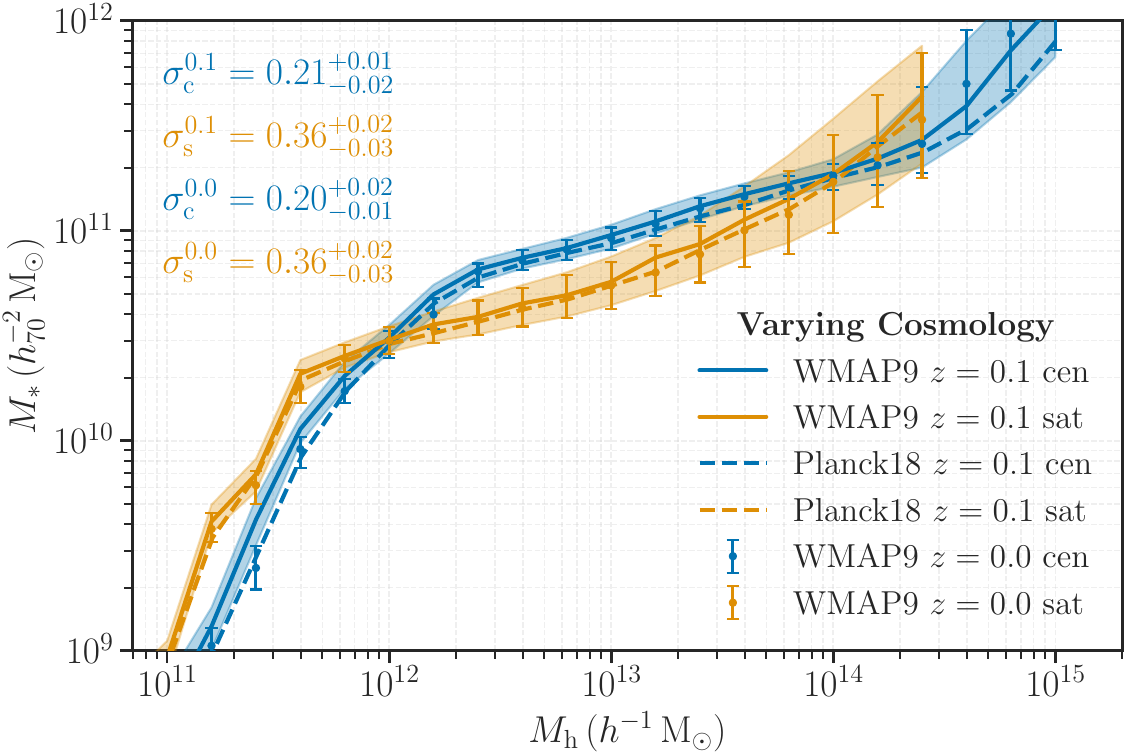}
    \caption{SHMRs constrained by fitting $\nwprp$ measurements under different assumed cosmologies using the fiducial non-parametric model. Results are shown for Planck18 and WMAP9 cosmologies at $z=0.1$. We also include fits obtained using the WMAP9 cosmology at $z=0$, which, when interpreted at $z=0.1$, effectively mimic a cosmology with $\Omega_{\rm m}=0.216$ and $\sigma_8=0.87$.}
    \label{fig:wmap_shmr}
\end{figure}

\begin{figure*}
    \centering
    \includegraphics[width=\textwidth]{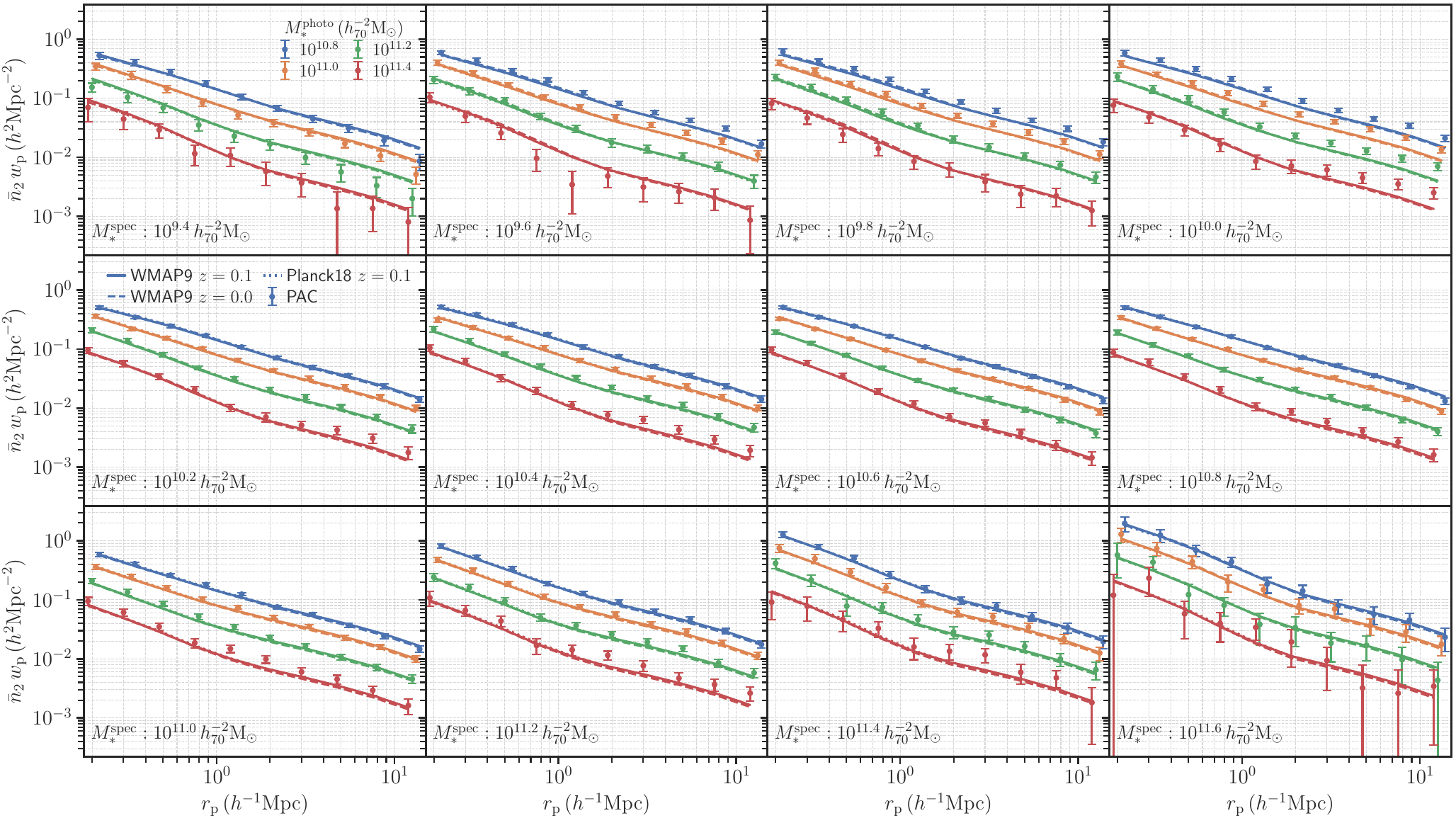}
    \caption{Examples of fits to observational $\nwprp$ measurements using different cosmologies, shown for several stellar-mass bins. Results are shown for Planck18 and WMAP9 cosmologies at $z=0.1$. We also include fits obtained using the WMAP9 cosmology at $z=0$, which, when interpreted at $z=0.1$, effectively mimic a cosmology with $\Omega_{\rm m}=0.216$ and $\sigma_8=0.87$.}
    \label{fig:wmap_fitting}
\end{figure*}

\subsection{Effects of the assumed cosmology}
Another factor worth examining is the assumed Planck18 cosmology used in both the measurements and the modelling. A set of tensions has recently emerged between Planck18 results and several late-time probes \citep{2021A&A...646A.140H,2022ApJ...934L...7R,2022PhRvD.105b3520A,2023PhRvD.108l3519D,2025PhRvD.112h3515A}. It is therefore important to assess how sensitive our results are to the assumed cosmology, and whether adopting a different cosmology could relieve the $\sim15\%$ internal tension identified above.

Here we restrict attention to variations within flat $\Lambda$CDM. Since the dependence on $H_0$ is absorbed into our unit conventions, and the luminosity-distance dependence on $\Omega_{\rm m}$ is very weak at $z\sim0.1$, the dominant cosmology dependence in our analysis comes from how the halo and subhalo distributions respond to changes in $\Omega_{\rm m}$ and $\sigma_8$. 

To gain intuition for the cosmological dependence, we first consider a simplified case for halo auto‐correlations on linear scales, where $\bar{n}_{\rm h}\Wp \propto n_{\rm h} b_{\rm h}^2$. We adopt the \citet{2008ApJ...688..709T} halo mass function for $n_{\rm h}$ and the \citet{2010ApJ...724..878T} model for $b_{\rm h}$. In Figure~\ref{fig:cosmo_depend}, we show how $n_{\rm h} b_{\rm h}^2$, $n_{\rm h}$, and $b_{\rm h}^2$ depend on $\Omega_{\rm m}$ and $\sigma_8$ at fixed $M_{\rm vir}$, quantified by $\partial \ln F / \partial \ln p$, where $F$ denotes the quantity of interest and $p$ the cosmological parameter, evaluated around the Planck18 cosmology. From the upper panels, $n_{\rm h}$ increases with $\Omega_{\rm m}$ over the full mass range, with a slightly stronger sensitivity at high masses. Its dependence on $\sigma_8$ is strongly positive at the high‐mass end but becomes mildly negative below $M_{\rm vir}\sim10^{12.5}\, \msh$. The trends are opposite for $b_{\rm h}^2$, which decreases with both $\Omega_{\rm m}$ and $\sigma_8$, as a higher abundance implies lower peak height and thus lower bias. The behaviour of $n_{\rm h} b_{\rm h}^2$ reflects a combination of these trends: it generally increases with $\Omega_{\rm m}$ and decreases with $\sigma_8$, except at the very massive end. To illustrate parameter degeneracies, the lower panels show the degenerate direction
$\theta_D = \theta_G + 90^\circ$, where $\theta_G$ is the gradient direction in the $(\Omega_{\rm m}\,\sigma_8)$ plane, along with the gradient amplitude $|G|$. The effective power‐law exponent in a combined dependence $\sigma_8\,\Omega_{\rm m}^{x}$ is given by $x = \tan(\theta_D)$, where we restrict $\theta_D\in[-90^\circ,90^\circ]$ so that $\theta_D$ and $x$ share the same sign. A negative $\theta_D$ indicates that the quantity changes in the same direction with respect to both $\sigma_8$ and $\Omega_{\rm m}$. We find that $\theta_D$ varies with halo mass for all three quantities. Thus, measurements of one of these quantities at multiple $M_{\rm vir}$ values can break the $\Omega_{\rm m}$–$\sigma_8$ degeneracy. Moreover, combining different observables at a single mass scale is also effective, given their distinct $\theta_D$ behaviours. A related analysis for mass‐limited samples is presented in \citet{2024MNRAS.530.4203X}.

In practice, obtaining unbiased measurements of $M_{\rm vir}$ is challenging. A more observationally robust quantity is the halo rank, which only depends on the relative ordering in mass and can be estimated using many mass proxies. In that case, the directly accessible statistics are $b_{\rm h}^2$ and $n_{\rm h}$. To examine whether these two are sufficient to constrain cosmology, we show in Figure~\ref{fig:cosmo_depend_ncum} the dependence of $b_{\rm h}^2$ on $\sigma_8$ and $\Omega_{\rm m}$ as a function of cumulative number density $n_{\rm cum}$, ranked from the most massive haloes. At fixed $n_{\rm cum}$, $b_{\rm h}^2$ decreases with increasing $\sigma_8$ while showing only weak dependence on $\Omega_{\rm m}$, and the degenerate direction $\theta_D$ varies with $n_{\rm cum}$. This demonstrates that even without absolute halo mass calibration, measurements of $b_{\rm h}^2$ at several number‐density thresholds can still constrain cosmology and effectively break the $\sigma_8$–$\Omega_{\rm m}$ degeneracy. 

\begin{figure*}
    \centering
    \includegraphics[width=\textwidth]{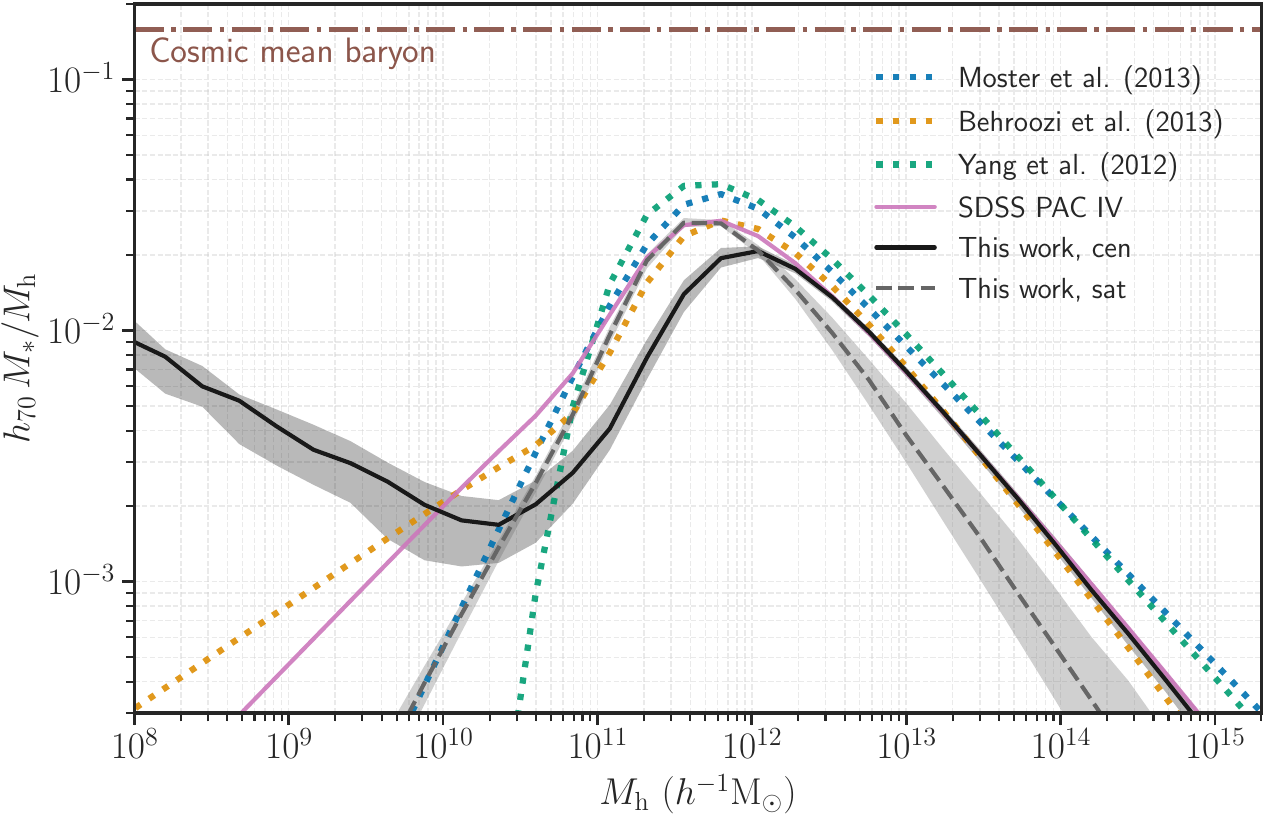}
    \caption{Comparison of our fiducial parametric SHMRs with previous studies. Results from \citet{2012ApJ...752...41Y}, \citet{2013ApJ...770...57B}, \citet{2013MNRAS.428.3121M}, and \citetalias{2023ApJ...944..200X} are shown. All halo masses are converted to the virial definition $M_{\rm vir}$ using the concentration model of \citet{2021MNRAS.506.4210I}. }
    \label{fig:compare_SHMR}
\end{figure*}

\begin{figure}
    \centering
    \includegraphics[width=\columnwidth]{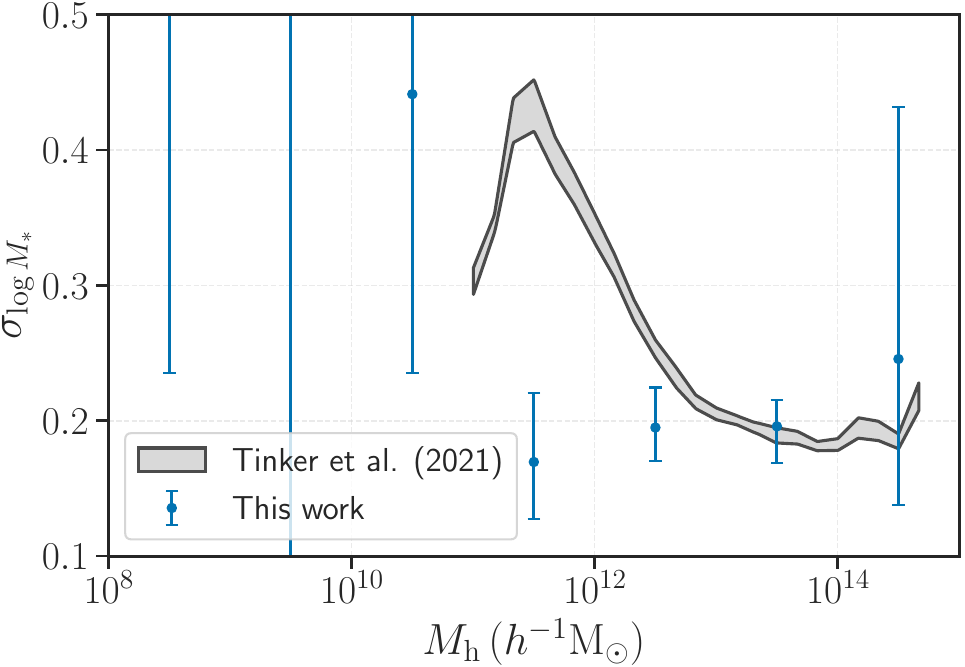}
    \caption{Comparison of the scatter in the central SHMR from our varying-scatter model with that from \citet{2021ApJ...923..154T}. Results are from the parametric $\beta_\mathrm{c}$ free model. The $1\sigma$ interval for \citet{2021ApJ...923..154T} is obtained by converting their reported 95\% confidence range assuming Gaussian errors.}
    \label{fig:compare_scatter}
\end{figure}

However, this does not hold if only the measurements of $n_{\rm h} b_{\rm h}^2$ are available. In Figure~\ref{fig:cosmo_depend_nwp}, we show the $n_{\rm h}$–$b_{\rm h}^2$ relations for several cosmologies varying $\sigma_8$ and $\Omega_{\rm m}$, together with contours of constant $n_{\rm h} b_{\rm h}^2$. With only $n_{\rm h} b_{\rm h}^2$ measurements, the intersections between the $n_{\rm h}$–$b_{\rm h}^2$ relations and the constant $n_{\rm h} b_{\rm h}^2$ contours indicate that one can always select points that match a given set of $n_{\rm h} b_{\rm h}^2$ values for any cosmology. Therefore, $n_{\rm h} b_{\rm h}^2$ alone cannot constrain cosmology.

Now we return to our $\nwprp$ measurements in stellar-mass bins. Although scatter in the SHMR and the presence of satellites introduce additional complexity by mixing halo ranks, these effects are constrained within our modelling framework, so the qualitative conclusions on cosmological dependence should still hold. Therefore, if an incorrect cosmology is assumed when modelling the data, it remains possible to find a SHAM model that fits all $\nwprp$ measurements, as suggested by Figure~\ref{fig:cosmo_depend_nwp}. However, when we separately examine $\n$ and $\Wp$, such a model must necessarily mispredict both quantities in order to match $\nwprp$, because $\n$ and $\Wp$ cannot be simultaneously reproduced under an incorrect cosmology according to Figure~\ref{fig:cosmo_depend_ncum}. Thus, an incorrect assumed cosmology may be a potential source of the $\sim15\%$ systematic offset observed in the GSMF.

To assess the cosmology dependence of our modelling, we additionally use a simulation from the \textsc{CosmicGrowth} suite \citep{2007ApJ...657..664J,2019SCPMA..6219511J}, which adopts a nine-year WMAP–like cosmology \citep{2013ApJS..208...19H} with $\Omega_{\rm m}=0.268$ and $\sigma_8=0.830$, i.e.\ $\sim10\%$ lower $\Omega_{\rm m}$ and slightly higher $\sigma_8$ than in Jiutian. Subhaloes in \textsc{CosmicGrowth} are identified using \textsc{HBT+}, and orphan galaxies are corrected following the method of \citet{2025JCAP...12..009X}, consistent with the Jiutian treatment.

To isolate the impact of cosmology, we measure $\nwprp$ and $\Wp$ in \textsc{CosmicGrowth} at a snapshot near $z=0.1$, using the same fiducial parametric SHMR calibrated in Planck18. Examples for several stellar-mass bins are shown in Figures~\ref{fig:test_sys_nwp} and \ref{fig:test_sys_wp}, compared against Planck18 predictions. The WMAP9–like cosmology generally yields slightly lower $\nwprp$ and slightly higher $\Wp(\rp)$, primarily due to the lower $\Omega_{\rm m}$. These differences are much smaller than those induced by maximal assembly bias ($\rho=1$) under the same cosmology. 

We then fit the WMAP9–like mock in Planck18 cosmology using the non-parametric SHMR model. The MAP SHMRs are compared with the truth in Figure~\ref{fig:AB_cosmo_SHMR}(b). We find that adopting Planck18 cosmology to fit a WMAP9–like universe does not introduce significant SHMR bias: the inferred mean central SHMR is only slightly lower, well within the observational uncertainties. The derived GSMF is compared with the true one in Figure~\ref{fig:AB_cosmo_SMF}(c). The agreement is very good, with only a slight excess in a few stellar-mass bins. This indicates that a $\sim10\%$ change in $\Omega_{\rm m}$ does not substantially alter the inferred results.

Nevertheless, we fit our $\nwprp$ measurements with $M_{*}>10^{9.0}\,\ms$ in \textsc{CosmicGrowth} at $z\sim0.1$ using the non-parametric SHMR model. The resulting SHMRs are presented in Figure~\ref{fig:wmap_shmr}, and the inferred GSMFs are compared with the Jiutian results in Figure~\ref{fig:AB_cosmo_SMF}(a). Example fits for several stellar-mass bins are shown in Figure~\ref{fig:wmap_fitting}. The WMAP9–like cosmology yields slightly higher SHMRs and slightly lower GSMFs than Planck18, reducing the GSMF discrepancy relative to \citetalias{2025MNRAS.540.1635X} from $\sim15\%$ to $\sim10\%$, but the tension remains. 

Although some combination of cosmology variation and assembly bias could further mitigate the systematics, we find that fitting the observations using a WMAP9–like cosmology evaluated at $z=0$ yields a surprisingly good match to both $\nwprp$ and the GSMF. The corresponding SHMR and GSMF are shown in Figures~\ref{fig:wmap_shmr} and \ref{fig:AB_cosmo_SMF}(a). In this case, the GSMF agrees almost perfectly with \citetalias{2025MNRAS.540.1635X} across the full mass range, including $M_*\simeq10^{11.6}\,\ms$ where the Planck18-based result shows clear tension. Conversely, fitting this mock with a Planck18-based model introduces a $>10\%$ systematic offset in GSMF, as illustrated in Figure~\ref{fig:AB_cosmo_SMF}(c). If this represents the true structure growth at $z=0.1$, it is approximately equivalent to a universe with $\Omega_{\rm m}\simeq0.216$ and $\sigma_8\simeq0.867$ \citep{2010MNRAS.405..143A}, i.e.\ a cosmology with a notably low matter density. 

Before drawing such a striking conclusion, this raises an additional concern: whether the effective redshift $z_{\rm eff}$ of our $\nwprp$ measurements has been correctly determined, given that modelling at $z=0$ and $z=0.1$ already produces noticeable differences. Defining a single meaningful $z_{\rm eff}$ is challenging, due to the broad redshift completeness variations shown in Figure~\ref{fig:SM_limits} and the complex redshift dependence of the photometric background across stellar-mass bins. A more practical approach is to switch to directly modelling the GSMF and $\Wp$ from DESI BGS, since PAC provides advantages mainly at the low-mass end, while the GSMF tension appears at high masses where BGS is highly complete and better understood. Unlike PAC, $\Wp$ from BGS is affected by fibre collisions on small scales, but several established correction methods exist \citep{1998ApJ...494....1J,2017MNRAS.472.1106B,2019ApJ...872...26Y,2025JCAP...01..127L}. Nevertheless, given that for $M_*>10^{10.0}\,\ms$ the completeness for our measurements typically extends to $z\approx0.2$ (Figure~\ref{fig:SM_limits}), the corresponding clustering constraints should have $z_{\rm eff}>0.1$—though this needs to be explicitly verified.

Another potential source of systematic uncertainty is baryonic effects on structure formation, particularly on small scales, which has recently received extensive attention in the weak-lensing community \citep[e.g.][]{2006ApJ...640L.119J,2022MNRAS.516.5355A}. Although its impact on clustered tracers such as haloes and galaxies is generally expected to be weaker than on the matter field \citep{2025arXiv251204080C}, it may still affect our modelling and warrants further investigation.

In conclusion, although modelling $\nwprp$ measurements under an incorrect cosmology can lead to biased inferences of GSMF and $\Wp$, the resulting SHMRs appear to remain relatively robust, as indicated in Figure~\ref{fig:wmap_shmr}. 

\section{Comparison and Implications}\label{sec:compare}
In this section, we compare our results with previous studies and discuss their implications.

\subsection{Comparison of SHMR}
In Figure~\ref{fig:compare_SHMR}, we compare our mean central SHMR from the fiducial parametric model with results from \citet{2013ApJ...770...57B}, \citet{2013MNRAS.428.3121M}, \citet{2012ApJ...752...41Y}, and \citetalias{2023ApJ...944..200X}. Because \citet{2013MNRAS.428.3121M} adopt $M_{200c}$ rather than $M_{\rm vir}$, their halo masses are converted to $M_{\rm vir}$ using the concentration model of \citet{2021MNRAS.506.4210I}. We do not perform detailed stellar-mass cross-calibration across studies—as was done in \citetalias{2025MNRAS.540.1635X}—because the original catalogues used in some of the earlier works are not publicly available. Consequently, residual systematic differences in stellar-mass estimates between studies may remain.

Comparing with \citetalias{2023ApJ...944..200X}, which assumed a single SHMR for both centrals and satellites, the inferred relation in that work closely follows the central SHMR at the high-mass end while being more consistent with the satellite relation at intermediate masses. Behaviour at $M_{\rm h}<10^{10.0}\,\msh$ reflects extrapolation, as only $\nwprp$ measurements with $M_*>10^{8.0}\,\ms$ were used there. Because that analysis relied exclusively on high-mass spectroscopic data—where the contribution from low-mass centrals to $\nwprp$ is small—a unified SHMR could not be ruled out by the available measurements. Once low-mass spectroscopic bins are included, such a model no longer reproduces the $\nwprp$ involving low-mass spectroscopic and photometric galaxies. Even so, the agreement of \citetalias{2023ApJ...944..200X} with the central (high-mass) and satellite (intermediate-mass) relations obtained here demonstrates good overall consistency between the PAC–based results.

Compared with other studies, the \citet{2013ApJ...770...57B} SHMR shows behaviour similar to \citetalias{2023ApJ...944..200X}, while the relations from \citet{2012ApJ...752...41Y} and \citet{2013MNRAS.428.3121M} lie systematically above both \citet{2013ApJ...770...57B} and our result over the full mass range, likely reflecting differences in stellar-mass calibration. None of the previous works exhibit the upturn at $M_{\rm h}\lesssim10^{10.0}\,\msh$, most likely because constraints at this scale were either absent or based entirely on extrapolation. This highlights the key advantage of PAC measurements in probing the low-mass regime.

\begin{figure*}
    \centering
    \includegraphics[width=\textwidth]{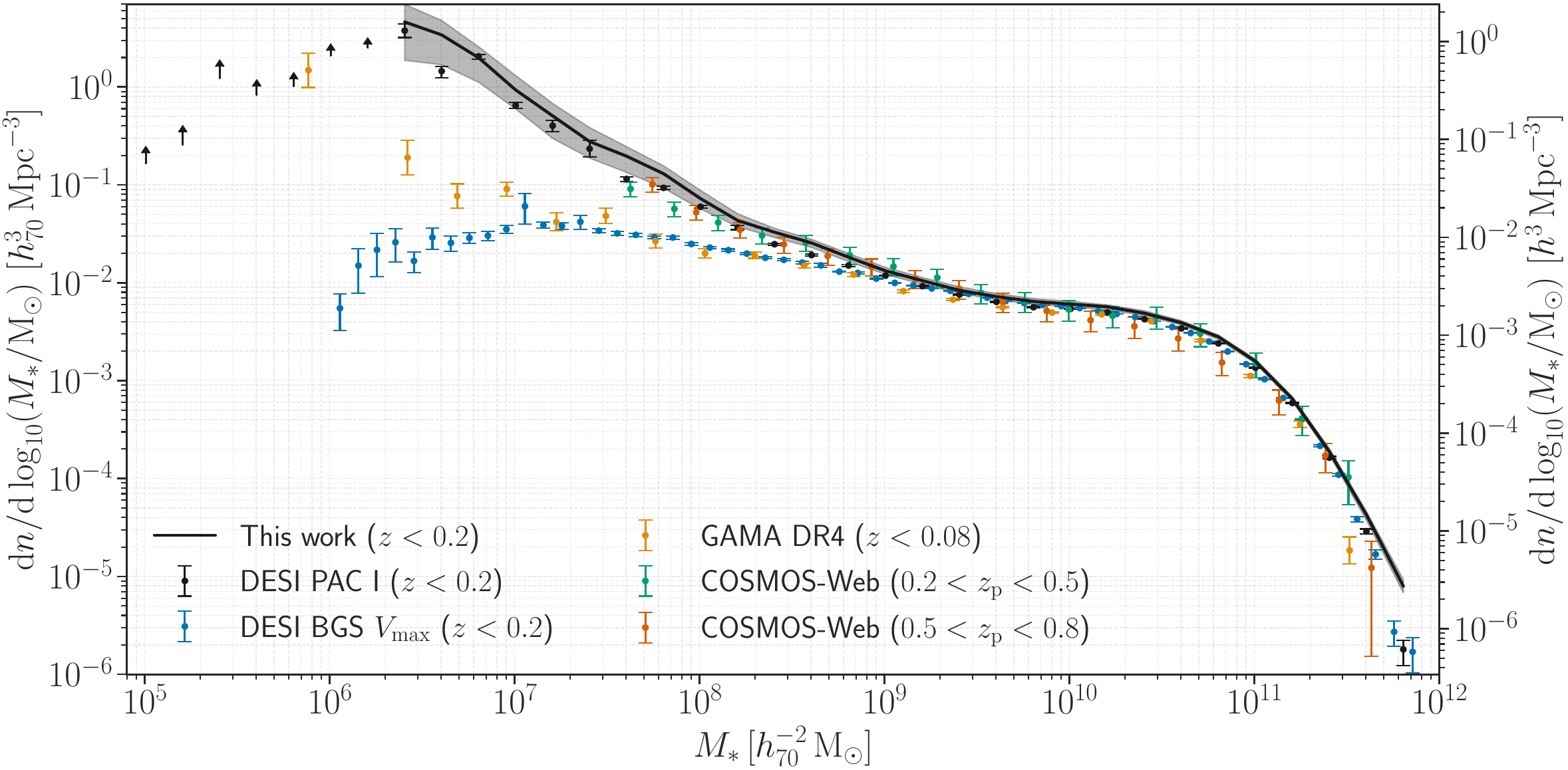}
    \caption{Comparison of the GSMF from our fiducial non-parametric model with previous measurements. Results from \citetalias{2025MNRAS.540.1635X} at $z<0.2$, GAMA DR4 \citep{2022MNRAS.513..439D} at $z<0.08$, DESI BGS DR1 using the $V_{\rm max}$ method \citep{2025arXiv251101803M} at $z<0.2$, and COSMOS-Web \citep{2025A&A...695A..20S} at higher redshifts ($0.2<z_{\rm p}<0.5$ and $0.5<z_{\rm p}<0.8$) are included. Stellar masses for GAMA DR4 and COSMOS-Web are calibrated to our mass scale following \citetalias{2025MNRAS.540.1635X}.}
    \label{fig:compare_smf}
\end{figure*}

\begin{figure*}
    \centering
    \includegraphics[width=\textwidth]{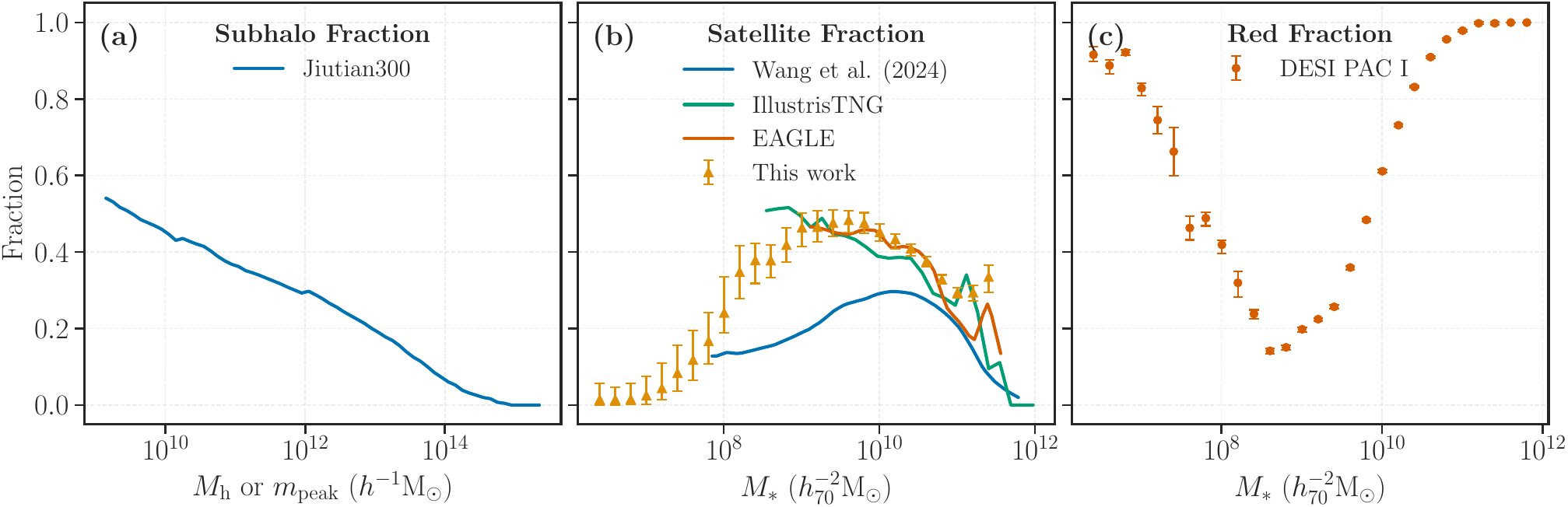}
    \caption{(a) Subhalo fraction from Jiutian-300 as a function of (sub)halo mass. 
(b) Satellite fraction as a function of stellar mass from the fiducial non-parametric model in this work. Results from \citet{2024ApJ...971..119W}, based on a DESI Y1 group catalogue \citep{2021ApJ...909..143Y}, as well as from the IllustrisTNG \citep{2018MNRAS.473.4077P} and EAGLE \citep{2015MNRAS.446..521S} hydrodynamic simulations, are shown for comparison. 
(c) Red-galaxy fraction as a function of stellar mass from \citetalias{2025MNRAS.540.1635X}.}
    \label{fig:fsat_fred}
\end{figure*}

In addition to the mean SHMR, Figure~\ref{fig:compare_scatter} compares the mass-dependent scatter inferred from our non-parametric varying-scatter model with that from \citet{2021ApJ...923..154T}, based on the SDSS group catalogue. Since \citet{2021ApJ...923..154T} report $95\%$ confidence intervals, we convert them to $1\sigma$ assuming Gaussian errors for consistency. At $M_{\rm vir}\in[10^{11.0},10^{15.0}]\,\msh$, \citet{2021ApJ...923..154T} find that the scatter increases toward lower halo masses, whereas our inferred scatter remains relatively constant. The lower-mass regime is not probed in \citet{2021ApJ...923..154T} due to data limitations, and in our case remains poorly constrained because of the strong degeneracy between the scatter and the low-mass slope of the mean relation.

\subsection{Comparison of GSMF}
In Figure \ref{fig:compare_smf}, we compare the GSMF from our fiducial non-parametric model with previous measurements. Local constraints include \citetalias{2025MNRAS.540.1635X} at $z<0.2$, GAMA DR4 \citep{2022MNRAS.513..439D} at $z<0.08$, and DESI BGS $V_{\rm max}$ measurements \citep{2025arXiv251101803M} at $z<0.2$. We also include COSMOS-Web results \citep{2025A&A...695A..20S} at higher redshifts ($0.2<z_{\rm p}<0.5$ and $0.5<z_{\rm p}<0.8$), which currently allow direct counting of dwarf galaxies well beyond the local volume. Stellar masses from GAMA DR4 and COSMOS-Web are calibrated onto our mass scale following the procedure in \citetalias{2025MNRAS.540.1635X}.

Our GSMF is quite consistent with all measurements at $M_{*}>10^{9.0}\,h_{70}^{-2}\,\ms$. However, as reported in \citetalias{2025MNRAS.540.1635X}, DESI BGS and GAMA show lower number densities for galaxies with $M_{*}<10^{9.0}\,h_{70}^{-2}\,\ms$, which is attributed to the presence of a local void \citep{2008ApJ...676..184T,2010Natur.465..565P}. This interpretation is supported by the COSMOS-Web results, which agree with the \citetalias{2025MNRAS.540.1635X} GSMF. In \citetalias{2025MNRAS.540.1635X} and this work, the local GSMF is measured using the much deeper DECaLS sample compared to DESI BGS and GAMA, enabling dwarf galaxies beyond the local void to be observed. The consistency between the GSMF derived in this work and that of \citetalias{2025MNRAS.540.1635X} further confirms this picture and indicates that the galaxy-bias assumptions adopted in \citetalias{2025MNRAS.540.1635X} are reasonable. In addition, we note that the decline in the BGS $V_{\rm max}$ GSMF around $10^{7.0}\,h^{-2}_{70}\ms$ is caused by the loss of high mass-to-light–ratio galaxies in these bins, because BGS is flux-limited rather than mass–limited. As a result, the $V_{\rm max}$ method becomes affected by incompleteness at these stellar masses.
  
\subsection{Satellite fraction and red-galaxy fraction}
Based on the red/blue GSMFs from \citetalias{2025MNRAS.540.1635X} and the central/satellite GSMFs from this work (Figure~\ref{fig:fiducial_smf}), it is useful to examine the corresponding red-galaxy fraction $f_{\rm red}$ and satellite fraction $f_{\rm sat}$. These are shown in Figure~\ref{fig:fsat_fred}. The red fraction decreases with stellar mass down to $M_*\sim10^{9.0}\,h^{-2}_{70}\,\ms$, then rises again toward higher masses, approaching unity at both the low- and high-mass ends, with a minimum of $\sim$15\%. By contrast, the satellite fraction increases with stellar mass up to $M_*\sim10^{9.0}\,h^{-2}_{70}\,\ms$, then declines toward higher masses, peaking at $\sim$50\%. Combining these two fractions, we find that red dwarfs and red massive galaxies are predominantly centrals. In the intermediate stellar–mass range ($10^{8.0}\,h^{-2}\ms < M_* < 10^{10.0}\,h^{-2}\ms$), the fact that $f_{\rm sat} > f_{\rm red}$ implies that a substantial fraction of satellites remain blue. A more detailed investigation of environmental and mass quenching can also be done within the PAC framework as in \citet[][SDSS PAC VII]{2025ApJ...984..193Z}, which will be pursued in future work. Separately modelling blue and red galaxies can also provide an additional validation of the galaxy-bias assumptions adopted in \citetalias{2025MNRAS.540.1635X}.

We also compare the satellite fraction derived in this work with previous studies. We find systematically higher satellite fractions across most stellar masses than those reported by \citet{2024ApJ...971..119W}, which are based on a DESI Y1 group catalogue \citep{2021ApJ...909..143Y}. This difference may arise from varying definitions of satellite galaxies in observations and simulations. We further compare our results with the IllustrisTNG and EAGLE hydrodynamic simulations and find good agreement down to $10^{9.0}\,h^{-2}_{70}\,\ms$. For reference, we also show the subhalo fraction at fixed (sub)halo mass in Figure~\ref{fig:fsat_fred}(a). The satellite fraction reflects a mixture of subhalo fractions across halo-mass bins, weighted by the central and satellite SHMRs. Because we find that the red fraction is sensitive to the adopted classification criteria, we do not present detailed comparisons with other studies without a consistent calibration.

The dominance of red galaxies at the low-mass end is notable, and the fact that they are primarily centrals makes the result even more interesting, because the dominant quenching mechanism for such systems is unclear. A characteristic stellar mass is $M_*\simeq10^{7.6}\,h^{-2}_{70}\ms$, where the red population begins to dominate. These galaxies reside in haloes with mean masses near the low-mass characteristic scale $M_2$ of the central SHMR--the scale at which the upturn feature emerges (Figure~\ref{fig:fiducial_shmr}). This halo mass scale also coincides with the onset of reionization suppression in the HOF models (Figure~\ref{fig:HOF}).

Together, these observations motivate a coherent {\it hypothesis}: prior to reionization, star formation in low-mass haloes may have been substantially more efficient than commonly assumed, allowing some small haloes to form relatively massive stellar systems before being quenched by the rising UV background. These quenched remnants would naturally appear as the central red dwarfs dominating the low-mass GSMF today. Such objects do arise in current galaxy formation models, but typically with lower stellar masses \citep{2018ApJ...863..123B}. This scenario offers a unified explanation for both the prevalence of red central dwarfs and the elevated stellar-to-halo mass ratios inferred at the low-mass end. 

This \emph{hypothesis} may also help explain the large discrepancy between the GSMFs derived from PAC and BGS. If the suppression is indeed driven by the Local Void, the number density of early-formed galaxies can vary much more strongly with environment than the late-formed ones, since halo bias is significantly larger at earlier times, as also indicated by halo assembly bias analyses \citep{2005MNRAS.363L..66G,2006ApJ...652...71W,2007ApJ...657..664J}. This picture can further account for the difference in the red fraction: in void regions, few galaxies form early enough to be quenched by reionization, and the population is therefore likely dominated by blue galaxies, in contrast to our results in an average environment. The similarity between the GSMFs from BGS and GAMA and the blue GSMF in \citetalias{2025MNRAS.540.1635X} suggests that these surveys primarily miss the early-formed red population. This may explain the discrepancy between our study and \citet{2026Inpre}, who found that blue galaxies dominate the low-mass end. Because \citet{2026Inpre} used a much shallower sample ($r < 19.5$) than this work ($r \lesssim 22.5$), their results primarily reflect the satellite red fraction in the Local Void. If environmental effects are indeed as important as discussed above, the discrepancy likely arises because the two studies probe different galaxy populations, although differences in methodology may also contribute. More dedicated studies are required to quantify these effects. 

For the discrepancy between central and satellite SHMRs, ultimately, improved measurements and modelling are required to verify it. However, if our \emph{hypothesis} is correct—that these galaxies formed predominantly at very early times—an upturn in the satellite SHMR is not necessarily expected. Because dynamical times are short at high redshift and mergers are often major, the merger time-scales for satellites can be very short. As a result, many early-formed systems may not survive as satellites to $z=0.1$. Whether an upturn appears in the satellite SHMR therefore depends on the survival fraction of these early-formed galaxies. If only a small fraction remain, the satellite SHMR need not exhibit a corresponding upturn.

Based solely on the indications from the PAC analysis, it would be premature to draw firm conclusions. A direct test is needed. We plan to identify available spectroscopy for these central red dwarf galaxies in archival data; if a significant fraction exhibit low metallicities, this would provide much stronger evidence that they formed early, prior to reionization.

\begin{figure}
    \centering
    \includegraphics[width=\columnwidth]{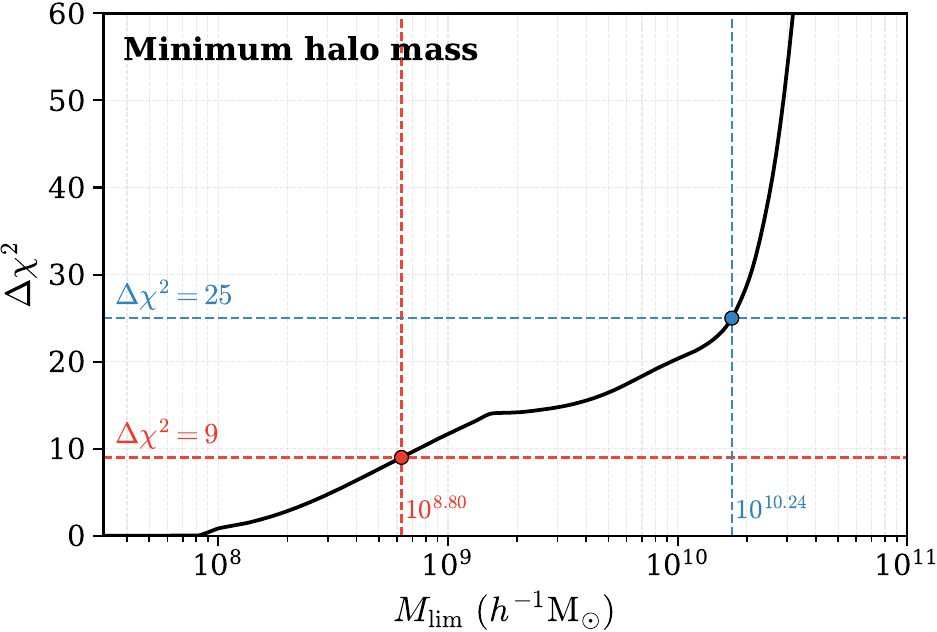}
    \caption{Profile-likelihood constraint on the minimum halo mass. The curve shows the $\Delta\chi^2$ of the best fit to the $\nwprp$ measurements as a function of the imposed minimum halo mass in the Planck18 cosmology at $z\simeq0.1$, relative to the case with no minimum-mass cutoff. At each $M_{\rm lim}$, all other parameters of the fiducial parametric SHAM model are re-optimized. The horizontal dashed lines indicate the Gaussian-equivalent $3\sigma$ and $5\sigma$ thresholds, while the vertical dashed lines mark the corresponding upper limits on $M_{\rm lim}$. }
    \label{fig:dchi2_Mlim}
\end{figure}

\subsection{Minimum halo mass}
As indicated by our SHMR constraints, dark matter haloes with masses down to $\sim10^{8.0}$-$10^{9.0}\,\msh$ are required to host the observed galaxies and reproduce the $\nwprp$ measurements. It is therefore informative to quantify the upper bounds on the minimum halo mass allowed by the data, as this can place constraints on dark matter models \citep{2019Galax...7...81Z}. 

To quantify this, we impose a series of minimum halo masses, $M_{\rm lim}$, in the Jiutian simulation (Planck18 cosmology, $z=0.1$) and refit the fiducial parametric SHAM model at each fixed $M_{\rm lim}$, profiling over all other model parameters. We define $\Delta\chi^2=-2[\ln\mathcal{L}_{\rm prof}(M_{\rm lim})-\ln\mathcal{L}_{\rm prof}(\mathrm{no\ cutoff})]$. Figure~\ref{fig:dchi2_Mlim} shows the resulting profile-likelihood constraint as a function of $M_{\rm lim}$. We find $\Delta\chi^2=9$ and $25$ at $M_{\rm lim}=10^{8.80}\,\msh$ and $10^{10.24}\,\msh$, respectively. Treating $M_{\rm lim}$ as a single effective degree of freedom and adopting the asymptotic Gaussian likelihood-ratio approximation, these correspond to Gaussian-equivalent $3\sigma$ and $5\sigma$ upper limits on the minimum halo mass capable of hosting galaxies. Including the suppression of halo occupation predicted by reionization models would shift the inferred limits toward lower masses; therefore, these values should be interpreted as conservative upper bounds on the minimum halo mass.

In this work, we only use measurements down to $10^{6.4}\,h_{70}^{-2}\,\ms$. As shown in Figure~\ref{fig:compare_smf}, \citetalias{2025MNRAS.540.1635X} also provides lower limits on the number densities extending to $10^{5.0}\,h_{70}^{-2}\,\ms$. Including these constraints would likely further reduce the upper bound on minimum halo mass and will be considered in future work.

\section{Conclusion}\label{sec:con}
In this paper, we measure the excess surface density $\nwprp$ of photometrically selected galaxies from DECaLS around spectroscopically selected galaxies from the DESI Y1 BGS for 349 combinations of photometric and spectroscopic stellar-mass bins using the Photometric Objects Around Cosmic Webs (PAC) method \citep{2011ApJ...734...88W,2022ApJ...925...31X,2025MNRAS.540.1635X}. We model 349 $\nwprp$ measurements using a stellar mass-halo mass (SHMR)-based subhalo abundance matching (SHAM) framework applied to two high-resolution $N$-body simulations, Jiutian-1G ($M_{\rm h}>10^{10.3}\msh$) and Jiutian-300 ($10^{8.7}\msh<M_{\rm h}<10^{10.3}\msh$). The model is further extended to $M_{\rm h}<10^{8.7}\msh$ using an analytic halo mass function and assuming a constant halo bias at low masses. Leveraging the extensive information contained in the PAC data, we constrain the SHMR down to halo masses of order $10^{8.0}\,\msh$ and stellar mass down to $10^{6.4}\,h_{70}^{-2}\ms$. The extended model affects only the modelling of the two lowest stellar-mass bins at $M_*<10^{6.6}\,h_{70}^{-2}\ms$ and does not affect the main results. The main conclusions are summarized below:
\begin{itemize}
    \item We model the mean SHMR for central and satellite galaxies separately, using both non-parametric and parametric approaches. The two methods yield consistent results and both reveal an upturn in the central SHMR at $M_{\rm h}\lesssim10^{10.0}\,\msh$, indicating that the SFE increases toward lower halo masses. 
    \item Allowing the scatter in the SHMR to vary with (sub)halo mass, we constrain the scatter at $M_{\rm h}>10^{11.0}\,\msh$, where it remains relatively constant. At $M_{\rm h}<10^{11.0}\,\msh$, the scatter becomes strongly degenerate with the slope of the mean relation and cannot be robustly determined using PAC alone. Nonetheless, the upturn in the central mean SHMR persists in the varying–scatter model, albeit with larger uncertainties.
    \item After accounting for reionization effects on the HOF of low-mass haloes, we find that with H\,\textsc{i} cooling alone, there are not enough luminous haloes to match the $\nwprp$ measurements in the lowest stellar-mass bin, even under an extreme assumption of galaxy assembly bias. Once assembly bias is included, the \citetalias{2023MNRAS.524.2290N} H$_2$–cooling model yields fits to the PAC data that are comparable though still slightly worse than those of the fiducial fully occupied model.
    \item Although we do not include galaxy assembly bias in our SHAM modelling in this paper, we test its potential impact by considering the most extreme case where, at fixed halo mass, galaxies in earlier–formed haloes strictly have larger stellar masses. Incorporating this bias increases $\nwprp$ for a given SHMR. Fitting a mock with the most extreme assembly bias using a model ignoring assembly bias, we find that the inferred central mean SHMR remains largely unchanged, while the satellite mean SHMR is overestimated and the scatters are underestimated.
    \item The GSMF derived from our SHAM model is generally consistent with that from \citetalias{2025MNRAS.540.1635X}, but with overall values higher by $\sim15\%$. This confirms that the galaxy bias assumption adopted in \citetalias{2025MNRAS.540.1635X} is broadly reasonable. However, it also reveals a $\sim15\%$ internal inconsistency in our SHAM framework: our model cannot simultaneously match both $\n$ and $\Wp(\rp)$ across all stellar–mass bins.
    \item We find that incorporating maximal assembly bias ($\rho=1$) can reduce the discrepancy by only $\sim5\%$, and switching the assumed cosmology from Planck18 to WMAP9 (with $\sim10\%$ lower $\Omega_{\rm m}$) yields a similar $\sim5\%$ reduction. Using the WMAP9 model evaluated at $z=0$ instead of $z=0.1$ removes the discrepancy entirely, effectively corresponding to a cosmology with $\Omega_{\rm m}\simeq0.216$ and $\sigma_8\simeq0.867$. This outcome implies either that the assumed cosmology is incorrect or that a more accurate determination of the effective redshift $z_{\rm eff}$ of the measurements is required.
    \item Combining $f_{\rm sat}(M_*)$ measured in this work with $f_{\rm red}(M_*)$ from \citetalias{2025MNRAS.540.1635X}, we find that the satellite fraction is very low at the low-mass end, implying that the red dwarf galaxies dominating this regime in \citetalias{2025MNRAS.540.1635X} are primarily centrals. Notably, three characteristic scales coincide: red galaxies begin to dominate at $M_*\simeq10^{7.6}\,h^{-2}\ms$ (\citetalias{2025MNRAS.540.1635X}), the impact of reionization on halo occupation becomes significant at $M_{\rm h}\simeq10^{10}\,\msh$ (reionization model in \citet{2025ApJ...983L..23N}), and this same point ($10^{10}\,\msh,10^{7.6}\,h^{-2}\ms$) marks the onset of the upturn in the central mean SHMR (this work). These trends motivate a {\it hypothesis} in which star-formation efficiency is higher than previously thought before reionization, allowing relatively large stellar masses to form in low–mass haloes by the end of reionization. Subsequent quenching by the UV background would then produce the red dwarf population observed today. While such galaxies do appear in current galaxy-formation models, they typically have much lower stellar masses.
    \item Applying a minimum halo–mass cutoff to our fiducial SHMR model, we find $\Delta\chi^2=9$ and $25$ at $M_{\rm lim}=10^{8.80}\,\msh$ and $10^{10.24}\,\msh$, respectively, corresponding to $3\sigma$ and $5\sigma$ upper limits on the minimum mass of dark matter haloes under a Gaussian likelihood assumption.    
\end{itemize}

As we are now exploring a previously inaccessible regime of the galaxy–halo connection, our results provide new insight into several open problems, including the nature of dark matter, the physics of reionization, and the formation and quenching of dwarf galaxies.

Although significant progress has been achieved with DESI and DECaLS using the PAC method, there is clear room for improvement on both the observational and modelling sides. Observationally, DECaLS is not very deep; Stage-IV imaging surveys such as LSST \citep{2019ApJ...873..111I}, \emph{Euclid} \citep{2011arXiv1110.3193L}, CSS-OS \citep{2019ApJ...883..203G}, and Roman \citep{2015arXiv150303757S} are ongoing or forthcoming, and are all at least $\sim3$ magnitudes deeper than DECaLS. These surveys will enable PAC measurements to extend to much lower stellar masses, and higher-order PAC observables (e.g.\ three-point correlations) can provide additional constraints to break the degeneracy between the low-mass SHMR slope and scatter. On the modelling side, developing efficient compression schemes for the $\Wp$ tables that incorporate secondary halo properties will allow a more general treatment of galaxy assembly bias.

\section*{Acknowledgements}
We thank Jiaxin Han for helpful discussions regarding the Jiutian subhalo catalogues. We are grateful to Xiaolin Luo for assistance with downloading the Jiutian data. We thank Zheng Zheng, Jeremy Tinker and Siyi Xu for their valuable comments and great help during the DESI Collaboration Wide Review. We thank the anonymous referees for their helpful comments, which improved the analysis.

K.X. is supported by the funding from the Center for Particle Cosmology at U Penn. Y.P.J. is supported by NSFC (12595311,12133006), by National Key R\&D Program of China (2023YFA1607801), and by 111 project No. B20019. S.B. is supported by the UK Research and Innovation (UKRI) Future Leaders Fellowship [grant number MR/V023381/1]. This work made use of the Gravity Supercomputer at the Department of Astronomy, Shanghai Jiao Tong University. This work used the DiRAC@Durham facility managed by the Institute for Computational Cosmology on behalf of the STFC DiRAC HPC Facility (www.dirac.ac.uk). The equipment was funded by BEIS capital funding via STFC capital grants ST/K00042X/1, ST/P002293/1, ST/R002371/1 and ST/S002502/1, Durham University and STFC operations grant ST/R000832/1. DiRAC is part of the National e-Infrastructure. 

This material is based upon work supported by the U.S. Department of Energy (DOE), Office of Science, Office of High-Energy Physics, under Contract No. DE–AC02–05CH11231, and by the National Energy Research Scientific Computing Center, a DOE Office of Science User Facility under the same contract. Additional support for DESI was provided by the U.S. National Science Foundation (NSF), Division of Astronomical Sciences under Contract No. AST-0950945 to the NSF’s National Optical-Infrared Astronomy Research Laboratory; the Science and Technology Facilities Council of the United Kingdom; the Gordon and Betty Moore Foundation; the Heising-Simons Foundation; the French Alternative Energies and Atomic Energy Commission (CEA); the National Council of Humanities, Science and Technology of Mexico (CONAHCYT); the Ministry of Science, Innovation and Universities of Spain (MICIU/AEI/10.13039/501100011033), and by the DESI Member Institutions: \url{https://www.desi.lbl.gov/collaborating-institutions}.

The DESI Legacy Imaging Surveys consist of three individual and complementary projects: the Dark Energy Camera Legacy Survey (DECaLS), the Beijing-Arizona Sky Survey (BASS), and the Mayall z-band Legacy Survey (MzLS). DECaLS, BASS and MzLS together include data obtained, respectively, at the Blanco telescope, Cerro Tololo Inter-American Observatory, NSF’s NOIRLab; the Bok telescope, Steward Observatory, University of Arizona; and the Mayall telescope, Kitt Peak National Observatory, NOIRLab. NOIRLab is operated by the Association of Universities for Research in Astronomy (AURA) under a cooperative agreement with the National Science Foundation. Pipeline processing and analyses of the data were supported by NOIRLab and the Lawrence Berkeley National Laboratory. Legacy Surveys also uses data products from the Near-Earth Object Wide-field Infrared Survey Explorer (NEOWISE), a project of the Jet Propulsion Laboratory/California Institute of Technology, funded by the National Aeronautics and Space Administration. Legacy Surveys was supported by: the Director, Office of Science, Office of High Energy Physics of the U.S. Department of Energy; the National Energy Research Scientific Computing Center, a DOE Office of Science User Facility; the U.S. National Science Foundation, Division of Astronomical Sciences; the National Astronomical Observatories of China, the Chinese Academy of Sciences and the Chinese National Natural Science Foundation. LBNL is managed by the Regents of the University of California under contract to the U.S. Department of Energy. The complete acknowledgments can be found at \url{https://www.legacysurvey.org/}.

Any opinions, findings, and conclusions or recommendations expressed in this material are those of the author(s) and do not necessarily reflect the views of the U. S. National Science Foundation, the U. S. Department of Energy, or any of the listed funding agencies.

The authors are honored to be permitted to conduct scientific research on I'oligam Du'ag (Kitt Peak), a mountain with particular significance to the Tohono O’odham Nation.

\section*{Data Availability}
All raw data used in this paper are publicly available. Data points for each figure are available in a machine readable form: \url{https://doi.org/10.5281/zenodo.18943406}. Other data will be shared upon reasonable request to the corresponding author.



\bibliographystyle{mnras}
\bibliography{example} 




\appendix

\section{Validating the extrapolation of halo abundance and clustering}\label{sec:extra}

\begin{figure}
    \centering
    \includegraphics[width=\columnwidth]{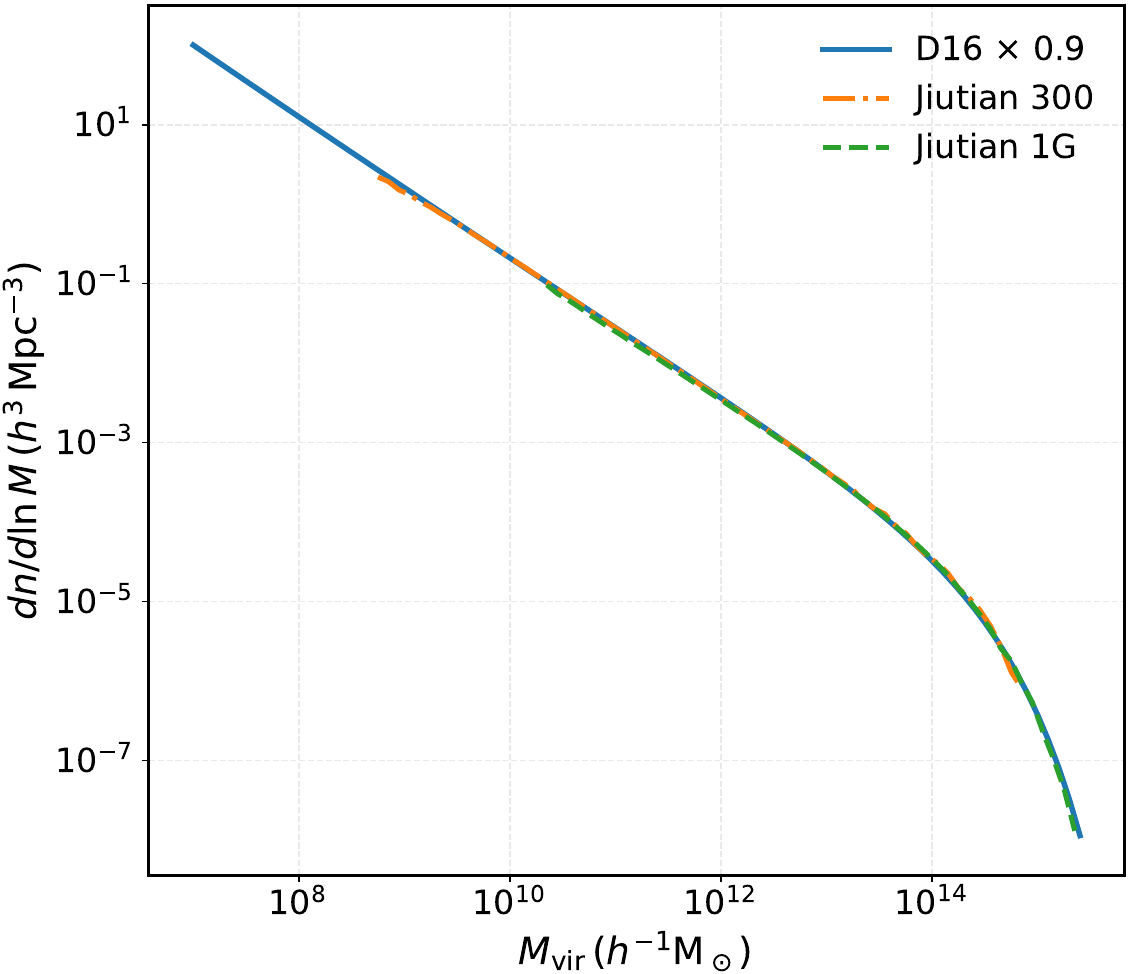}
    \caption{Comparison of the HMFs from Jiutian-1G and Jiutian-300 with those from a modified \citet{2016MNRAS.456.2486D} model, rescaled by a factor of 0.9 to achieve the best match to the Jiutian results.}
    \label{fig:HMF}
\end{figure}

\begin{figure*}
    \centering
    \includegraphics[width=\textwidth]{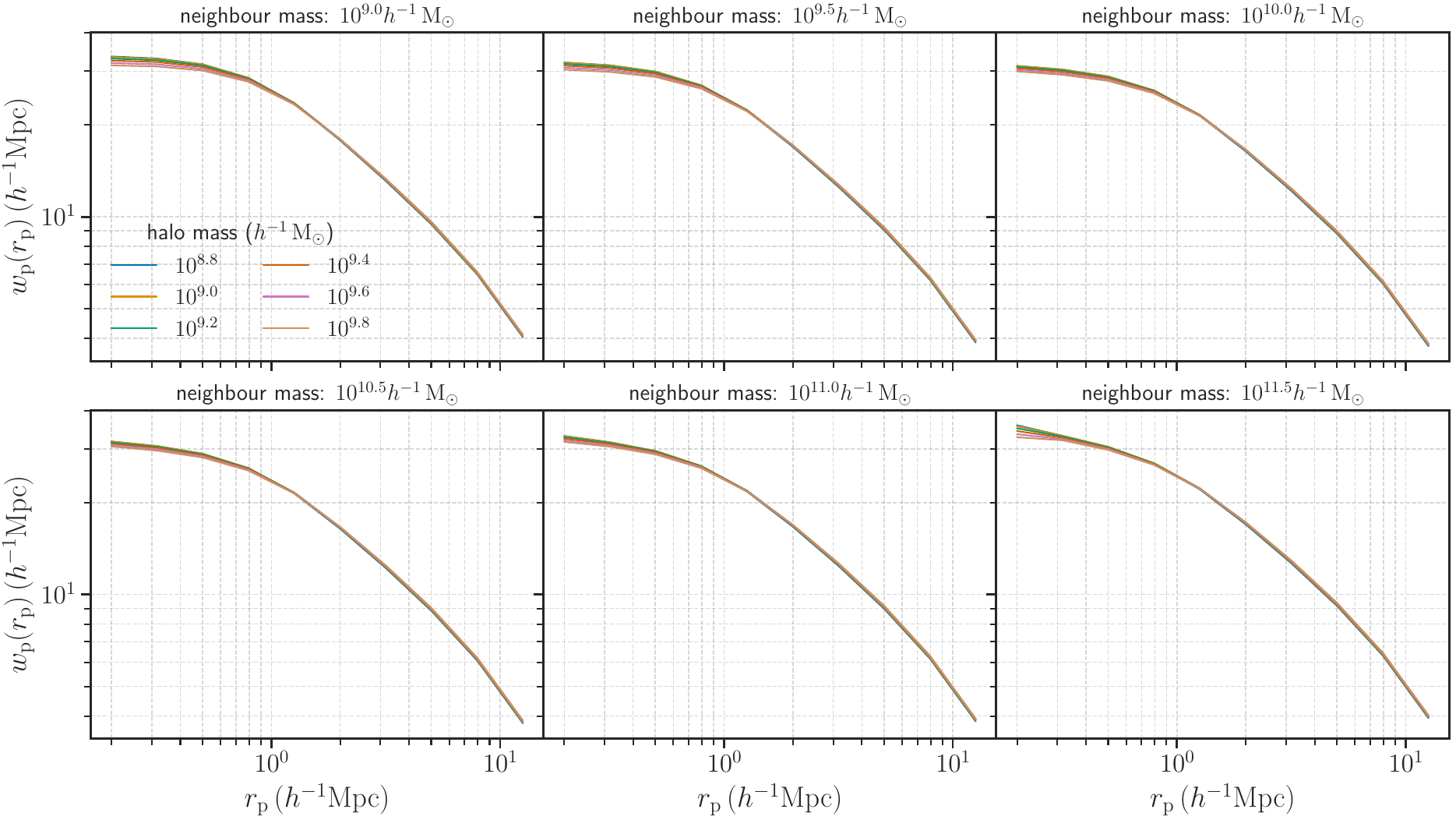}
    \caption{Comparison of $w_{\rm p}(r_{\rm p})$ in six low-mass halo bins within $M_{\rm vir}\in[10^{8.7},10^{9.9}]\,\msh$, used to validate the constant-bias assumption at lower masses. Results are shown for cross-correlations with neighbouring (sub)haloes at six different masses. This tests the combination of $w_{\rm p}^{\rm hh}(r_{\rm p})$ and $w_{\rm p}^{\rm hs}(r_{\rm p})$.}
    \label{fig:wp_all}
\end{figure*}

\begin{figure*}
    \centering
    \includegraphics[width=\textwidth]{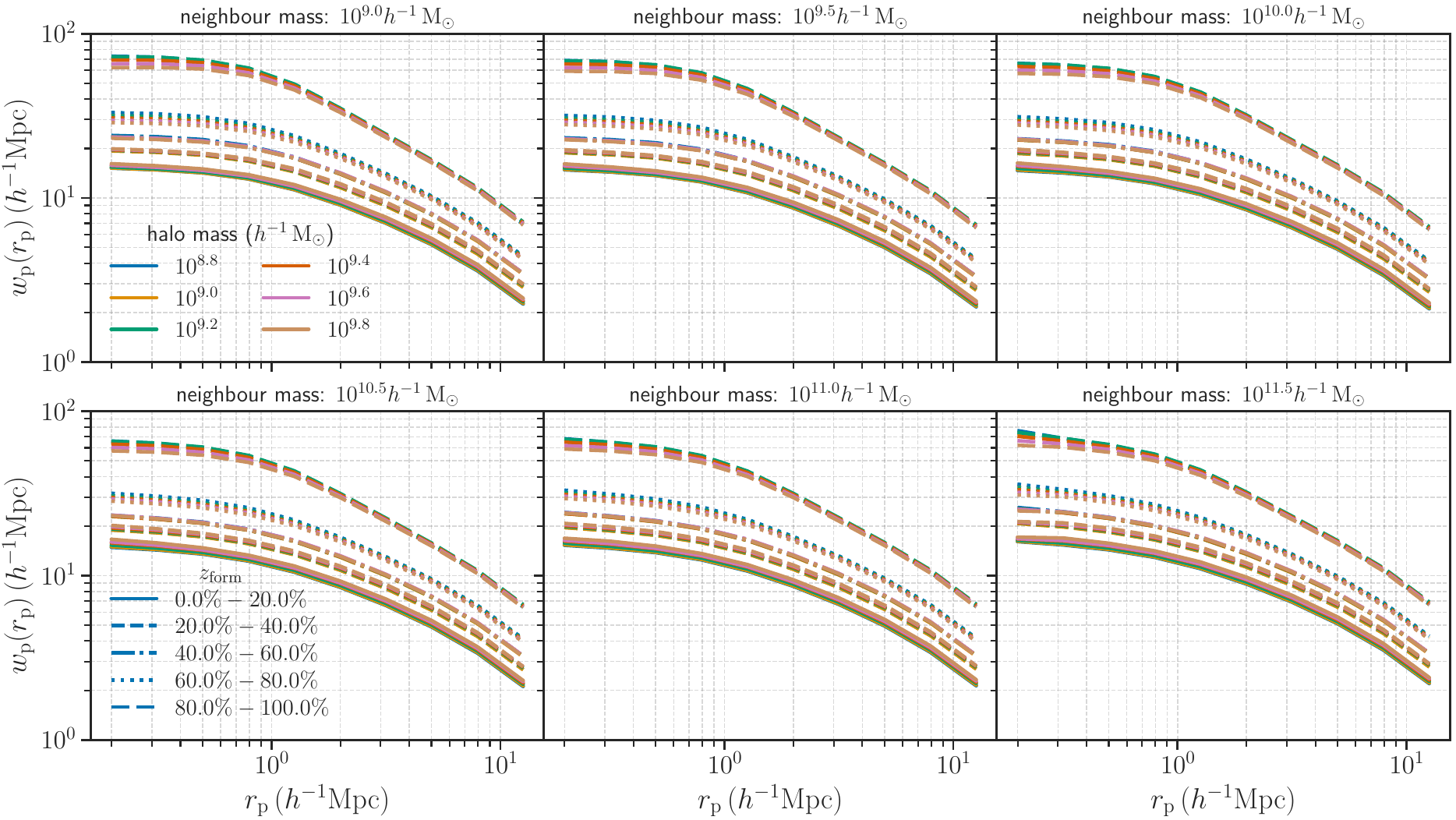}
    \caption{Comparison of $w_{\rm p}(r_{\rm p})$ in six low-mass halo bins within $M_{\rm vir}\in[10^{8.7},10^{9.9}]\,\msh$ and five $z_{\rm form}$ fraction bins. Results are shown for cross-correlations with neighbouring (sub)haloes at six different masses.}
    \label{fig:wp_ab}
\end{figure*}

In this section, we validate the extrapolations of halo abundance and clustering used in the SHAM modelling.

In Figure~\ref{fig:HMF}, we compare the HMFs from Jiutian-1G and Jiutian-300 with those from a modified \citet{2016MNRAS.456.2486D} model, rescaled by a factor of 0.9 to best match the Jiutian results. After rescaling, the \citet{2016MNRAS.456.2486D} model agrees well with the Jiutian simulations, supporting its use for extrapolating the halo abundance below $M_{\rm vir}<10^{8.7}\,\msh$. A systematic mismatch of order 10\% is common when comparing HMFs from different simulations and fitting functions \citep{2023A&A...671A.100E}, primarily due to differences in halo finders.

In Figure~\ref{fig:wp_all}, we test the extrapolation of $\Wp^{\rm hh}(r_{\rm p})$ and $\Wp^{\rm hs}(r_{\rm p})$ under the assumption of an approximately constant low-mass halo bias. We compare $\Wp(r_{\rm p})$ for haloes in the mass range $[10^{8.7},10^{9.9}]\,\msh$ with their neighbouring haloes and subhaloes at fixed masses. We find that $\Wp(r_{\rm p})$ changes very little across halo masses in this range, and the conclusion holds for different choices of neighbouring objects. We therefore expect the same behaviour to extend to lower masses, and the extrapolation of $\Wp^{\rm hh}(r_{\rm p})$ and $\Wp^{\rm hs}(r_{\rm p})$ below $M_{\rm vir}<10^{8.7}\,\msh$ using measurements from haloes with $M_{\rm vir}\in[10^{8.7},10^{9.5}]\,\msh$ is reasonable.

In Figure~\ref{fig:wp_ab}, we extend the test in Figure~\ref{fig:wp_all} to examine the extrapolation of $\Wp^{\rm hh}(r_{\rm p})$ and $\Wp^{\rm hs}(r_{\rm p})$ toward lower halo masses at fixed $z_{\rm form}$ percentile. At a given $z_{\rm form}$ percentile, $\Wp(r_{\rm p})$ varies by only $\sim10\%$ across halo masses within $[10^{8.7},10^{9.9}]\,\msh$. Therefore, extrapolating $\Wp(r_{\rm p})$ to lower masses at the same $z_{\rm form}$ percentile is justified.

\section{Additional photometric artificial test}\label{sec:fracflux}
\begin{figure}
    \centering
    \includegraphics[width=\linewidth]{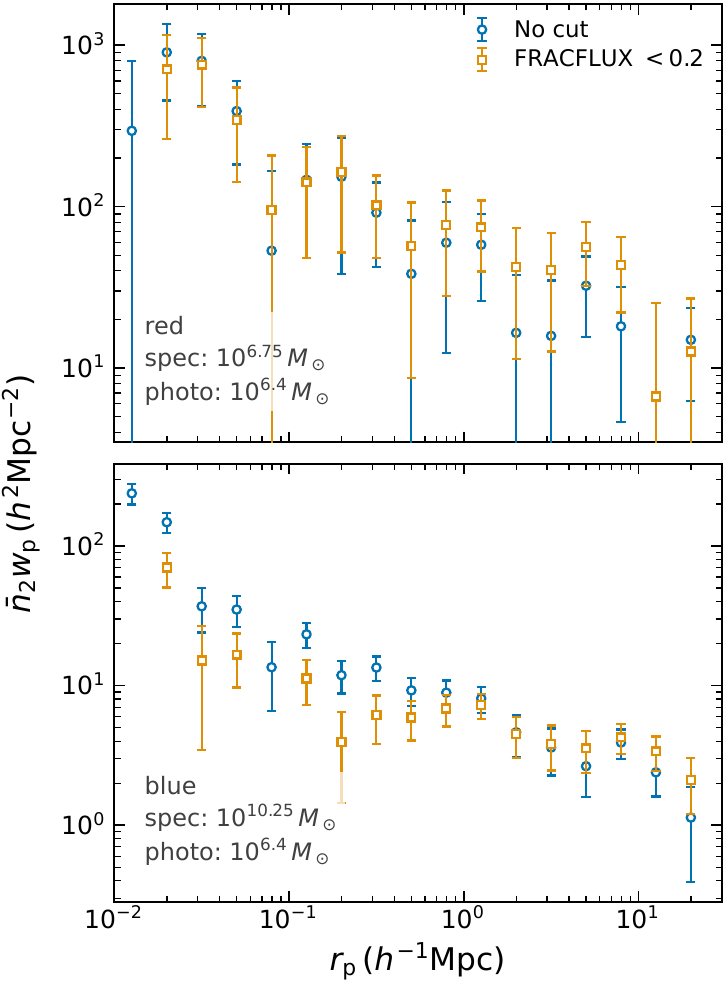}
    \caption{
    Comparison of $\bar{n}_2 w_{\rm p}$ measurements with and without a conservative {\tt FRACFLUX} cut. 
    The upper and lower panels show red and blue DECaLS galaxies, respectively, for the indicated spectroscopic and photometric stellar-mass bins. 
    Open circles show the fiducial measurements without the {\tt FRACFLUX} cut, while open squares show the results after applying ${\tt FRACFLUX}<0.2$. 
    The red-galaxy measurements are nearly unchanged, while the blue-galaxy measurements show differences mainly on small scales. 
    The agreement on larger scales indicates that the inferred $\bar{n}_2$ is not significantly affected by photometric artifacts or by the {\tt FRACFLUX} selection.}
    \label{fig:fracflux}
\end{figure}

In Figure~\ref{fig:fracflux}, we perform an additional test for photometric artifacts using {\tt FRACFLUX}. We compare the fiducial measurements with the results after applying ${\tt FRACFLUX}<0.2$ for the lowest-mass ($M_*=10^{6.4}\ms$) red and blue DECaLS samples. For red galaxies, there is almost no difference between the two measurements. For blue galaxies, some differences appear on small scales, which may be caused either by residual contamination or by a change in the nonlinear bias of real galaxies induced by the {\tt FRACFLUX} selection. However, the two measurements are consistent on larger scales, indicating that the change in $\bar{n}_2$ is small, since low-mass galaxies should have similar $w_{\rm p}$ on linear scales.

\section{convergence diagnostics of the fiducial non-parametric model}\label{sec:diagnostics}

\begin{figure}
    \centering
    \includegraphics[width=\linewidth]{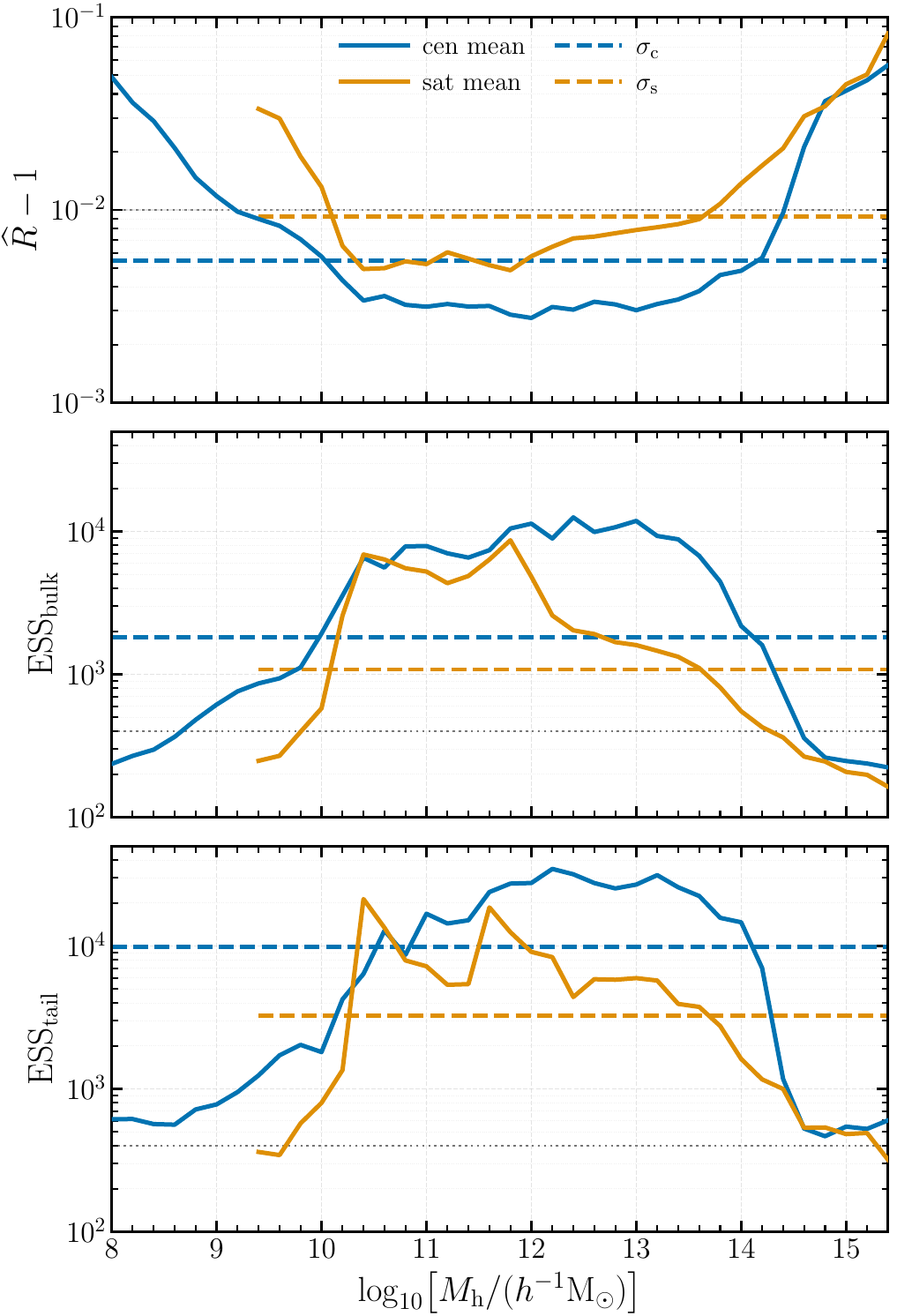}
    \caption{Convergence diagnostics for the fiducial non-parametric SHMR model. From top to bottom, the panels show the rank-normalized split $\hat{R}-1$, bulk effective sample size, and tail effective sample size for the central- and satellite-galaxy mean relations and their scatter parameters. Solid curves denote the parameters describing the mean SHMRs, while dashed curves denote $\sigma_{\rm c}$ and $\sigma_{\rm s}$. The horizontal dotted lines indicate $\hat{R}-1=0.01$ in the upper panel and an effective sample size of 400 in the lower two panels.}
    \label{fig:diagnostics}
\end{figure}

To assess the convergence of the fiducial non-parametric model, Figure~\ref{fig:diagnostics} shows the rank-normalized split $\hat{R}-1$, bulk effective sample size, and tail effective sample size for the parameters describing the central- and satellite-galaxy mean relations and their scatters.

\section{PAC measurements and fiducial model fits}\label{sec:fits}
\begin{figure*}
    \centering
    \includegraphics[width=\textwidth]{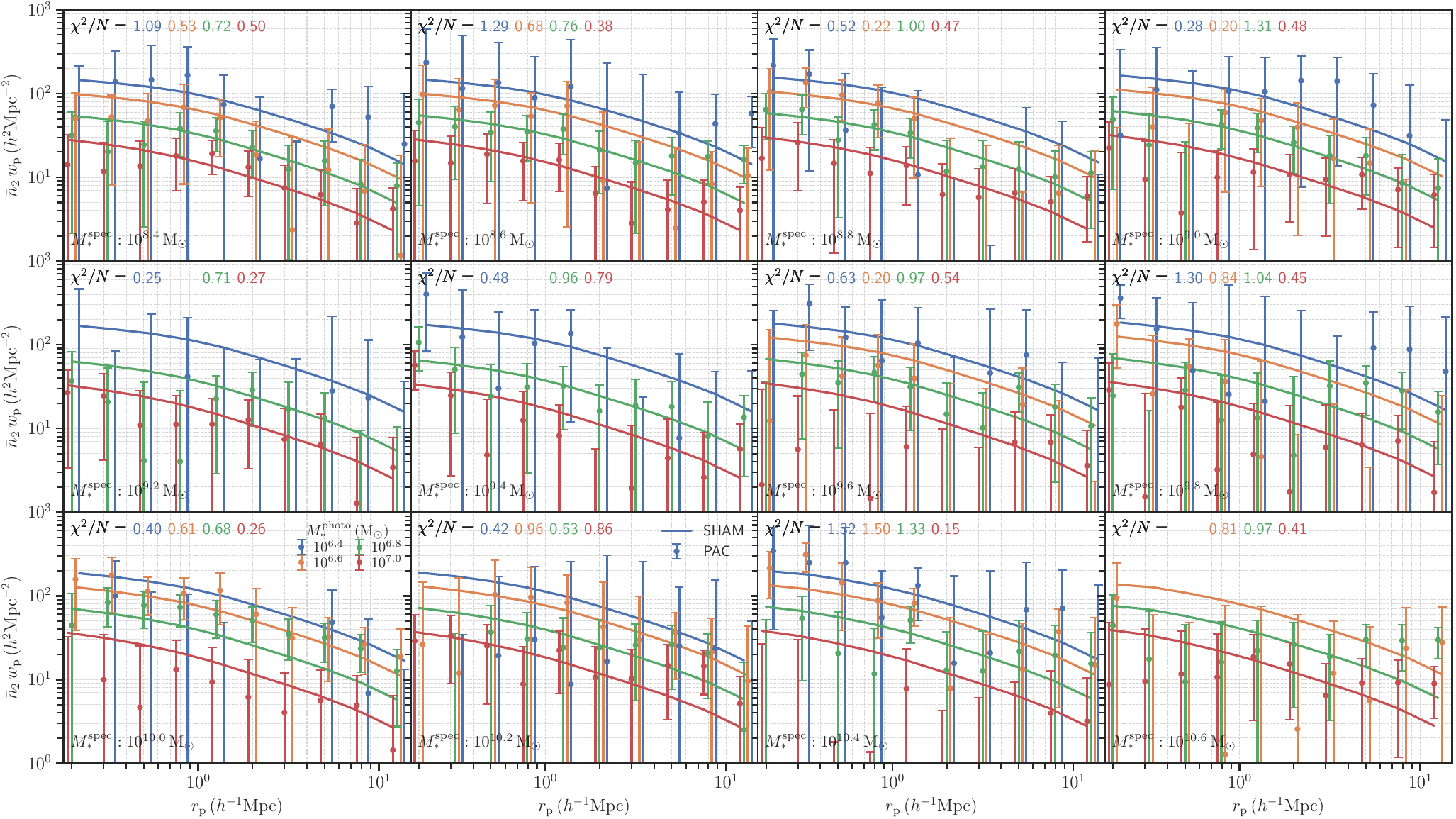}
    \caption{PAC measurements and best-fit predictions from the non-parametric fiducial model. Each panel shows the results for one $M_*^{\rm spec}$ bin and four $M_*^{\rm photo}$ bins. Data points are slightly shifted horizontally by factors of 1.10, 1.05, 1.00, and 0.95 for clarity. This figure illustrates the case with $M_*^{\rm photo}\in[10^{6.3},10^{7.1}]\,\ms$. Reduced $\chi^2$ values for each measurement are listed.}
    \label{fig:fits1}
\end{figure*}

\begin{figure*}
    \centering
    \includegraphics[width=\textwidth]{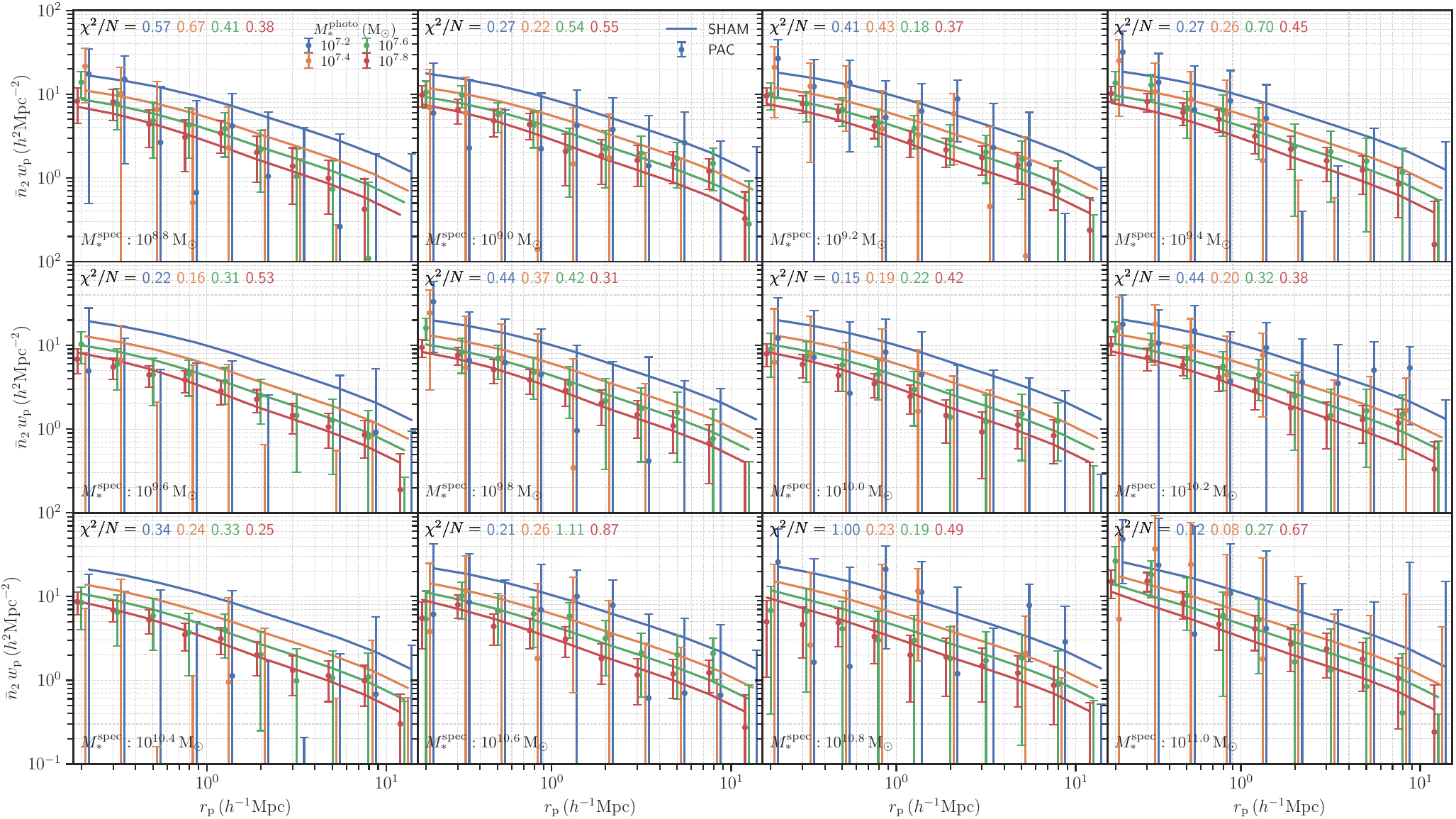}
    \caption{Same as Figure~\ref{fig:fits1}, but for the case with $M_*^{\rm photo}\in[10^{7.1},10^{7.9}]\,\ms$.}
    \label{fig:fits2}
\end{figure*}

\begin{figure*}
    \centering
    \includegraphics[width=\textwidth]{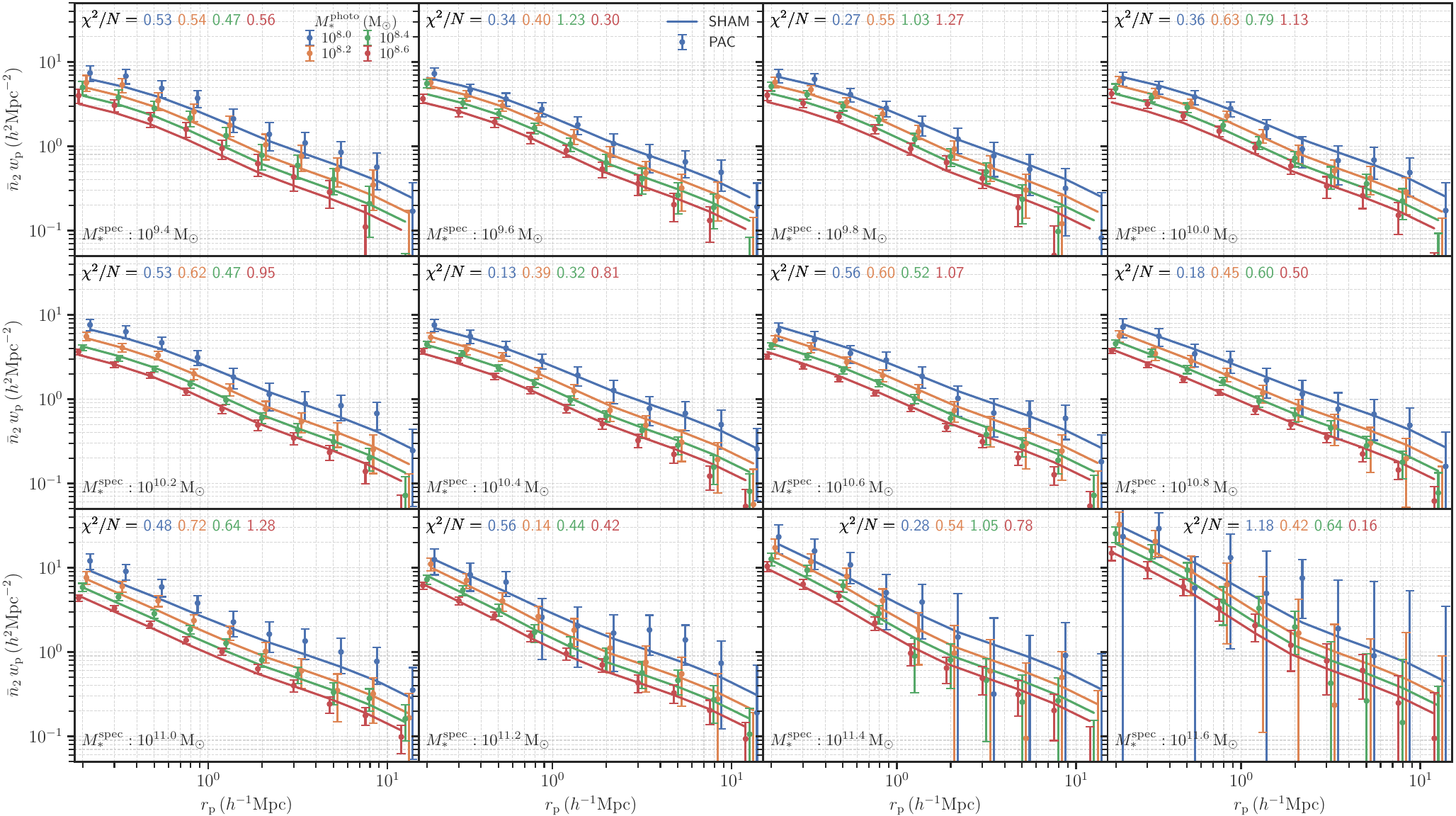}
    \caption{Same as Figure~\ref{fig:fits1}, but for the case with $M_*^{\rm photo}\in[10^{7.9},10^{8.7}]\,\ms$.}
    \label{fig:fits3}
\end{figure*}

\begin{figure*}
    \centering
    \includegraphics[width=\textwidth]{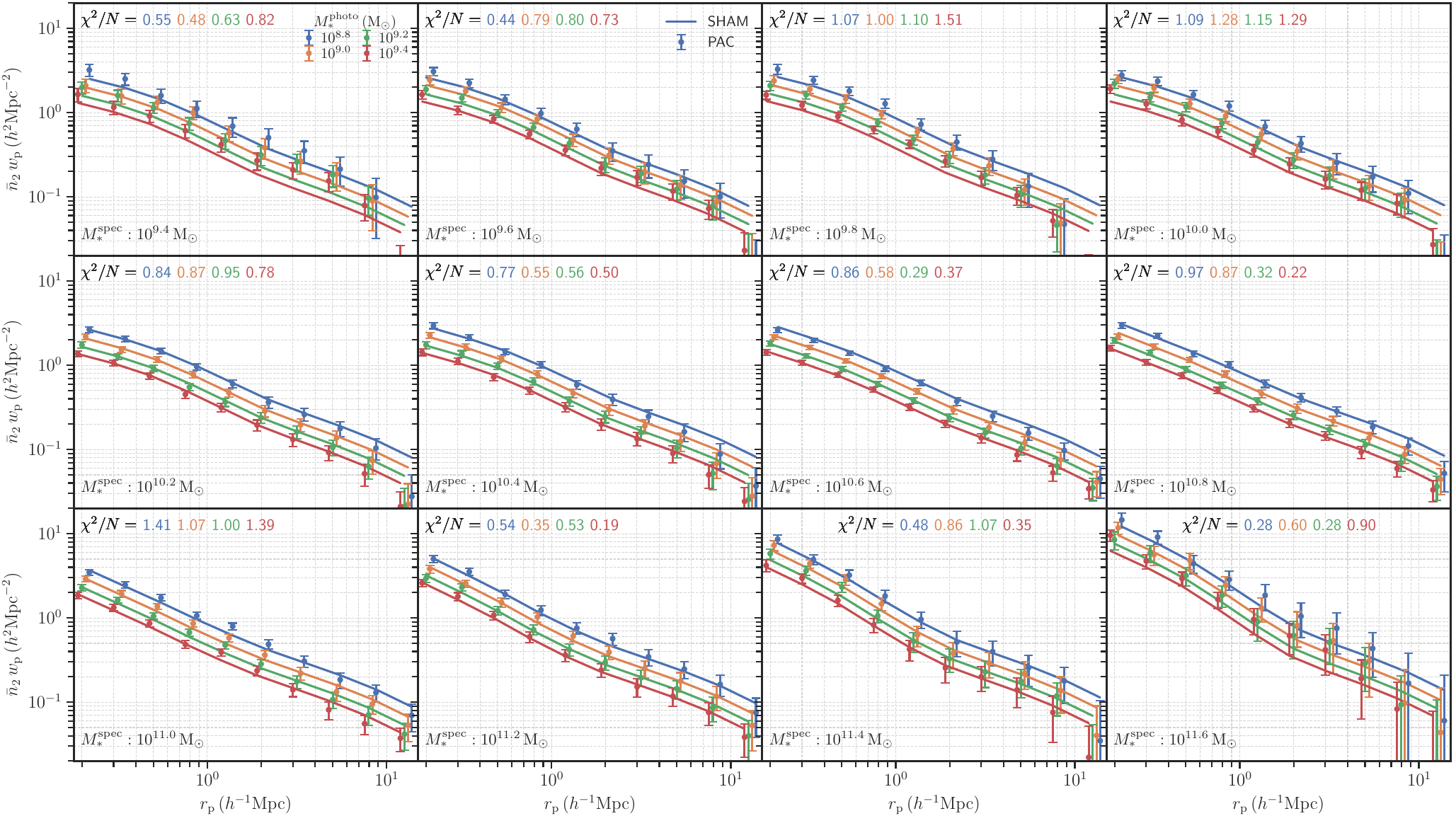}
    \caption{Same as Figure~\ref{fig:fits1}, but for the case with $M_*^{\rm photo}\in[10^{8.7},10^{9.5}]\,\ms$.}
    \label{fig:fits4}
\end{figure*}

\begin{figure*}
    \centering
    \includegraphics[width=\textwidth]{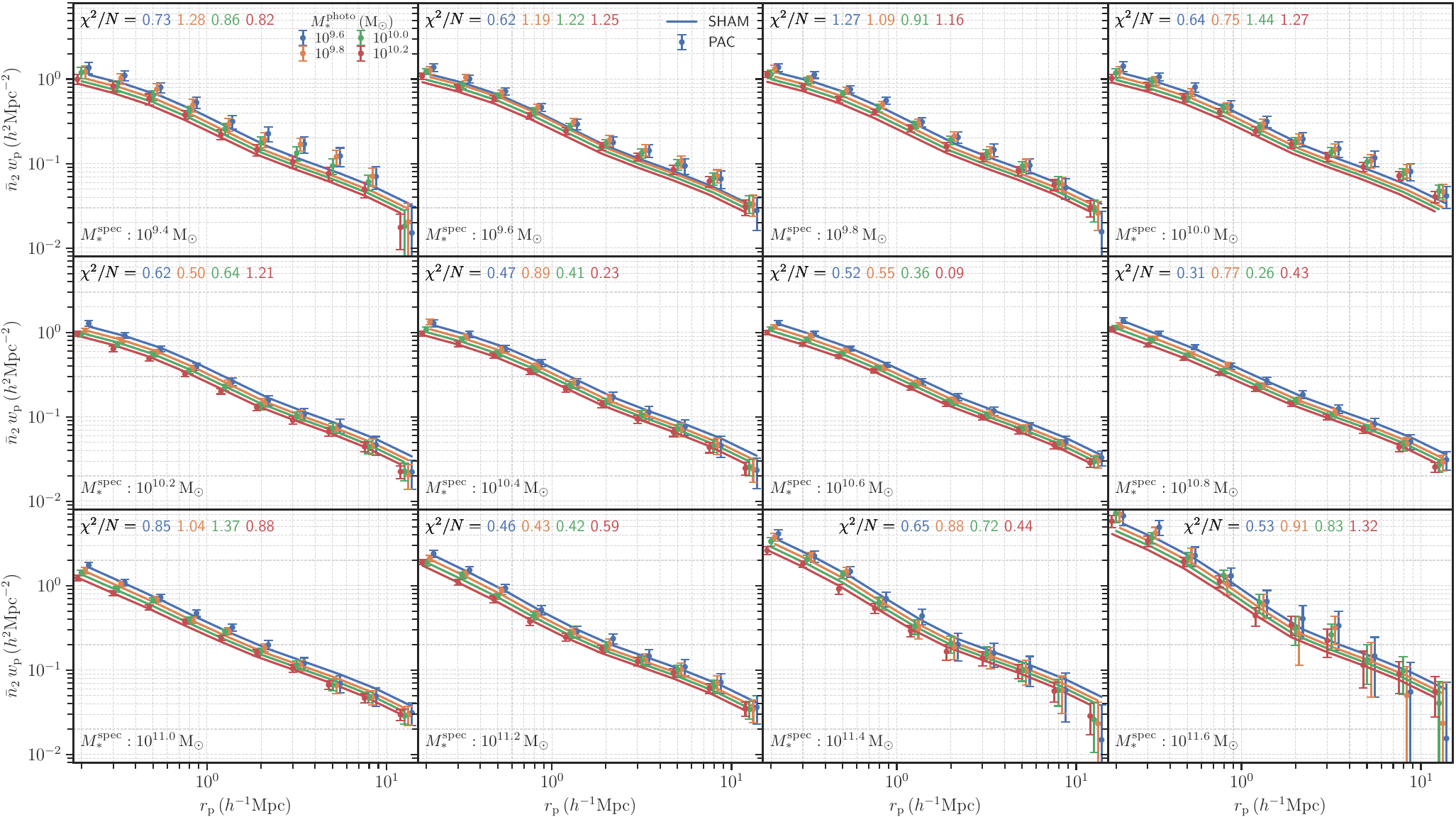}
    \caption{Same as Figure~\ref{fig:fits1}, but for the case with $M_*^{\rm photo}\in[10^{9.5},10^{10.3}]\,\ms$.}
    \label{fig:fits5}
\end{figure*}

\begin{figure*}
    \centering
    \includegraphics[width=\textwidth]{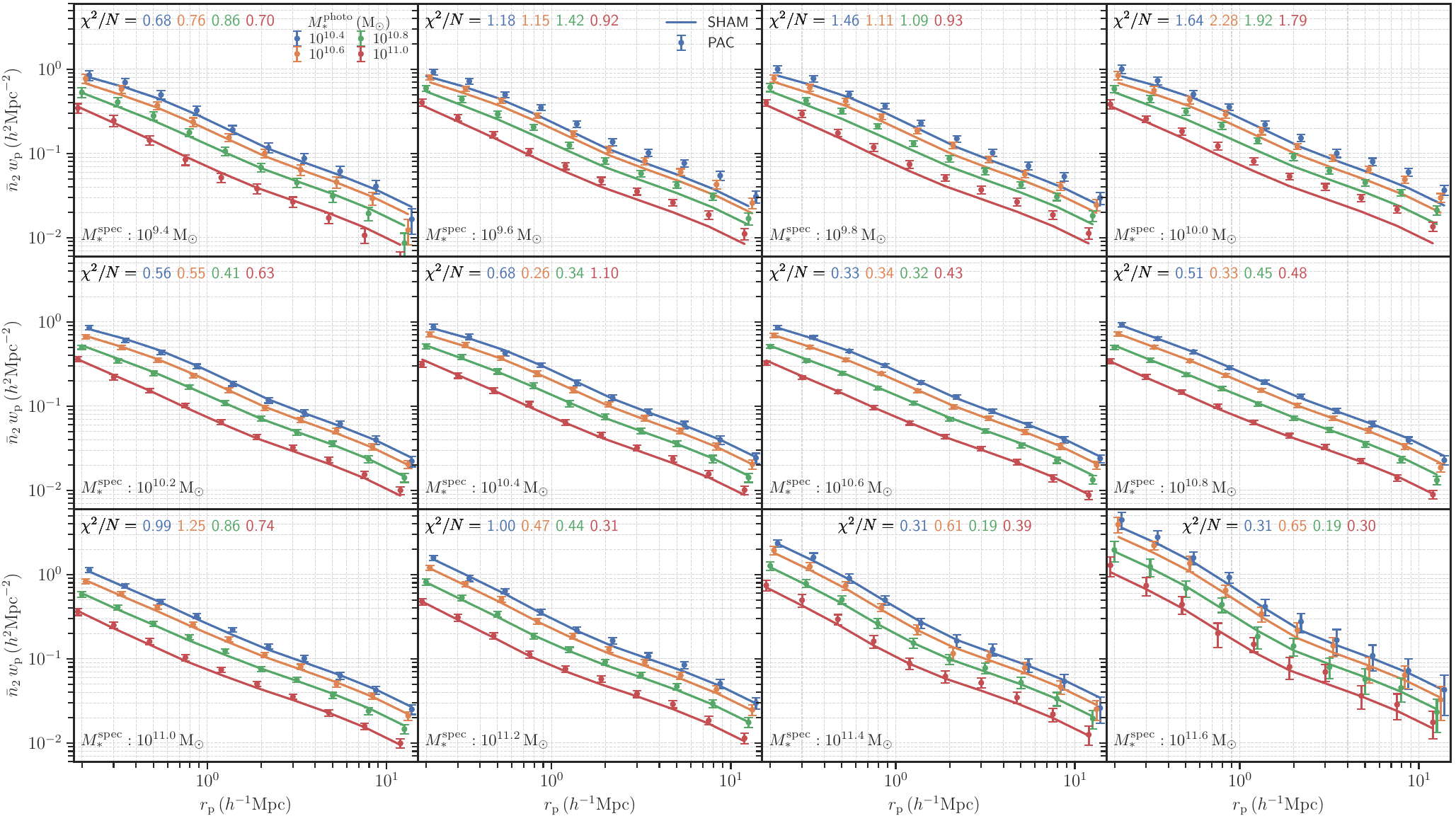}
    \caption{Same as Figure~\ref{fig:fits1}, but for the case with $M_*^{\rm photo}\in[10^{10.3},10^{11.1}]\,\ms$.}
    \label{fig:fits6}
\end{figure*}

\begin{figure*}
    \centering
    \includegraphics[width=\textwidth]{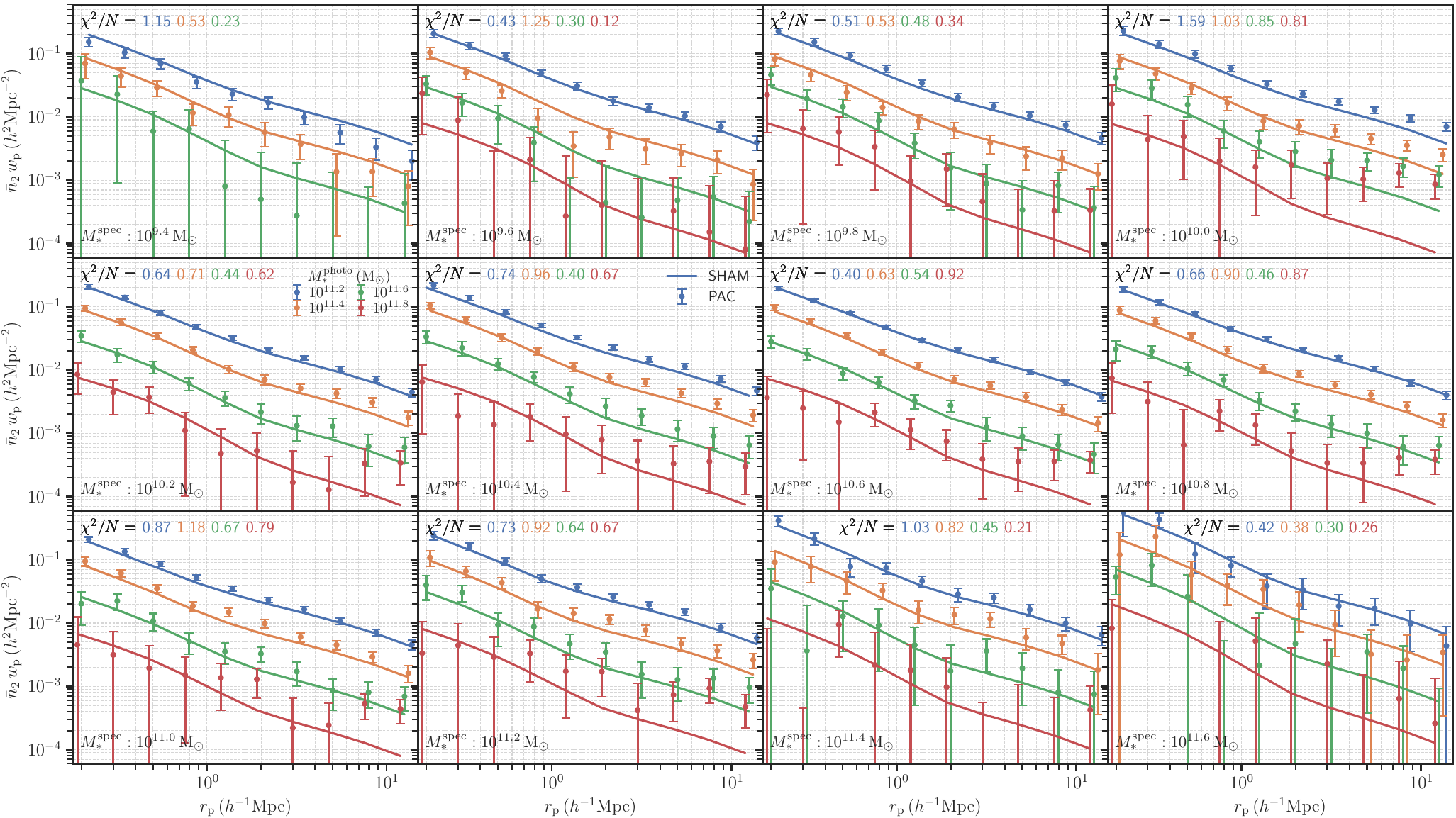}
    \caption{Same as Figure~\ref{fig:fits1}, but for the case with $M_*^{\rm photo}\in[10^{11.1},10^{11.9}]\,\ms$.}
    \label{fig:fits7}
\end{figure*}

In Figures~\ref{fig:fits1}–\ref{fig:fits7}, we present the full set of $\sim349$ $\nwprp$ measurements together with the best-fit predictions from the non-parametric fiducial model.

\section{Author Affiliations}\label{sec:author}
\scriptsize
$^{1}$ Center for Particle Cosmology, Department of Physics and Astronomy, University of Pennsylvania, Philadelphia, PA 19104, USA\\
$^{2}$ Institute for Computational Cosmology, Department of Physics, Durham University, South Road, Durham DH1 3LE, UK\\
$^{3}$ State Key Laboratory of Dark Matter Physics, Tsung-Dao Lee Institute \& School of Physics and Astronomy, Shanghai Jiao Tong University, Shanghai 201210, P.R. China\\
$^{4}$ Lawrence Berkeley National Laboratory, 1 Cyclotron Road, Berkeley, CA 94720, USA\\
$^{5}$ Department of Physics, Boston University, 590 Commonwealth Avenue, Boston, MA 02215, USA\\
$^{6}$ Dipartimento di Fisica ``Aldo Pontremoli'', Universit\`a degli Studi di Milano, Via Celoria 16, I-20133 Milano, Italy\\
$^{7}$ INAF-Osservatorio Astronomico di Brera, Via Brera 28, 20122 Milano, Italy\\
$^{8}$ Department of Physics \& Astronomy, University College London, Gower Street, London, WC1E 6BT, UK\\
$^{9}$ Institut d'Estudis Espacials de Catalunya (IEEC), c/ Esteve Terradas 1, Edifici RDIT, Campus PMT-UPC, 08860 Castelldefels, Spain\\
$^{10}$ Institute of Space Sciences, ICE-CSIC, Campus UAB, Carrer de Can Magrans s/n, 08913 Bellaterra, Barcelona, Spain\\
$^{11}$ Instituto de F\'{\i}sica, Universidad Nacional Aut\'{o}noma de M\'{e}xico, Circuito de la Investigaci\'{o}n Cient\'{\i}fica, Ciudad Universitaria, Cd. de M\'{e}xico C.~P.~04510, M\'{e}xico\\
$^{12}$ Departamento de F\'isica, Universidad de los Andes, Cra. 1 No. 18A-10, Edificio Ip, CP 111711, Bogot\'a, Colombia\\
$^{13}$ Observatorio Astron\'omico, Universidad de los Andes, Cra. 1 No. 18A-10, Edificio H, CP 111711 Bogot\'a, Colombia\\
$^{14}$ Institute of Cosmology and Gravitation, University of Portsmouth, Dennis Sciama Building, Portsmouth, PO1 3FX, UK\\
$^{15}$ University of Virginia, Department of Astronomy, Charlottesville, VA 22904, USA\\
$^{16}$ Fermi National Accelerator Laboratory, PO Box 500, Batavia, IL 60510, USA\\
$^{17}$ Department of Astronomy, The University of Texas at Austin, 2515 Speedway, Stop C1400, Austin, TX 78712, USA\\
$^{18}$ NSF NOIRLab, 950 N. Cherry Ave., Tucson, AZ 85719, USA\\
$^{19}$ Department of Physics, Southern Methodist University, 3215 Daniel Avenue, Dallas, TX 75275, USA\\
$^{20}$ Sorbonne Universit\'{e}, CNRS/IN2P3, Laboratoire de Physique Nucl\'{e}aire et de Hautes Energies (LPNHE), FR-75005 Paris, France\\
$^{21}$ Departament de F\'{i}sica, Serra H\'{u}nter, Universitat Aut\`{o}noma de Barcelona, 08193 Bellaterra (Barcelona), Spain\\
$^{22}$ Institut de F\'{i}sica d’Altes Energies (IFAE), The Barcelona Institute of Science and Technology, Edifici Cn, Campus UAB, 08193, Bellaterra (Barcelona), Spain\\
$^{23}$ Instituci\'{o} Catalana de Recerca i Estudis Avan\c{c}ats, Passeig de Llu\'{\i}s Companys, 23, 08010 Barcelona, Spain\\
$^{24}$ Department of Physics and Astronomy, Siena College, 515 Loudon Road, Loudonville, NY 12211, USA\\
$^{25}$ Department of Physics and Astronomy, University of Waterloo, 200 University Ave W, Waterloo, ON N2L 3G1, Canada\\
$^{26}$ Perimeter Institute for Theoretical Physics, 31 Caroline St. North, Waterloo, ON N2L 2Y5, Canada\\
$^{27}$ Waterloo Centre for Astrophysics, University of Waterloo, 200 University Ave W, Waterloo, ON N2L 3G1, Canada\\
$^{28}$ Instituto de Astrof\'{i}sica de Andaluc\'{i}a (CSIC), Glorieta de la Astronom\'{i}a, s/n, E-18008 Granada, Spain\\
$^{29}$ Departament de F\'isica, EEBE, Universitat Polit\`ecnica de Catalunya, c/Eduard Maristany 10, 08930 Barcelona, Spain\\
$^{30}$ Department of Physics and Astronomy, Sejong University, 209 Neungdong-ro, Gwangjin-gu, Seoul 05006, Republic of Korea\\
$^{31}$ Abastumani Astrophysical Observatory, Tbilisi, GE-0179, Georgia\\
$^{32}$ Department of Physics, Kansas State University, 116 Cardwell Hall, Manhattan, KS 66506, USA\\
$^{33}$ Faculty of Natural Sciences and Medicine, Ilia State University, 0194 Tbilisi, Georgia\\
$^{34}$ CIEMAT, Avenida Complutense 40, E-28040 Madrid, Spain\\
$^{35}$ University of Michigan, 500 S. State Street, Ann Arbor, MI 48109, USA\\
$^{36}$ National Astronomical Observatories, Chinese Academy of Sciences, A20 Datun Road, Chaoyang District, Beijing, 100101, P.~R.~China\\


\bsp	
\label{lastpage}
\end{document}